\documentclass[12pt]{article}
\usepackage{axodraw}
\usepackage{epsf}
\usepackage{rotating}


\begin{document}

\def\thefootnote{\fnsymbol{footnote}}

\vspace*{-1.5cm}

\noindent
DESY T-00-01 \\
MZ-TH/00--53 \\ 
PSI-PR-00-17 \\

\vspace*{1.5cm}

\begin{center}

{\large\sc {\bf THE STANDARD MODEL: \\ 
PHYSICAL BASIS AND SCATTERING EXPERIMENTS}}  

\vspace{1cm}

{\sc H.~Spiesberger$^1$, M.~Spira$^2$ and P.M.~Zerwas$^3$}

\vspace{0.5cm}

$^1$ Institut f\"ur Physik, Johannes-Gutenberg-Universit\"at, \\
D-55099 Mainz, Germany

\vspace*{0.3cm}

$^2$ Paul-Scherrer-Institut, \\
CH--5232 Villigen PSI, Switzerland 

\vspace*{0.3cm}

$^3$ DESY, Deutsches Elektronen-Synchroton, \\
D-22603 Hamburg, Germany

\end{center}

\vspace{2cm}

\begin{center}
{\sc Abstract}
\end{center}
\noindent
We present an introduction into the basic concepts of the Standard
Model, i.e.\ the gauge theories of the forces and the Higgs mechanism
for generating masses. The Glashow-Salam-Weinberg theory of the
electroweak interactions will be described in detail. The key
experiments are reviewed, including the precision tests at high
energies. Finally, the limitations and possible physics areas beyond the
Standard Model are discussed.

\vfill
\footnoterule
\noindent
{\footnotesize To be published in "Scattering", P.\ Sabatier {\it ed.},
  Academic Press, London (2000).} 

\def\thefootnote{\arabic{footnote}}
\setcounter{footnote}{0}
\newpage
\tableofcontents
\newpage


\section{\sc Prologue}

A most fundamental element of physics is the reduction principle. The
large variety of macroscopic forms of matter can be traced back,
according to this principle, to a few microscopic constituents which
interact by a small number of basic forces. The reduction principle has
guided the unraveling of the structure of physics from the macroscopic
world through atomic and nuclear physics to particle physics. The laws
of Nature are summarized in the Standard Model of particle physics (Gla
61, Sal 68, Wei 67, Fri 72). All experimental observations are
compatible with this model at a level of very high accuracy. Not all
building blocks of the model, however, have been experimentally
established so far. In particular, the Higgs mechanism for generating
the masses of the fundamental particles (Hig 64, Eng 64, Gur 64) which
is a cornerstone of the system, still lacks experimental verification up
to now, even though indirect indications support this mechanism quite
strongly.\\ 

Even if all the elements of the Standard Model will be established
experimentally in the near future, the model cannot be considered the
{\it ultima ratio} of matter and forces. Neither the fundamental
parameters, masses and couplings, nor the symmetry pattern can be
derived; these elements are merely built into the model by hand.
Moreover, gravity with a structure quite different from the electroweak
and strong forces, is not coherently incorporated in the theory.\\

Despite this criticism, the Standard Model provides a valid framework
for the description of Nature, probed from microscopic scales of order
$10^{-16}$ cm up to cosmological distances of order $10^{+28}$ cm. The
model therefore ranks among the greatest achievements of mankind in
understanding Nature.\\

\noindent
The Standard Model consists of three components: 

\noindent
{\bf 1.} The basic \underline{constituents of matter} are leptons and
quarks (Gel 64, Zwe 64) which are realized in three families of
identical structure: \\
\hspace*{4cm}\mbox{
\begin{tabular}{llll}
\\
\underline{leptons:} \hspace*{0.5cm} & $\nu_e$ \hspace*{0.5cm} 
& $\nu_{\mu}$
\hspace*{0.5cm} & $\nu_{\tau}$ \\
         & $e^-$ & $\mu^-$ & $\tau^-$ \\[0.5ex]
\underline{quarks:} & $u$ & $c$ & $t$ \\
                    & $d$ & $s$ & $b$ \\
\\
\end{tabular}
} \\
The entire ensemble of these constituents has been identified
experimentally. The least known properties of these constituents are the
profile of the top quark, the mixing among the lepton states and the
quark states, and in particular, the structure of the neutrino
sector. \\

\noindent
{\bf 2.} Four different \underline{forces} act between the leptons and
quarks:
\begin{center}
\begin{picture}(100,90)(20,-20)
\ArrowLine(0,0)(25,25)
\ArrowLine(25,25)(0,50)
\Photon(25,25)(75,25){3}{6}
\ArrowLine(75,25)(100,50)
\ArrowLine(100,0)(75,25)
\put(-100,48){\underline{electromagnetic:}}
\put(45,10){$\gamma$}
\end{picture}
\begin{picture}(100,90)(-100,-20)
\ArrowLine(0,0)(25,25)
\ArrowLine(25,25)(0,50)
\Gluon(25,25)(75,25){3}{6}
\ArrowLine(75,25)(100,50)
\ArrowLine(100,0)(75,25)
\put(-80,48){\underline{strong:}}
\put(45,10){$g$}
\end{picture} \\
\begin{picture}(100,90)(20,-20)
\ArrowLine(0,0)(25,25)
\ArrowLine(25,25)(0,50)
\Photon(25,25)(75,25){3}{6}
\ArrowLine(75,25)(100,50)
\ArrowLine(100,0)(75,25)
\put(-100,48){\underline{weak:}}
\put(35,10){$W^\pm,Z$}
\end{picture}
\begin{picture}(100,90)(-100,-20)
\ArrowLine(0,0)(25,25)
\ArrowLine(25,25)(0,50)
\ZigZag(25,25)(75,25){3}{6}
\ArrowLine(75,25)(100,50)
\ArrowLine(100,0)(75,25)
\put(-80,48){\underline{gravitational:}}
\put(45,10){$G$}
\end{picture} 
\end{center}
The electromagnetic and weak forces are unified in the Standard Model.
The fields associated with these forces, as well as the fields
associated with the strong force, are spin-1 fields, describing the
photon $\gamma$, the electroweak gauge bosons $W^\pm$ and $Z$, and the
gluons $g$. The interactions of the force fields with the fermionic
constituents of matter as well as their self-interactions are described
by Abelian and non-Abelian $SU(3)\times SU(2)\times U(1)$ gauge theories
(Wey 29, Yan 54). The experimental exploration of these fundamental
gauge symmetries is far advanced in the sector of lepton/quark-gauge
boson interactions, yet much less is known so far from experiment about
the self-interactions of the force fields. The gravitational interaction
is mediated by a spin-2 field, describing the graviton $G$, with a
character quite different from spin-1 gauge fields. The gravity sector
is attached {\it ad hoc} to the other sectors of the Standard Model, not
properly formulated yet as a quantum phenomenon. \\

\begin{figure}[tb]
\begin{center} 
\begin{picture}(200,170)(0,10)
\LinAxis(0,0)(180,0)(2,10,0,0,0.8)
\LinAxis(0,0)(0,160)(2,10,-0,0,0.8)
\Line(100,-3)(100,3)
\SetScale{2.}
\ArrowLine(89,0)(91,0)
\ArrowLine(0,79)(0,81)
\SetScale{100.} 
\SetWidth{0.006}
\Curve{( 0.000000000000000000 , 0.757807107935890234 )
 ( 0.020000000000000000 , 0.757200983498678815 )
 ( 0.040000000000000000 , 0.755384065176691766 )
 ( 0.060000000000000000 , 0.752360717938870827 )
 ( 0.080000000000000000 , 0.748138216733451933 )
 ( 0.100000000000000006 , 0.742726746487966216 )
 ( 0.120000000000000009 , 0.736139402109238339 )
 ( 0.140000000000000013 , 0.728392188483387937 )
 ( 0.160000000000000003 , 0.719504020475829731 )
 ( 0.180000000000000021 , 0.709496722931271306 )
 ( 0.200000000000000011 , 0.698395030673716333 )
 ( 0.220000000000000029 , 0.686226588506462454 )
 ( 0.240000000000000019 , 0.673021951212100955 )
 ( 0.260000000000000009 , 0.658814583552519095 )
 ( 0.280000000000000027 , 0.643640860268896997 )
 ( 0.300000000000000044 , 0.627540066081710646 )
 ( 0.320000000000000007 , 0.610554395690729779 )
 ( 0.339999999999999969 , 0.592728953775018552 )
 ( 0.360000000000000042 , 0.574111754992935985 )
 ( 0.380000000000000060 , 0.554753723982135294 )
 ( 0.400000000000000022 , 0.534708695359564223 )
 ( 0.419999999999999984 , 0.514033413721465049 )
 ( 0.440000000000000058 , 0.492787533643374243 )
 ( 0.460000000000000020 , 0.471033619680123528 )
 ( 0.480000000000000038 , 0.448837146365838047 )
 ( 0.500000000000000000 , 0.426266498213938305 )
 ( 0.520000000000000018 , 0.403392969717138616 )
 ( 0.540000000000000036 , 0.380290765347448212 )
 ( 0.560000000000000053 , 0.357036999556170687 )
 ( 0.579999999999999960 , 0.333711696773904054 )
 ( 0.600000000000000089 , 0.310397791410540580 )
 ( 0.619999999999999996 , 0.287181127855267504 )
 ( 0.640000000000000013 , 0.264150460476566429 )
 ( 0.660000000000000031 , 0.241397453622212821 )
 ( 0.679999999999999938 , 0.219016681619277481 )
 ( 0.700000000000000067 , 0.197105628774124991 )
 ( 0.720000000000000084 , 0.175764689372414906 )
 ( 0.739999999999999991 , 0.155097167679101006 )
 ( 0.760000000000000120 , 0.135209277938431432 )
 ( 0.779999999999999916 , 0.116210144373949220 )
 ( 0.800000000000000044 , 0.098211801188491321 )
 ( 0.820000000000000062 , 0.081329192564189620 )
 ( 0.839999999999999969 , 0.065680172662470393 )
 ( 0.860000000000000098 , 0.051385505624054063 )
 ( 0.880000000000000115 , 0.038568865568955904 )
 ( 0.900000000000000022 , 0.027356836596485609 )
 ( 0.920000000000000040 , 0.017878912785247100 )
 ( 0.939999999999999947 , 0.010267498193139056 )
 ( 0.960000000000000075 , 0.004657906857354410 )
 ( 0.980000000000000093 , 0.001188362794380730 )
 ( 1.00000000000000000 , 0.000000000000000000 )
 ( 1.02000000000000002 , 0.001236862449288654 )
 ( 1.04000000000000004 , 0.005045904096617600 )
 ( 1.06000000000000005 , 0.011576988875652300 )
 ( 1.08000000000000007 , 0.020982890699352618 )
 ( 1.10000000000000009 , 0.033419293459972764 )
 ( 1.12000000000000011 , 0.049044791029061805 )
 ( 1.14000000000000012 , 0.068020887257462942 )
 ( 1.15999999999999992 , 0.090511995975313702 )
 ( 1.18000000000000016 , 0.116685440992046738 )
 ( 1.20000000000000018 , 0.146711456096388609 )
 ( 1.21999999999999997 , 0.180763185056360215 )
 ( 1.23999999999999999 , 0.219016681619277759 )
 ( 1.26000000000000001 , 0.261650909511750740 )
 ( 1.28000000000000003 , 0.308847742439684547 )
 ( 1.30000000000000004 , 0.360791964088277795 )
 ( 1.32000000000000006 , 0.417671268122023653 )
 ( 1.34000000000000008 , 0.479676258184711046 )
 ( 1.35999999999999988 , 0.547000447899421705 )
 ( 1.38000000000000012 , 0.619840260868533277 )
 ( 1.40000000000000013 , 0.698395030673717110 )
 ( 1.41999999999999993 , 0.782867000875938079 )
 ( 1.44000000000000017 , 0.873461325015458145 )
 ( 1.46000000000000019 , 0.970386066611831799 )
 ( 1.47999999999999998 , 1.07385219916390673 )
 ( 1.50000000000000000 , 1.18407360614982915 )
 ( 1.52000000000000024 , 1.30126708102703659 )}
\SetScale{1.} \SetWidth{0.5}
\put(14,140){$V[\varphi]$}
\put(160,14){$|\varphi|$}
\put(90,-20){$v/\sqrt{2}$}
\end{picture}
\vspace*{1cm}
\end{center}
\caption[]{\label{fg:0} \it The Higgs potential of the Standard Model.}
\end{figure}
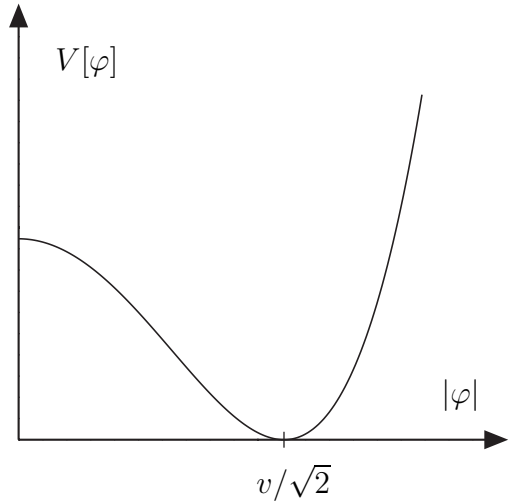

\noindent
{\bf 3.} The third component of the Standard Model is the
\underline{Higgs mechanism} (Hig 64, Eng 64, Gur 64). In this sector of
the theory, scalar fields interact with each other in such a way that
the ground state acquires a non-zero field strength, breaking the
electroweak symmetries spontaneously. The potential describing these
self-interactions is displayed in Fig.\ \ref{fg:0}. The interaction
energies of electroweak gauge bosons, leptons and quarks with this field
manifest themselves as non-zero masses of these particles. If this
picture is correct, a scalar particle, the Higgs boson, should be
observed with a mass of less than about 700 GeV, the final {\it
  experimentum crucis} of the Standard Model. \\

Experimental efforts extending over more than a century, have been
crucial in developing these basic ideas to a coherent picture. The first
elementary particle discovered at the end of the 19th century was the
electron (Wie 97, Tho 97, Kau 97,97a), followed later by the other
charged leptons, the $\mu$ (And 37) and $\tau$ leptons (Per 75). The
first species of weakly interacting neutrinos, $\nu_e$, was found in the
fifties (Rei 53), the others, $\nu_{\mu}$ (Dan 62) and $\nu_{\tau}$ (Pol
00), one and five decades later. The up, down and strange quarks were
``seen'' first in deep-inelastic electron- and neutrino-nucleon
scattering experiments (Fri 72a, Eic 73), the discovery of the charm
quark (Aub 74, Aug 74) marked what is called ``November revolution'' of
particle physics. The bottom quark of the third family was isolated in
the 70's (Her 77) while the discovery of the top quark followed only
recently (Abe 95, Aba 95).

The photon as the quantum associated with the electromagnetic field, was
discovered when the photo-electric effect was interpreted theoretically
(Ein 05), while the heavy electroweak bosons $W^\pm$, $Z$ have first
been isolated in $p\bar{p}$ collisions (Arn 83, Ban 83, Arn 83a, Bag
83).  Gluons as the carriers of the strong force were discovered in the
fragmented form of hadron jets, generated in $e^+e^-$ annihilation at
high energies (Bra 79, Bar 79, Ber 79).\\

Future experimental activities will focus, in the framework of the
Standard Model, on the properties of the top quark, the non-Abelian
gauge symmetry structure of the self-interactions among the force
fields, and last not least, on the search of Higgs bosons and, if
discovered, on the analysis of its properties.  This experimental
program is a continuing task at the existing collider facilities LEP2,
HERA and Tevatron, and it will extend to the next-generation facilities,
the $pp$ collider LHC (ATL 99, CMS 94), a prospective $e^+e^-$ linear
collider (Zer 99, Acc 98, Mur 96) with beam energies in the TeV range,
and a prospective muon-collider (Ank 99, Aut 99).  Other experimental
facilities will lead to a better understanding of the neutrino
properties and map out the quark mixings.\\


\section{\sc Introduction}


\subsection{\sc The path to the standard model of the electroweak
  interactions} 

The weak interactions of the elementary particles have been discovered
in $\beta$-decay processes. They are described by an effective
Lagrangian of current $\times$ current type (Fer 34), in which the weak
currents are coherent superpositions of charged vector and axial-vector
currents, accounting for the violation of parity. For the $\mu$-decay
process $\mu^-\rightarrow e^-\bar \nu_e \nu_\mu$ the Lagrangian is
defined as
\begin{equation}
{\cal L}=\frac{G_F}{\sqrt 2}\left[\bar \nu_\mu \gamma_\lambda \left( 
1-\gamma_5 \right)\mu \right] \left[\bar e \gamma_\lambda 
\left( 1-\gamma_5 \right) \nu_e \right ] 
\end{equation}
The overall strength of the interaction is measured by the Fermi
coupling constant
\begin{equation}
G_F \simeq 1.17 \times 10^{-5}~{\rm GeV}^{-2}
\end{equation}
which carries the dimension of [mass$]^{-2}$.\\

The Fermi theory of the weak interactions can only be interpreted as an
effective low-energy theory which cannot be extended to arbitrarily high
energies. Applying this theory to the scattering process $\nu_\mu e^-
\rightarrow \mu^- \nu_e$ at high energies, the scattering amplitude
rises indefinitely with the square of the energy $E_{\rm cms} =
\sqrt{s}$ in the center-of-mass system of the colliding particles:
\begin{equation}
f_0[\nu_{\mu} e^- \rightarrow \mu^- \nu_e] = \frac{G_Fs}{2\sqrt 2\pi} 
\end{equation}
However, as an $S$-wave scattering amplitude, $f_0$ must fulfill the
unitarity condition $|f_0|^2\leq \Im m f_0$, leading to the upper limit
$|\Re e f_0|\leq 1/2$. The theory can therefore not be applied at
energies in excess of
\begin{equation}
s\leq \sqrt 2 \pi/G_F \sim (600~{\rm GeV})^2 \, .
\end{equation}

A simple line of arguments allows us to deduce the structure of the weak
interactions from unitarity constraints (Lle 73, Cor 74) applied to a
set of high-energy scattering processes.  In this way the existence of
charged and neutral vector bosons can be predicted, as well as the
existence of a scalar particle, together with the properties of their
interactions.\\

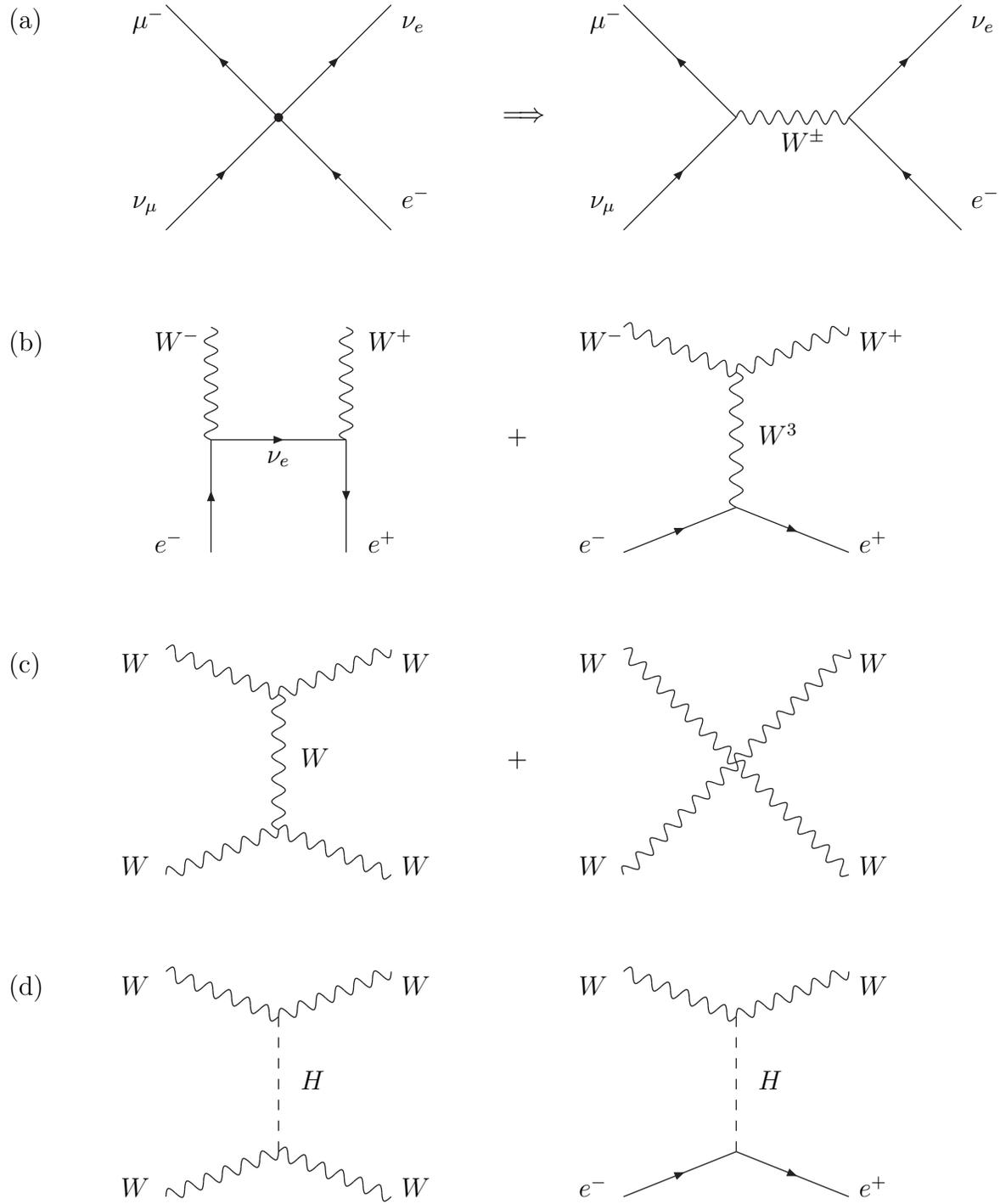
\begin{figure}[hbtp]
\begin{center}
\begin{picture}(100,100)(50,0)
\ArrowLine(0,0)(50,50)
\ArrowLine(50,50)(0,100)
\ArrowLine(100,0)(50,50)
\ArrowLine(50,50)(100,100)
\Vertex(50,50){2}
\put(-70,90){(a)}
\put(-15,10){$\nu_\mu$}
\put(-15,90){$\mu^-$}
\put(105,90){$\nu_e$}
\put(105,10){$e^-$}
\put(150,48){$\Longrightarrow$}
\end{picture}
\begin{picture}(100,100)(-50,0)
\ArrowLine(0,0)(50,50)
\ArrowLine(50,50)(0,100)
\Photon(50,50)(100,50){3}{6}
\ArrowLine(100,50)(150,100)
\ArrowLine(150,0)(100,50)
\put(-15,10){$\nu_\mu$}
\put(-15,90){$\mu^-$}
\put(155,90){$\nu_e$}
\put(155,10){$e^-$}
\put(70,35){$W^\pm$}
\end{picture} \\[1.5cm]
\begin{picture}(100,100)(50,0)
\ArrowLine(20,0)(20,50)
\ArrowLine(20,50)(80,50)
\ArrowLine(80,50)(80,0)
\Photon(20,50)(20,100){3}{6}
\Photon(80,50)(80,100){3}{6}
\put(-70,90){(b)}
\put(-5,0){$e^-$}
\put(-5,90){$W^-$}
\put(90,90){$W^+$}
\put(90,0){$e^+$}
\put(152,48){$+$}
\put(45,40){$\nu_e$}
\end{picture}
\begin{picture}(100,100)(-50,0)
\ArrowLine(0,0)(50,20)
\ArrowLine(50,20)(100,0)
\Photon(50,20)(50,80){3}{6}
\Photon(50,80)(0,100){3}{6}
\Photon(50,80)(100,100){3}{6}
\put(-20,0){$e^-$}
\put(-20,90){$W^-$}
\put(105,90){$W^+$}
\put(105,0){$e^+$}
\put(60,48){$W^3$}
\end{picture} \\[1.5cm]
\begin{picture}(100,100)(50,0)
\Photon(0,0)(50,20){3}{6}
\Photon(50,20)(100,0){3}{6}
\Photon(50,20)(50,80){3}{6}
\Photon(50,80)(0,100){3}{6}
\Photon(50,80)(100,100){3}{6}
\put(-70,90){(c)}
\put(-20,0){$W$}
\put(-20,90){$W$}
\put(105,90){$W$}
\put(105,0){$W$}
\put(60,48){$W$}
\put(152,48){$+$}
\end{picture}
\begin{picture}(100,100)(-50,0)
\Photon(0,0)(50,50){3}{8}
\Photon(100,0)(50,50){3}{8}
\Photon(50,50)(0,100){3}{8}
\Photon(50,50)(100,100){3}{8}
\put(-20,0){$W$}
\put(-20,90){$W$}
\put(105,90){$W$}
\put(105,0){$W$}
\end{picture} \\[1.5cm]
\begin{picture}(100,100)(50,0)
\Photon(0,0)(50,20){3}{6}
\Photon(50,20)(100,0){3}{6}
\DashLine(50,20)(50,80){5}
\Photon(50,80)(0,100){3}{6}
\Photon(50,80)(100,100){3}{6}
\put(-70,90){(d)}
\put(-20,0){$W$}
\put(-20,90){$W$}
\put(105,90){$W$}
\put(105,0){$W$}
\put(60,48){$H$}
\end{picture}
\begin{picture}(100,100)(-50,0)
\ArrowLine(0,0)(50,20)
\ArrowLine(50,20)(100,0)
\DashLine(50,20)(50,80){5}
\Photon(50,80)(0,100){3}{6}
\Photon(50,80)(100,100){3}{6}
\put(-20,0){$e^-$}
\put(-20,90){$W$}
\put(105,90){$W$}
\put(105,0){$e^+$}
\put(60,48){$H$}
\end{picture} \\
\end{center}
\caption[]{ \label{fg:1} \it Diagrams for scattering processes with 
implications on unitarity constraints.}
\end{figure}

\noindent
\underline{{\bf a)} Charged $W^\pm$ Bosons.} The unitarity problem
described above can be solved by assuming the weak interactions to be
mediated by the exchange of a heavy charged vector boson $W^\pm$ (Yuk
35), cf.\ Fig.\ \ref{fg:1}(a).  The $W$ propagator damps the rise in the
energy of the scattering amplitude:
\begin{equation}
f_0[\nu_{\mu} e^- \rightarrow \mu^- \nu_e] \rightarrow 
 \frac{G_F s}{2\sqrt{2} \pi} \, \frac{M_W^2}{M_W^2 - s}
\end{equation}
which is compatible with the unitarity limit if the $W$-boson mass is
sufficiently light. Defining the dimensionless coupling between the
$W$-field and the weak current by $g_W/2\sqrt 2$, the connection to
Fermi's theory at low energies leads to the relation $G_F/\sqrt 2 =
g_W^2/8M_W^2$.  If the coupling $g_W$ is of the same order as the
electromagnetic $e$, the mass of the $W$ boson is close to 100 GeV.  The
weak interactions in this picture are therefore not really weak but
their strength is reduced only by the short-range character of the
$W$-exchange mechanism at low energies. The interaction is effectively
weak since it is confined to distances of order $\lambda_W=M^{-1} _W$
where $\lambda_W$ is the small Compton wavelength of the $W$ boson. \\

\noindent
\underline{{\bf b)} Neutral $W^3$ Boson.} Induced by the $\ell\nu_\ell
W^\pm$ couplings, the theory predicts the production of $W^+W^-$ pairs
in $e^+e^-$ annihilation. With the ingredients introduced up to this
point\footnote {Since the same argument can also be derived from $\nu
  \bar{\nu}$ annihilation to $W^+W^-$ pairs, the electromagnetic
  interactions can be disregarded in the present context.}, this process
is mediated by the exchange of a neutrino, cf.\ Fig.\ \ref{fg:1}(b).
When the $W$ bosons in the final state are polarized longitudinally,
their wave-functions $e_\mu(k) = (k^3, 0, 0, k^0)/M_W$ grow linearly
with the energy. The scattering amplitude for the process
$e^+e^-\rightarrow W^+W^-$, if mediated solely by $\nu_e$ exchange,
therefore grows quadratically for high energies, and it violates the
unitarity limit eventually. This divergence can be damped by the
exchange of a doubly charged lepton in the $u$-channel, or else by the
exchange of a neutral vector boson $W^3$ in the $s$-channel.  Following
the second branch, a trilinear coupling of the three $W$ bosons,
$W^{\pm} = (W^1 \mp iW^2)/\sqrt{2}$ and $W^3$, must be introduced with
strength $g_W \varepsilon_{klm}$. The couplings $g_W I^k_{ab}$ between
the leptons $a,b$ and the $W$ bosons $k$, and the trilinear
self-couplings $g_W \varepsilon_{klm}$ of the $W$ bosons must fulfill
the consistency conditions
\begin{equation}
\left[I^k, I^l \right]=i\varepsilon_{klm} I^m
\end{equation}
to restore unitarity at high energies.\\

\noindent
\underline{{\bf c)} $W$ Self-Interactions.} As a result of the trilinear
couplings among the $W$ bosons, the $W$ bosons can scatter
quasi-elastically, $WW\rightarrow WW$, cf.\ Fig.\ \ref{fg:1}(c). The
amplitude for the scattering of longitudinally polarized $W$ bosons,
built-up by virtual $W$ exchanges, grows as the fourth power of
$\sqrt{s}$ for high energies. This leading divergence is canceled by
introducing a quadrilinear coupling among the $W$ bosons which must be
of second order in $g_W$, and the dependence on the charge indices given
by the tensors
\begin{equation}
\lambda_{klmn} = g_W^2
\varepsilon _{klp^\prime}\varepsilon_{p^\prime mn}
\end{equation}
However, unitarity is not yet completely restored for asymptotic
energies since the amplitude still grows quadratically in the energy.\\

\noindent
\underline{{\bf d)} The Higgs Boson.} Since all intrinsic mechanisms to
render a massive vector-boson theory conform with the requirement of
unitarity at high energies have been exhausted, only two paths are left
for solving this problem. The $WW$ scattering amplitude may either be
damped by introducing strong interactions between the $W$ bosons at high
energies, or a new particle must be introduced, the scalar Higgs boson
$H$, the exchange of which interferes destructively with the exchange of
vector-bosons, Fig.\ \ref{fg:1}(d). In fact, if the $HWW$ coupling is
defined by $g_W M_W$, the scattering amplitude approaches for energies
far above all masses involved, the asymptotic limit
\begin{equation}
f_0 [W_LW_L \rightarrow W_LW_L] \rightarrow \frac{G_F M_H^2}{4
  \sqrt{2} \pi}
\end{equation}
which fulfills the unitarity requirement for sufficiently small values
of the Higgs-boson mass M$_H$.

The same argument applies to fermion-antifermion annihilation to
longitudinally polarized $W$ bosons. For non-zero fermion mass $m_f$,
the annihilation amplitude, based on Fig.\ \ref{fg:1}(b), grows as
$m_f\sqrt{s}$ indefinitely.  The rise is damped by the destructive
Higgs-boson exchange in Fig.\ \ref{fg:1}(d). This damping mechanism is
operative only if the coupling of the Higgs boson to a source particle
grows as the mass of the particle.

By extending the analysis to the process $WW \rightarrow HH$ and to
amplitudes involving 3-particle final states, $WW \rightarrow WW+H$
and $WW \rightarrow HHH$, the unitarity requirements can be exploited
to determine the quartic $W$-Higgs interactions and the Higgs
self-interaction potential. The general form of the potential is
constrained to be of quadrilinear type with the coefficients fixed
uniquely by the mass of the Higgs boson and the scale of the $WWH$ 
coupling.\\

\noindent
\underline{In Summary}. The consistent formulation of the weak
interactions as a theory of fields interacting weakly up to high
energies leads us to a vector-boson theory complemented by a scalar
Higgs field which couples to other particles proportional to the masses
of the particles.\\

The assumption that the particles remain weakly interacting up to very
high energies, is a prerequisite for deriving the relative strengths
of the weak to the electromagnetic coupling.\\


\subsection{\sc The theoretical base} 

The structure of the electroweak system that has emerged from the
requirement of asymptotic unitarity, can theoretically be formulated as
a gauge field theory. The fundamental forces of the Standard Model, the
electromagnetic (Dir 27, Jor 28, Hei 29, Tom 46, Sch 48, Fey 49) and the
weak forces (Gla 61, Sal 68, Wei 67) as well as the strong forces (Fri
72, Fri 73, Gro 73, Pol 73), are mediated by gauge fields. This concept
could consistently be extended to massive gauge fields by introducing
the Higgs mechanism (Hig 64, Eng 64, Gur 64) which generates masses
without destroying the underlying gauge symmetries of the theory. \\


\subsubsection{\sc Gauge theories}

Gauge field theories (Wey 29, Yan54) are invariant under gauge
transformations of the fermion fields: $\psi \rightarrow S\psi$. $S$ is
either a phase factor for Abelian transformations or a unitary matrix
for non-Abelian transformations acting on multiplets of fermion fields
$\psi$.  To guarantee the invariance under local transformations for
which $S$ depends on the space-time point $x$, the usual space-time
derivatives $\partial_{\mu}$ must be extended to covariant derivatives
$D_{\mu}$ which include a new vector field $V_{\mu}$:
\begin{equation}
i\partial_{\mu} \rightarrow iD_{\mu} = i\partial_{\mu} - gV_{\mu}
\end{equation}
$g$ defines the universal gauge coupling of the system. The gauge field
$V_{\mu}$ is transformed by a rotation plus a shift under local gauge
transformations:
\begin{equation}
V_{\mu} \rightarrow SV_{\mu}S^{-1} + ig^{-1}[\partial_{\mu} S]S^{-1}
\end{equation}
By contrast, the curl $F$ of $V_{\mu}$, $F_{\mu\nu} = -i g^{-1}
[D_{\mu}, D_{\nu}]$, is just rotated under gauge transformations.\\

The Lagrangian which describes the system of spin-1/2 fermions and
vectorial gauge bosons for massless particles, can be cast into the
compact form:
\begin{equation}
{\cal L}[\psi,V]=\bar \psi i \!\not\!\! D \psi -\frac{1}{2} {\rm Tr}
F^2 
\end{equation}
It incorporates the following interactions: \\

\hspace*{2cm}\mbox{
\begin{tabular}{ll}
{\it fermions-gauge bosons} & 
$-g\bar \psi \! \not\! V \psi$ \\
{\it three-boson couplings} &
$ig {\rm Tr}(\partial_\nu V_\mu - \partial_\mu V_\nu)[V_\mu,V_\nu]$ \\
{\it four-boson couplings} & 
$\frac{1}{2} g^2 {\rm Tr} [V_\mu,V_\nu]^2$  
\end{tabular}}\\

\noindent
These types of interactions coincide exactly with the interactions
derived from the unitarity requirements for fermion and vector boson
fields interacting weakly up to asymptotic energies.\\


\subsubsection{\sc The Higgs mechanism}

If mass terms for gauge bosons and for left/right-chiral fermions are
introduced by hand, they destroy the gauge invariance of the theory.
This problem has been solved by means of the Higgs mechanism (Hig 64,
Eng 64, Gur 64) in which masses are introduced into gauge theories in a
consistent way.  The solution of the problem is achieved at the expense
of a new fundamental degree of freedom, the Higgs field, which is a
scalar field.\\ 

Scalar fields $\varphi$ can interact with each other so that the ground
state of the system, corresponding to the minimum of the
self-interaction potential
\begin{equation}
V=\frac{\lambda}{2}\left[ \left| \varphi\right| ^2-\frac{v^2}{2}\right] ^2
\end{equation}
is realized for a non-zero value of the field strength\footnote{Since
  the fixing of the ground-state value of $\varphi$ destroys the gauge
  symmetry in the scalar sector before the interaction with gauge field
  is switched on, this is reminiscent of spontaneous symmetry breaking.}
$\varphi \rightarrow v/\sqrt 2$, cf.\ Fig.\ \ref{fg:0}. The interaction
energies of massless gauge bosons and fermions with the Higgs field in
the ground state can be re-interpreted as the gauge-boson and fermion
masses.\\ 

The vector bosons are coupled to the ground-state Higgs field by means
of the covariant derivative, giving rise to the value $M_V^2 =
g^2v^2/4$ of the vector-boson mass.

By contrast, the interaction between fermion fields and the Higgs field
is of Yukawa type 
\begin{equation}
{\cal L}_Y = g_f \bar ff\varphi
\end{equation}
Replacing the Higgs field by its ground state value, $\varphi
\rightarrow v/\sqrt{2}$, one obtains the mass term $g_f v/\sqrt 2 \bar
ff$, from which one can read off the fermion mass $m_f = g_f
v/\sqrt{2}$.\\ 

As a result, the rules derived in a heuristic way from asymptotic
unitarity are borne out naturally in the Higgs mechanism. Thus, the
Higgs mechanism provides a microscopic picture for generating the masses
in a theoretically consistent massive gauge field theory.\\

In technical language, the Higgs mechanism leads to a renormalizable
gauge field theory including non-zero gauge-boson and fermion masses
(tHo 71, tHo 72). After fixing a small number of basic parameters which
must be determined experimentally, the theory is under strict
theoretical control, in principle to any required accuracy.\\


\section{\sc The Glashow-Salam-Weinberg Theory}

The Standard Model of electroweak and strong interactions is based on
the gauge group
\begin{equation}
G_{\rm SM} = SU(3)\times SU(2)\times U(1)
\end{equation}
of unitary gauge transformations. $SU(3)$ is the non-Abelian symmetry
group of the strong interactions (Fri 72). The gluonic gauge fields are
coupled to the color charges as formalized in quantum chromodynamics
(QCD).  $SU(2)$ is the non-Abelian electroweak-isospin group, to which
three $W$ gauge fields are associated. $U(1)$ is the Abelian hypercharge
group, the hypercharge $Y$ connected with the electric charge $Q$ and
the isospin $I_3$ by the relation $Y = 2(Q - I_3)$. The associated $B$
field and the neutral component of the $W$ triplet field mix to form the
photon field $A$ and the electroweak field $Z$.  The gauge theory of the
electroweak interactions based on the symmetry group $SU(2) \times U(1)$
is known as the Glashow-Salam-Weinberg
theory (Gla 61, Sal 68, Wei 67).\\


\subsection{\sc The electroweak interactions}

\subsubsection{\sc The matter sector}

The matter fields of the Standard Model are the leptons and quarks,
carrying spin-1/2. They are classified as left-handed isospin doublets
and right-handed isospin singlets\footnote{Right-handed neutrinos, even
  though they may formally be included, play a special r$\hat {\rm o}$le
  among the basic fermions. This sector will not be elaborated upon in
  the present context.}; moreover, quarks are color triplets.  This
symmetry pattern is realized in the first, second and third generation
of the fermions in identical form: \\
\begin{center}
\begin{tabular}{cccccccc}
$\displaystyle \left[ {\nu_e \atop e^-} \right]_L$ & 
$\displaystyle {\nu_{eR}\atop e^-_R}$ & $\qquad$ &
$\displaystyle \left[ {\nu_\mu \atop\mu^-} \right]_L$ & 
$\displaystyle {\nu_{\mu R}\atop \mu^-_R}$ & $\qquad$ & $\displaystyle
\left[ {\nu_\tau \atop \tau^- } \right]_L$ & $\displaystyle
{\nu_{\tau R} \atop \tau^-_R}$ \\ [5ex] 
$\displaystyle \left[ {u \atop d} \right]_L$ & 
$\displaystyle {u_R\atop d_R}$ & $\qquad$ &
$\displaystyle \left[ {c \atop s} \right]_L$ &
$\displaystyle {c_R\atop s_R}$ & $\qquad$ & $\displaystyle \left[ {t
    \atop b } \right]_L$ & $\displaystyle {t_R \atop b_R}$\\[3ex]
\end{tabular}
\end{center} 

\noindent
The left-handed down-type quark states are Cabibbo-Kobayashi-Maskawa
mixtures of the mass eigenstates (Cab 63, Kob 73). \\

This symmetry structure cannot be derived within the Standard Model.
However, the experimental observations are incorporated in a natural
way.  The different isospin assignment to left-handed and right-handed
fields allows for maximal parity violation in the weak interactions.
Given the assignments of electric charge, hypercharge and isospin, three
color degrees of freedom are needed in the quark sector to cancel
anomalies and to render the gauge-field theory renormalizable.  The same
symmetry pattern is needed in each of the three generations to suppress
flavor-changing neutral-current interactions to the level excluded by
experimental analyses.  Moreover, at least three generations must be
realized in Nature to incorporate $CP$ violation in the Standard
Model.\\


\subsubsection{\sc The gauge sector}

The symmetries associated with isospin, hypercharge and color are
realized as local gauge symmetries. The corresponding spin-1 gauge
fields are the following vector fields: \\
\hspace*{2cm}\mbox{
\begin{tabular}{llll}
$SU(2)$ & {\it isospin}      & $W_\mu^i$ & isotriplet $i=1,2,3$ \\
$U(1)$  & {\it hypercharge}  & $B_\mu$   & \\
$SU(3)$ & {\it color}        & $G_\mu^a$ & gluon color octet $a=1,\dots,8$ 
\end{tabular} \\
} \\

\noindent
The non-Abelian $SU(2)$ isospin and $SU(3)$ color fields interact among 
each other in trilinear and quadrilinear vertices.\\


\subsubsection{\sc The Higgs sector}

To combine left-handed doublets and right-handed singlets in the
fermion-Higgs Yukawa interaction, the Higgs field must be an isodoublet
field $\varphi= [\varphi^0,\varphi^-]$.

The value of the field in the ground state is determined by the minimum
of the self-interaction potential $V(\varphi)$. A field component $H$
which describes small oscillations about the ground state defines the
physical Higgs field. Thus the scalar isodoublet field may be
parametrized as:
\begin{equation}
\varphi = U \left[ {0 \atop (v+H)/\sqrt{2}} \right]
\end{equation}
where the $SU(2)$ matrix $U$ incorporates the three remaining Goldstone
degrees of freedom besides the physical field $H$. \\


\subsubsection{\sc Interactions}

The interactions of the Standard Model are summarized by three terms in
the basic Lagrangian\footnote{We will not work out the full Lagrangian
  needed in calculations of higher-order corrections. This would require
  additional terms for the gauge fixing and the ghost sector.}:
\begin{equation}
{\cal L}={\cal L}_{gauge}+{\cal L}_{fermions}+{\cal L}_{Higgs}
\end{equation}
The first term is built up by the gauge fields and their
self-interactions:
\begin{equation}
{\cal L}_{gauge}=-\frac{1}{4}W^i_{\mu\nu}W^i_{\mu\nu}-
\frac{1}{4}B_{\mu\nu}B_{\mu\nu}-
\frac{1}{4}G^a_{\mu\nu}G^a_{\mu\nu}
\end{equation}
with the field strengths
\begin{eqnarray}
 W^i_{\mu\nu}&=&\partial_\nu W^i_\mu -\partial_\mu W^i_\nu -g_W
\epsilon^{ijk}W^j_\mu W^k_\nu \\
 B_{\mu\nu}&=&\partial_\nu B_\mu - \partial_\mu B_\nu \\
 G^a_{\mu\nu}&=&\partial_\nu G^a_\mu -\partial_\mu G^a_\nu -g_s
f^{abc}G^a_\mu G^b_\nu   
\end{eqnarray}
The tensors $\epsilon^{ijk}$ and $f^{abc}$ are the $SU(2)$ and $SU(3)$
structure constants, $g_W$ and $g_s$ are the weak-isospin and the strong
coupling, respectively. \\

The second term summarizes the fermion-gauge boson couplings 
\begin{equation}
{\cal L}_{fermion}=\sum \bar f\,i\not\!\! D\,f
\end{equation}
with the sum running over the left- and right-handed field components of
the leptons and quarks. Depending on the fermion species, the covariant
derivative takes the form
\begin{equation}
iD_\mu=i\partial_\mu + g_W I^iW^i_\mu - g_W^\prime \frac{Y}{2}B_\mu +
g_s T^aG_\mu^a
\end{equation}
where the hypercharge coupling is denoted by $g_W^{\prime}$.\\

Finally, the Higgs Lagrangian contains the Higgs-gauge boson
interactions generated by the covariant derivative, the Higgs-fermion
Yukawa couplings and the potential of the Higgs self-interactions:
\begin{equation}
{\cal L}_{Higgs}=|D_\mu\varphi|^2+g_f^d \bar f_L^d \varphi f_R^d 
+ g_f^u \bar f_L^u \tilde\varphi f_R^u + {\rm h.c.} -
\frac{\lambda}{2}\left[ |\varphi|^2-\frac{v^2}{2}\right] ^2
\label{eq:laghiggs}
\end{equation}
The field $\varphi$ can generate the masses for down-type leptons and
quarks $f^d$, while the field $\tilde\varphi=i\tau_2\varphi^*$ is the
charge-conjugated Higgs field which generates the masses of the up-type
fermions $f^u$.

The Lagrangian ${\cal L}$ summarizes the laws of physics for the three
basic interactions, the electromagnetic, the weak and the strong
interactions between the leptons and the quarks, and it predicts the
form of the self-interactions between the gauge fields.  Moreover, the
specific form of the Higgs interactions generates the masses of the
fundamental particles, the leptons and quarks, the gauge bosons and the
Higgs boson itself, and it predicts the interactions of the Higgs
particle.\\


\subsection{\sc Masses and mass eigenstates of particles}

In the unitary gauge the mass terms are extracted by substituting
$\varphi \rightarrow [0,v/\sqrt 2]$ in the basic Higgs Lagrangian
(\ref{eq:laghiggs}). The apparent $SU(2)$ symmetry seems to be lost
thereby, but only superficially so and remaining present in hidden form;
the resulting Lagrangian preserves an apparent local $U(1)$ gauge
symmetry which is identified with the electromagnetic gauge symmetry:
$SU(2) \times U(1) \rightarrow U(1)_{\rm em}$. \\

\noindent
{\underline{Gauge Bosons:}} The mass matrix of the gauge bosons in the
basis $(\vec W, B)$ takes the form
\begin{equation}
{\cal M}^2_V = \frac{1}{4} v^2
  \left(
  \begin{array}{cccc}
g_W^2 & & & \\
 & g_W^2 & & \\
 & & g_W^2 & g_Wg_W^\prime \\ 
 & & g_Wg_W^\prime & g_W^{\prime 2}
  \end{array} 
  \right)
\end{equation}
After diagonalization the fields are assigned the following mass
eigenvalues: \\

\hspace*{2cm}\mbox{
\begin{tabular}{ll}
{\it charged weak bosons} $W^\pm$ & 
 $M^2_{W^\pm} = \frac{1}{4}g_W^2v^2$ \\
{\it neutral weak boson} $Z$      & 
 $M^2_Z =\frac{1}{4} (g_W^2+g_W^{\prime 2})v^2$ \\ 
{\it photon} $\gamma$             &  
 $M^2_{\gamma}=0$ \\[3ex]
\end{tabular} }

\noindent
As eigenstates related to the two masses $M^2_{W^\pm}$ the charged
$W^\pm$ boson states may be defined as
\begin{equation}
W^\pm_\mu =\frac{1}{\sqrt 2}\left[ W_\mu^1\mp i W^2_\mu \right]  
\end{equation}
The specific form of the mass matrix leads to a vanishing eigenvalue, a
consequence of the residual $U(1)_{\rm em}$ gauge symmetry. The
associated eigenstate is the photon field which is a mixture of the
neutral isospin field $W^3$ and the neutral hypercharge field $B$ while
the orthogonal eigenstate corresponds to the $Z$ field:
\begin{eqnarray}
A_\mu &=& \sin \vartheta_W W^3_\mu + \cos \vartheta_W B_\mu \\
Z_\mu &=& \cos \vartheta_W W^3_\mu - \sin \vartheta_W B_\mu 
\end{eqnarray}
The electroweak mixing angle $\vartheta_W$ is defined by the ratio of
the $SU(2)$ and $U(1)$ couplings:
\begin{equation}
\tan \vartheta_W = g_W^\prime / g_W
\end{equation}
Experimentally the mixing angle turns out to be large, i.e.\ $\sin ^2
\vartheta_W \simeq 0.23$ . The fact that the experimental value for
$\sin ^2\vartheta_ W$ is far away from the limits 0 or 1, indicates a
large mixing effect. This supports the interpretation that the
electromagnetic and the weak interactions are indeed manifestations of a
unified electroweak interaction even though the underlying symmetry
group $SU(2)\times U(1)$ is not simple. This argument is strengthened
when the strong and electroweak symmetry group $SU(3)\times SU(2)\times
U(1)$ is unified to $SU(5)$: reduced to one single coupling, the
electroweak mixing angle is predicted at the unification point as $\sin
^2\vartheta_W = 3/8$. This value is renormalized to $\sim 0.2$ if the
couplings are evolved from the unification scale $\Lambda_U\sim 10^{16}$
GeV down to the electroweak scale $\Lambda_W \sim M_W$. It may therefore
be concluded that the electromagnetic and the weak interactions are
truly unified in the Glashow-Salam-Weinberg theory of the electroweak
interactions.\\ 

The ground-state value of the Higgs field is related to the Fermi
coupling constant. From the low-energy relation $G_F/\sqrt 2=g_W^2/8
M^2_W$ in $\beta$ decay and combined with the mass relation
$M_W^2 = g_W^2v^2/4$, the value of $v$ can be derived:
\begin{eqnarray}
v &=& \left[ 1/\sqrt 2 G_F \right]^{1/2}\\
\nonumber  &\simeq& 246\, {\rm GeV}   
\end{eqnarray}
The typical range for electroweak phenomena, defined by the weak masses
$M_W$ and $M_Z$, is of order 100 GeV. \\

\noindent
{\underline{Fermions:}} Both leptons as well as up-type and down-type
quarks are endowed with masses by means of the Yukawa interactions
with the Higgs ground state:
\begin{equation}
m_f=g_f \frac{v}{\sqrt 2}
\end{equation}
Though the masses of chiral fermion fields can be introduced in a
consistent way via the Higgs mechanism, the Standard Model does not
provide predictions for the experimental values of the Yukawa couplings
$g_f$ and, as a consequence, of the masses. A theory of the masses is
not available yet, even though interesting suggestions for the textures
of the mass matrices have been proposed, based on general matrix
symmetries. A deeper understanding may be expected from superstring
theories in which the Yukawa couplings are predictable numbers generated
by the string interactions.  

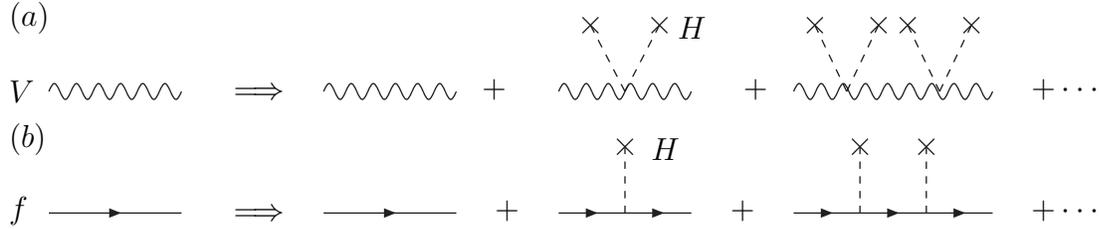
\begin{figure}[hbt]
\begin{center}
\begin{picture}(60,10)(80,55)
\Photon(0,25)(50,25){3}{6}
\put(70,22){$\Longrightarrow$}
\put(-15,21){$V$}
\put(-15,50){$(a)$}
\end{picture}
\begin{picture}(60,10)(40,55)
\Photon(0,25)(50,25){3}{6}
\put(60,23){$+$}
\end{picture}
\begin{picture}(60,10)(15,55)
\Photon(0,25)(50,25){3}{6}
\DashLine(25,25)(12,50){3}
\DashLine(25,25)(38,50){3}
\Line(9,53)(15,47)
\Line(9,47)(15,53)
\Line(35,53)(41,47)
\Line(35,47)(41,53)
\put(45,45){$H$}
\put(70,23){$+$}
\end{picture}
\begin{picture}(60,10)(-10,55)
\Photon(0,25)(75,25){3}{9}
\DashLine(20,25)(8,50){3}
\DashLine(20,25)(32,50){3}
\DashLine(55,25)(43,50){3}
\DashLine(55,25)(67,50){3}
\Line(5,53)(11,47)
\Line(5,47)(11,53)
\Line(29,53)(35,47)
\Line(29,47)(35,53)
\Line(40,53)(46,47)
\Line(40,47)(46,53)
\Line(64,53)(70,47)
\Line(64,47)(70,53)
\put(90,23){$+ \cdots$}
\end{picture} \\
\begin{picture}(60,80)(80,20)
\ArrowLine(0,25)(50,25)
\put(70,22){$\Longrightarrow$}
\put(-15,23){$f$}
\put(-15,50){$(b)$}
\end{picture}
\begin{picture}(60,80)(40,20)
\ArrowLine(0,25)(50,25)
\put(65,23){$+$}
\end{picture}
\begin{picture}(60,80)(15,20)
\ArrowLine(0,25)(25,25)
\ArrowLine(25,25)(50,25)
\DashLine(25,25)(25,50){3}
\Line(22,53)(28,47)
\Line(22,47)(28,53)
\put(35,45){$H$}
\put(65,23){$+$}
\end{picture}
\begin{picture}(60,80)(-10,20)
\ArrowLine(0,25)(25,25)
\ArrowLine(25,25)(50,25)
\ArrowLine(50,25)(75,25)
\DashLine(25,25)(25,50){3}
\DashLine(50,25)(50,50){3}
\Line(22,53)(28,47)
\Line(22,47)(28,53)
\Line(47,53)(53,47)
\Line(47,47)(53,53)
\put(90,23){$+ \cdots$}
\end{picture}  \\
\end{center}
\caption[]{\it \label{fg:2} Generating (a) gauge boson and (b) fermion
  masses through interactions with the scalar background field.} 
\end{figure}

In a physically more intuitive picture, the masses of gauge bosons and
fermions may be built up by (infinitely) repeated interactions of these
particles when propagating through the background Higgs field.
Interactions of the gauge fields with the scalar background field, Fig.\ 
\ref{fg:2}a, and Yukawa interactions of the fermion fields with the
background field, Fig.\ \ref{fg:2}b, shift the masses of these fields
from zero to non-zero values:
\begin{equation}
\begin{array}{lrclclcl}
\displaystyle
(a) \hspace*{2.0cm} &
\displaystyle
\frac{1}{q^2} & \Rightarrow & 
\displaystyle \frac{1}{q^2} + \sum_j \frac{1}{q^2}
\left[ \left( \frac{g_W v}{\sqrt{2}} \right)^2 \frac{1}{q^2} \right]^j & 
= &
\displaystyle \frac{1}{q^2-M_V^2} & 
: & 
\displaystyle M_V^2 = g_W^2 \frac{v^2}{4}
\\ \\
(b) &
\displaystyle
\frac{1}{\not \! q} & \Rightarrow & \displaystyle \frac{1}{\not \! q} +
\sum_j \frac{1}{\not \! q} \left[ \frac{g_fv}{\sqrt{2}} \frac{1}{\not
\! q} \right]^j & = & \displaystyle \frac{1}{\not \! q-m_f} & : &
\displaystyle m_f = g_f \frac{v}{\sqrt{2}}
\end{array}
\end{equation}
Thus generating masses in the Higgs mechanism is equivalent to the
Archimedes effect: objects in media weigh different from objects in the
vacuum.
 \\

\noindent
{\underline{The Higgs Boson:}} The mass of the Higgs boson is determined
by the curvature of the self-energy potential $V$:
\begin{equation}
M_H^2=\lambda v^2
\end{equation}
It cannot be predicted in the Standard Model since the quartic coupling
$\lambda$ is an unknown parameter. Nevertheless, stringent upper and
lower bounds can be derived from internal consistency conditions and
from extrapolations of the model to high energies.\\

The Higgs boson has been introduced as a fundamental particle to
render $2 \rightarrow 2$ and $2 \rightarrow 3$ scattering amplitudes
involving longitudinally polarized $W$ bosons compatible with
unitarity. Based on the general principle of time-energy uncertainty,
particles must decouple from a physical system if their mass grows
indefinitely. The mass of the Higgs particle must therefore be bounded
to restore unitarity in the perturbative regime. From the asymptotic
expansion of the elastic $S$-wave amplitude for $W_LW_L$ scattering
including $W$ and Higgs exchanges, ${\cal A}(W_LW_L \rightarrow
W_LW_L) \rightarrow G_F M_H^2 / 4\sqrt 2\pi$, it follows (Lee 77)
that
\begin{equation}
M_H^2\leq 2\sqrt 2\pi/G_F \sim (850\, {\rm GeV})^2
\end{equation}
Within the canonical formulation of the Standard Model, consistency
conditions therefore require a Higgs mass roughly below 1 TeV.
\begin{figure}[b]
\begin{center}
\begin{picture}(90,80)(60,-10)
\DashLine(0,50)(25,25){3}
\DashLine(0,0)(25,25){3}
\DashLine(50,50)(25,25){3}
\DashLine(50,0)(25,25){3}
\put(-15,45){$H$}
\put(-15,-5){$H$}
\put(55,-5){$H$}
\put(55,45){$H$}
\end{picture}
\begin{picture}(90,80)(10,-10)
\DashLine(0,50)(25,25){3}
\DashLine(0,0)(25,25){3}
\DashLine(75,50)(50,25){3}
\DashLine(75,0)(50,25){3}
\DashCArc(37.5,25)(12.5,0,360){3}
\put(-15,45){$H$}
\put(-15,-5){$H$}
\put(35,40){$H$}
\put(80,-5){$H$}
\put(80,45){$H$}
\end{picture}
\begin{picture}(50,80)(-40,2.5)
\DashLine(0,0)(25,25){3}
\DashLine(0,75)(25,50){3}
\DashLine(50,50)(75,75){3}
\DashLine(50,25)(75,0){3}
\ArrowLine(25,25)(50,25)
\ArrowLine(50,25)(50,50)
\ArrowLine(50,50)(25,50)
\ArrowLine(25,50)(25,25)
\put(-15,70){$H$}
\put(-15,-5){$H$}
\put(35,55){$t$}
\put(80,-5){$H$}
\put(80,70){$H$}
\end{picture}  \\
\setlength{\unitlength}{1pt}
\caption[]{\label{fg:3} \it Diagrams generating the evolution of
the Higgs self-interaction $\lambda$.}
\end{center}
\end{figure}
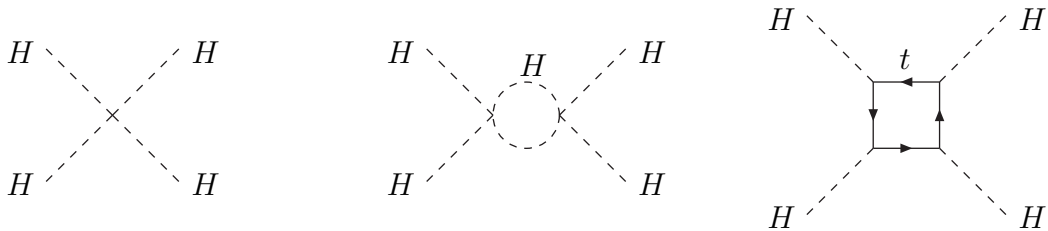

Quite restrictive bounds on the value of the Standard Model Higgs mass
follow from hypothetical assumptions on the energy scale $\Lambda$ up to
which the Standard Model can be extended before new physical phenomena
may emerge which are associated with strong interactions between the
fundamental particles.  The key to these bounds is the fact that quantum
fluctuations modify the self-interactions of the Higgs boson in such a
way that scattering processes, characterized by the energy scale $\mu$,
can still be described by the same form of interactions, yet with the
quartic coupling constant $\lambda$ replaced by an effective, energy
dependent coupling $\lambda(\mu)$. These quantum fluctuations are
described by Feynman diagrams as depicted in Fig.\ \ref{fg:3} (Cab 79,
Lin 86, She 89, Rie 97). The Higgs loop itself gives rise to an
indefinite increase of the coupling while the fermionic top-quark loop
drives, with increasing top mass, the coupling to smaller values,
finally even to values below zero. The variation of the effective
quartic Higgs coupling $\lambda(\mu)$ and the effective top-Higgs Yukawa
coupling $g_t(\mu)$ with energy may be written as
\begin{equation}
\begin{array}{lrl}
\displaystyle
\frac{d\lambda}{d{\log \mu^2}} = \frac{3}{8\pi^2}[\lambda^2 + \lambda
g_t^2-g_t^4] & \displaystyle
\quad \hspace{0.02cm} {\rm with} \quad & \lambda (v^2) = M_H^2/v^2
\nonumber \\[3ex]
\displaystyle
\frac{dg_t}{d{\log \mu^2}} = \frac{1}{32\pi^2}\left[ \frac{9}{2}g_t^3 -
  8g_tg_s^2\right] & \displaystyle
\quad {\rm with} \quad & g_t(v^2)=\sqrt 2 m_f /v
\end{array}
\end{equation}

For moderate top masses, the quartic coupling $\lambda$ rises
indefinitely, $d\lambda/d{\log \mu^2} \sim +\lambda^2$, and the coupling
becomes strong shortly before reaching the Landau pole: 
\begin{equation}
\lambda(\mu^2)=\frac{\lambda(v^2)}{1-\frac{3\lambda(v^2)}{8\pi^2}\log
\frac{\mu^2}{v^2}}
\end{equation}
Re-expressing the initial value of $\lambda$ by the Higgs mass, the
condition $\lambda(\Lambda)<\infty$, can be translated to an 
\underline{{\it upper bound}} on the Higgs mass:
\begin{equation}
M_H^2 {\raisebox{-0.13cm}{~\shortstack{$<$ \\[-0.07cm] $\sim$}}~}
 \frac{8\pi^2v^2}{3\log{(\Lambda^2/v^2)}} 
\end{equation}
This mass bound is related logarithmically to the energy $\Lambda$ up to
which the Standard Model is assumed to be valid. The maximal value of
$M_H$ for the minimal cut-off $\Lambda\sim 1$ TeV is given by $\sim 750$
GeV. This value is close to the estimate of $\sim 700$ GeV in lattice
calculations for $\Lambda \sim 1$ TeV, which allow the proper control of
non-perturbative effects near the boundary.

\begin{table}[b]
  \begin{center}
    \begin{tabular}{|l|c|} \hline
     $\Lambda$ & $M_H$ \\ \hline \hline
     1 TeV & 55 GeV $\leq M_H \leq$ 700 GeV \\ \hline
     $10^{19}$ GeV & 130 GeV $\leq M_H \leq$ 190 GeV \\ \hline
    \end{tabular}
    \caption{\it Higgs mass bounds for two values of the cut-off
      $\Lambda$.} 
    \label{tb:Higgsmass}
  \end{center}
\end{table}

A \underline{{\it lower bound}} on the Higgs mass can be based on the
requirement of vacuum stability (Cab 79, Lin 86, She 89, Rie 97, Alt
94).  Since top-loop corrections reduce $\lambda$ for increasing
top-Yukawa coupling, $\lambda$ becomes negative if the top mass becomes
too large. In this case, the self-energy potential would become deeply
negative and the ground state would not be stable any more. To avoid the
instability, the Higgs mass must exceed a minimal value for a given top
mass to balance the negative contribution. This lower bound depends on
the cut-off value $\Lambda$.

\begin{figure}[hbt]
\vspace*{0.5cm}
\hspace*{3.5cm}
\epsfxsize=9.5cm \epsfbox{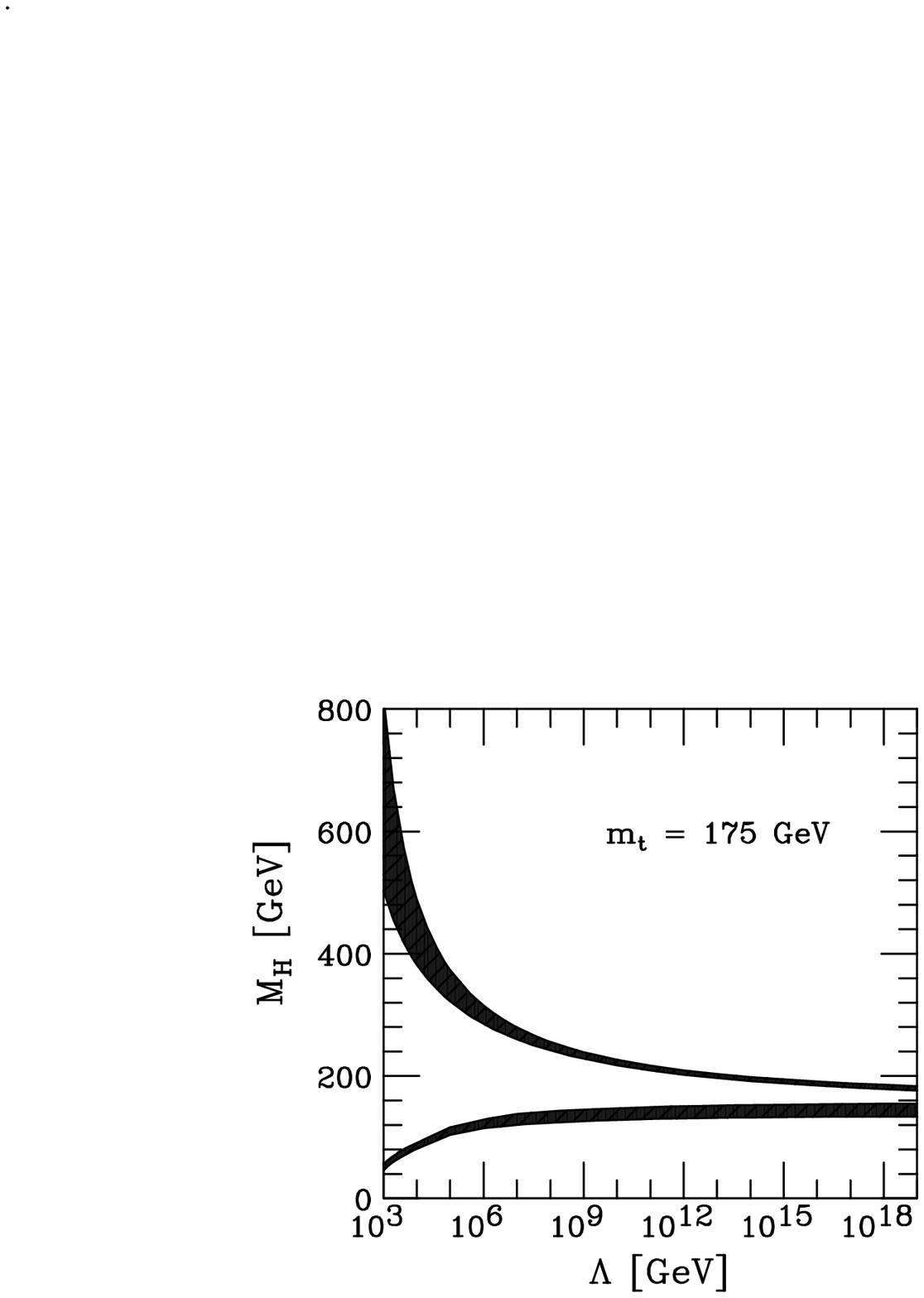}
\vspace*{-0.2cm}
\caption[]{\label{fg:4} \it Bounds on the mass of the Higgs boson in 
  the Standard Model. $\Lambda$ denotes the energy scale at which the
  Higgs-boson system of the Standard Model would become strongly
  interacting (upper bound); the lower bound follows from the
  requirement of vacuum stability; see Refs.\ (Cab 79, Lin 86, She 89,
  Rie 97, Alt 94).}
\end{figure}

\noindent
Only the leading contributions from $H$, $t$ and QCD loops are taken
into account. 

For any given $\Lambda$ the allowed values of $[m_t,M_H]$ pairs are
shown in Fig.\ \ref{fg:4}. For a top mass $m_t=175$ GeV, the allowed
Higgs mass values are collected in Table~\ref{tb:Higgsmass} for two
specific cut-off values $\Lambda$. If the Standard Model is assumed to
be valid up to the grand unification scale, the Higgs mass is restricted
to a narrow window between 130 and 190 GeV. The observation of a Higgs
mass above or below this window would demand a new strong interaction 
scale below the GUT scale.


\subsection{\sc Interactions between fermions and gauge bosons}

The basic Lagrangian for the interactions between leptons, quarks and
the electroweak gauge bosons $W^\pm$, $Z$ and $\gamma$ can be summarized
in the following condensed form:
\begin{eqnarray}
{\cal L}_{int}  = &-& \frac{g_W}{2\sqrt 2}\sum_k \bar f_k\gamma_\mu
(1-\gamma_5) \left(I^+W^+_\mu + I^-W^-_\mu\right) f_k 
\nonumber \\
&-& \frac{g_W}{4\cos\vartheta_W}\sum_k\bar f_k \gamma_\mu
\left(v_k - a_k \gamma_5\right) f_k Z_\mu 
\label{eq:Lint} \\
&-& e \sum_k q_k \bar f_k \gamma_\mu f_k A_\mu 
\nonumber
\end{eqnarray}
The first term describes the charged-current reactions, the second term
the neutral-current reactions, and the third term the parity-conserving
electromagnetic interactions. The coupling
\begin{equation}
e = g_W \sin\vartheta_W
\end{equation}
is the positron charge. $I^\pm$ are the isospin raising/lowering
matrices. The $SU(2)$ coupling $g_W$ is related to the Fermi coupling by
\begin{equation}
\frac{G_F}{\sqrt 2}=\frac{g_W^2}{8M^2_W}
\label{eq:mwgf}
\end{equation}
This relation follows from the local limit of the $W$ propagator
connecting the muonic and electronic currents in $\mu$ decay. The
relation (\ref{eq:mwgf}) will be modified by quantum effects, involving
the top-quark mass and the Higgs mass.

The vector and axial-vector charges of the second term of Eq.\ 
(\ref{eq:Lint}) are defined by the isospin $I_k^3$ and electric charge
$Q_k$ of the fermion field $f_k$:
\begin{eqnarray}
v_k &=& 2 I_k^3 - 4 Q_k\sin^2\vartheta_w \nonumber \\
a_k &=& 2 I^3_k  
\label{eq:vfaf}
\end{eqnarray}
$I_k^3 = \pm1/2$ for up- and down-type fields, respectively, and $Q_k
= 0$, $-1$ and $2/3$, $-1/3$ are the electric charges of the leptons
and quarks in units of the positron charge $e$. \\


\subsubsection{\sc Charged-current leptonic scattering processes}

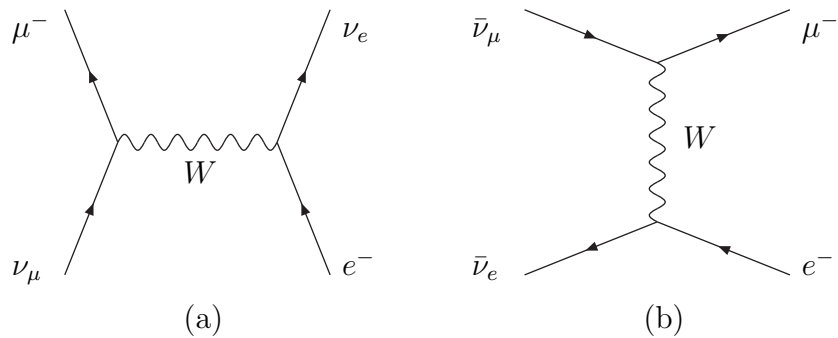
\begin{figure}[hbt]
\begin{center}
\begin{picture}(100,100)(20,10)
\ArrowLine(0,0)(20,50)
\ArrowLine(20,50)(0,100)
\Photon(20,50)(80,50){3}{6}
\ArrowLine(80,50)(100,100)
\ArrowLine(100,0)(80,50)
\put(-20,0){$\nu_\mu$}
\put(-20,90){$\mu^-$}
\put(105,0){$e^-$}
\put(105,90){$\nu_e$}
\put(45,35){$W$}
\put(45,-20){(a)}
\end{picture}
\begin{picture}(100,100)(-50,10)
\ArrowLine(50,20)(0,0)
\ArrowLine(100,0)(50,20)
\Photon(50,20)(50,80){3}{6}
\ArrowLine(0,100)(50,80)
\ArrowLine(50,80)(100,100)
\put(-20,0){$\bar \nu_e$}
\put(-20,90){$\bar \nu_\mu$}
\put(105,90){$\mu^-$}
\put(105,0){$e^-$}
\put(60,48){$W$}
\put(45,-20){(b)}
\end{picture} \\[8mm]
\end{center}
\caption[]{ \label{fg:5} \it The process $\nu_\mu e^- \rightarrow
  \nu_e\mu^-$.} 
\end{figure}

The process 
\begin{equation}
\nu_\mu e^- \rightarrow \nu_e\mu^-
\end{equation}
is a particularly instructive example for charged-current [${\cal CC}$]
processes. The reaction is mediated by the exchange of a $W$ boson in
the $t$-channel, cf.\ Fig.\ \ref{fg:5}a. If the total energy in the
center-of-mass system is small compared to the $W$ mass, the scattering
process is an $S$-wave reaction since only the left-handed components of
both incoming particles are active; the angular distribution, as a
result, is isotropic. The total cross section is given by
\begin{equation}
\label{cs}
\sigma [ \nu_\mu e^- \rightarrow \nu_e\mu^-] = \frac{G_F^2s}{\pi}
\end{equation}
in the intermediate-energy range $m_f\ll \sqrt s\ll M_W$ where all
fermion masses can be neglected.\\

This process may be contrasted with the reaction
\begin{equation}
\bar \nu_e e^- \rightarrow \bar \nu_\mu\mu^-
\end{equation}
which proceeds through the exchange of a $W$ boson in the $s$-channel,
Fig.\ \ref{fg:5}b.  Since the right-handed antineutrino in the initial
state interacts with the left-handed component of the electron, the
overall spin is one. Since backward scattering is forbidden by angular
momentum conservation, the angular dependence of the cross section must
be of the form $\sim (1+\cos\theta)^2$. As a result, the total cross
section is reduced by a factor 1/3 compared to (\ref{cs}):
\begin{equation}
\sigma [\bar \nu_e e^- \rightarrow \bar\nu_\mu\mu^-] =
\frac{1}{3}\frac{G_F^2s}{\pi} 
\end{equation}
These two cross sections are the prototypes for charged-current
reactions.\\


\subsubsection{\sc Deep-inelastic charged-current neutrino-nucleon
  scattering} 

\begin{figure}[bt]
\begin{center}
\begin{picture}(100,100)(0,10)
\ArrowLine(0,75)(50,75)
\ArrowLine(50,75)(100,100)
\ArrowLine(30,25)(50,25)
\ArrowLine(50,25)(100,0)
\ArrowLine(30,21)(50,21)
\Line(50,21)(100,21)
\ArrowLine(30,17)(50,17)
\Line(50,17)(100,17)
\Photon(50,25)(50,75){3}{6}
\Line(0,22)(30,22)
\Line(0,20)(30,20)
\GOval(24,21)(15,7)(0){0.7}
\put(-5,80){$\nu_e$}
\put(-5,8){${\cal N}$}
\put(105,0){$q$}
\put(90,100){$e^-$}
\put(55,47){$W$}
\end{picture}
\end{center}
\caption[]{ \label{fg:6} \it Neutrino-nucleon scattering.}
\end{figure}
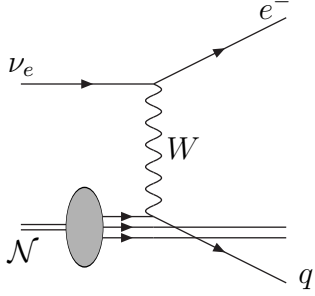

${\cal CC}$ interactions of neutrinos and quarks can be realized in
deep-inelastic neutrino-nucleon scattering at high
energies\footnote{This chapter will focus on properties of the
  electroweak interactions at intermediate energies $m_{\cal N} \ll
  \sqrt{s} \ll M_W$. The QCD aspects of deep-inelastic scattering are
  described in more generality in a different chapter of this volume.}.
The neutrino $\nu_\ell$ is transformed into a charged lepton $\ell = e,
\mu$ which is observed in the final state:
\begin{eqnarray}
\nu_\ell {\cal N}     & \rightarrow & \ell^- X\\
\bar\nu_\ell {\cal N} & \rightarrow & \ell^+ X
\end{eqnarray}
In QCD, the asymptotically free theory of the strong interactions,
these processes are built-up by the incoherent superposition of
neutrino-quark scattering processes (Bjo 70, Fey 72). In a simplified
picture, ignoring for the moment more complicated processes including
gluons, these are just the elastic scattering processes (cf.\ Fig.\ 
\ref{fg:6}):
\begin{eqnarray}
\nu_\ell d      & \rightarrow & \ell^-u\\
\bar \nu_\ell u & \rightarrow & \ell^+d 
\end{eqnarray}
\noindent
The first subprocess is mediated by the transfer of a $W^+$ boson from
the lepton system to the quark system in the $t$-channel while the
second subprocess is mediated by the exchange of a $W^-$ boson. The
additional strange quark targets, antiquark targets, etc. contribute
in a similar way.  According to the spin-0 and spin-1 rules described
above, the corresponding cross sections for intermediate energies are
given as
\begin{eqnarray}
\sigma [\nu_\ell d \rightarrow \ell^-u]     & = & \frac{G_F^2\hat
  s}{\pi} \\
\sigma [\bar\nu_\ell u \rightarrow \ell^+d] & = & \frac{G_F^2\hat
  s}{3\pi} 
\end{eqnarray}
The energy squared $\hat s=xs$ in the center-of-mass system is reduced
by the Bjorken factor $x$ in the neutrino-quark subsystem with respect
to the total cms-energy squared $s$ of the neutrino-nucleon system; $x$
is the fraction of nucleon energy carried by the struck quark in the
center-of-mass system. Denoting the density of up or down quarks in the
isoscalar nucleon state ${\cal N}=\frac{1}{2} (P+N)$ by $q(x)$, the
antiquarks by $\bar q(x)$, the neutrino-nucleon and the
antineutrino-nucleon cross section may be written as
\begin{eqnarray}
\sigma [\nu_\ell {\cal N} \rightarrow \ell^- X] &=&
\frac{G_F^2s}{\pi} \langle x\rangle_q \left(1 + 
 \frac{\langle x \rangle_{\bar{q}}}{3 \langle x\rangle_{q}}\right) \\
\sigma [\bar\nu_\ell {\cal N} \rightarrow \ell^+ X] &=&
\frac{G_F^2s}{3 \pi} \langle x\rangle_q \left(1 + 
 \frac{3 \langle x\rangle_{\bar{q}}}{\langle x \rangle_{q}}\right) 
\end{eqnarray}
with 
\[ 
\langle x\rangle_q = \int_0^1 dx\, x\, q(x) 
\qquad {\rm and} \qquad
\langle x\rangle_{\bar q} = \int_0^1 dx\, x\, \bar q(x) 
\]
measuring the overall momentum of the nucleon residing in the quarks and
antiquarks. The additional contributions due to strange quarks can
easily be included. This representation is valid for total energies
squared $s\ll M_W^2$.\\

The experimental analysis of deep-inelastic neutrino-nucleon and
antineutrino-nucleon scattering, complementing deep-inelastic scattering
of charged leptons mediated by photon-exchange, has led to a clear
picture of the basic constituents of matter: \\

\noindent
{\bf (i)} By observing the scaling behavior of the cross sections with
energy,
\begin{equation}
\sigma_\nu, \sigma_{\bar\nu} \propto s = 2m_{\cal N} E_{\nu, \bar{\nu}}
\end{equation}
it could be proved experimentally, that the nucleons are built-up by
light pointlike constituents. \\

\noindent
{\bf (ii)} Comparing the cross sections for antineutrino with neutrino
beams, it turns out that
\begin{equation}
\sigma_{\bar\nu}/\sigma_\nu \approx 1/3
\end{equation}
It follows from this observation that the constituent targets are
spin-1/2 fermions. Combining this observation with the information
obtained from measurements of $e{\cal N} \rightarrow eX$ scattering,
leads to the conclusion that they are fractionally charged quarks. \\

\noindent
{\bf (iii)} The quantities $\langle x\rangle_q$ and $\langle
x\rangle_{\bar q}$ measure the energy fraction residing in quarks and
antiquarks in a fast moving nucleon.  From the experimentally determined
values
\begin{eqnarray*}
  \langle x\rangle_q &\approx& 0.5        \\
  \langle x\rangle_{\bar q} &\approx& 0.05
\end{eqnarray*}
it can be concluded that most of the flavored constituents of a nucleon
are matter particles and only a small fraction consists of antimatter
particles. However, since $\langle x\rangle_q \approx 1/2$, only half of
the energy is carried by flavored constituents while the other half must
be attributed to non-flavored constituents which do not participate in
the electroweak interactions.  They can be identified with
flavor-neutral gluons which provide the binding between the quarks in
the nucleons.

With rising energy, the momentum transfer $Q^2$ from the leptons to the
quarks becomes so large that the $W$-boson exchange will not be a local
process anymore. When $Q^2$ is of the same order as $M_W^2$, the
$W$-boson propagates between the leptons and the quarks. This gives rise
to partial waves of higher angular momenta in the scattering processes
that affect the angular distributions of the final state leptons. The
scattering angle $\theta_*$ in the center-of-mass system of the $(\nu
q)$ pair is related to the relative energy transfer $y$ in the
laboratory frame by the formula $y=(1-\cos\theta_*)/2$. The differential
cross sections may therefore be written as
\begin{eqnarray}
\frac{d\sigma}{dy}[\nu_\ell d \rightarrow \ell^-u] & = & \frac{G_F^2\hat
  s}{\pi}\frac{1}{(1+Q^2/M_W^2)^2}\\[2ex]
\frac{d\sigma}{dy}[\bar\nu_\ell u \rightarrow \ell^+d] & = &
\frac{G_F^2\hat s}{\pi} \frac{(1-y)^2}{(1+Q^2/M_W^2)^2}
\end{eqnarray}
Moreover, as a result of QCD radiative corrections, the quark densities
$q$ and $\bar q$ become logarithmically dependent on the momentum
transfer $Q^2$. The total cross sections
\begin{equation}
\sigma [\nu_\ell {\cal N} \rightarrow \ell^-X] =
\frac{G_F^2s}{\pi}\int_0^1 \int_0^1 \frac{dx\,dy}{(1+Q^2/M_W^2)^2}
\left[ x\,q(x,Q^2) + x\,\bar q(x,Q^2)(1-y)^2 \right] 
\end{equation}
and
\begin{equation}
\sigma [\bar\nu_\ell {\cal N} \rightarrow \ell^+ X] =
\frac{G_F^2s}{\pi}\int_0^1 \int_0^1 \frac{dx\,dy}{(1+Q^2/M_W^2)^2}  
\left[ x\,q(x,Q^2) (1-y)^2+ x\,\bar q(x,Q^2) \right] 
\end{equation}
are therefore damped at high energies and, with $Q^2=xys$, they do not
rise any more linearly with $s$.

\begin{figure}[hbt]
\begin{center}
\hspace*{0cm}
\epsfxsize=10cm \epsfbox{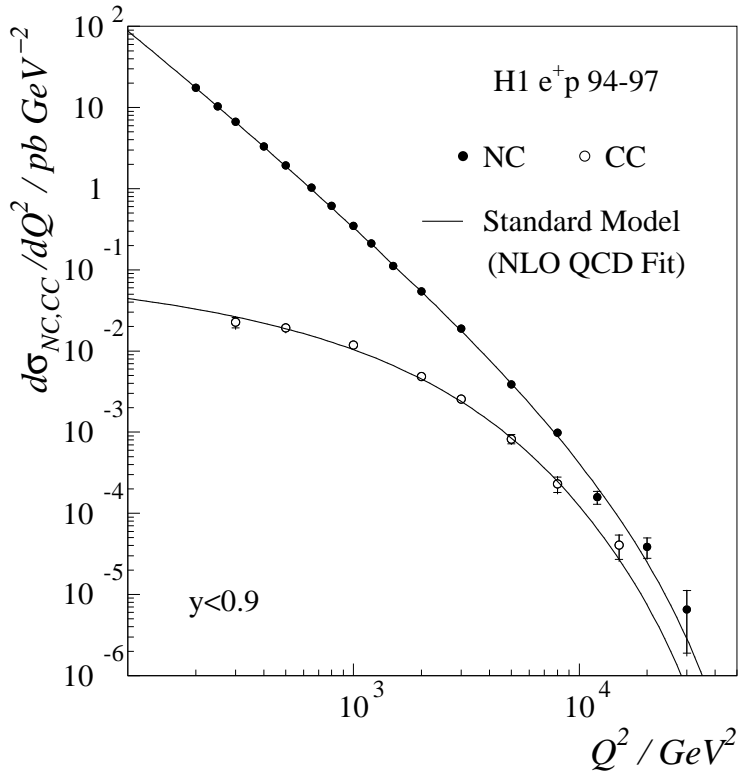}
\vspace*{-1.0cm}
\end{center}
\caption[]{\label{fg:7} \it Cross sections for deep inelastic ${\cal
    NC}$ and ${\cal CC}$ scattering of electrons and positrons at HERA, 
  cf.\ Ref.\ (Adl 00).}
\end{figure}

The damping is a consequence of the short-range character of the weak
force, restricted to a radius of the order of the Compton wave length
$\lambda_W = M_W^{-1}$ of the $W$ boson. Asymptotically the cross
sections approach the limit $\sigma \sim g_W^4\lambda^2_W\sim
G_F^2M_W^2$.  The large-$Q^2$ behavior of the ${\cal CC}$, as well as of
the ${\cal NC}$ cross sections in the equivalent processes $e^+P
\rightarrow \bar\nu_e X$ and $e^+P \rightarrow e^+ X$ has been observed
at HERA for the cms-energy $\sqrt s \simeq 300$ GeV which exceeds $M_W$
by nearly a
factor four, cf.\ Fig.\ \ref{fg:7} (Adl 00).\\

\subsubsection{\sc Neutral-current leptonic scattering processes}

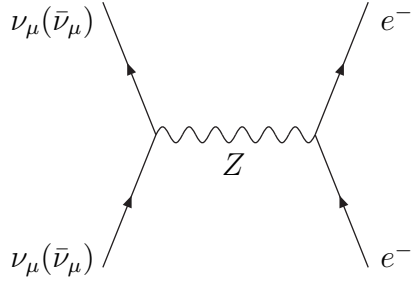
\begin{figure}[hbt]
\begin{center}
\begin{picture}(100,100)(0,0)
\ArrowLine(0,0)(20,50)
\ArrowLine(20,50)(0,100)
\Photon(20,50)(80,50){3}{6}
\ArrowLine(80,50)(100,100)
\ArrowLine(100,0)(80,50)
\put(-35,0){$\nu_\mu (\bar \nu_\mu)$}
\put(-35,90){$\nu_\mu (\bar \nu_\mu)$}
\put(105,0){$e^-$}
\put(105,90){$e^-$}
\put(45,35){$Z$}
\end{picture} \\
\end{center}
\caption[]{ \label{fg:8} \it Muon-(anti)neutrino electron scattering
  mediated by $Z$-exchange.} 
\end{figure}

The elastic scattering of muon-neutrinos or muon-antineutrinos on
electrons has been one of the classical experiments in which
neutral-current [${\cal NC}$] interactions have been established in the
electroweak sector of the Standard Model:
\begin{eqnarray}
\nu_\mu e^-     & \rightarrow & \nu_\mu e^-\\ 
\bar\nu_\mu e^- & \rightarrow & \bar\nu_\mu e^-  
\end{eqnarray}
These scattering processes are mediated solely by the exchange of a $Z$
boson, Fig.\ \ref{fg:8}. The scattering experiments can be performed by
shooting a beam of muon-neutrinos and antineutrinos on electrons in the
shells of atomic targets and observing the electron final state. The
observation of the single electron in the final state of
neutrino-electron scattering (Has 73) marked a break-through in the
development of particle physics, since it provided the first empirical
proof of the existence of weak neutral-current interactions.

\noindent
Combining the vector and axial-vector couplings to left- and
right-handed couplings,
\begin{equation}
\begin{array}{c}
C_L^i = \frac{1}{4}(v_i + a_i)\\
C_R^i = \frac{1}{4}(v_i - a_i)  
\end{array}
\end{equation}
the cross sections can be cast into the simple form
\begin{eqnarray}
\sigma \left[ \nu_\mu e^- \rightarrow \nu_\mu e^- \right] & = &
\frac{G_F^2s}{\pi}\left[C_L^2+\frac{1}{3}C_R^2\right] \\ 
\sigma \left[\bar\nu_\mu e^- \rightarrow \bar\nu_\mu
e^-\right]&=&\frac{G_F^2s}{\pi}\left[  \frac{1}{3}C_L^2+C_R^2\right]  
\end{eqnarray}
Detailed measurements of these neutral-current cross sections have been
exploited to determine the electroweak mixing angle
$\sin^2\vartheta_W$.\\ 


\subsubsection{\sc Deep-inelastic neutral-current scattering}

The analogous ${\cal NC}$ processes in deep-inelastic neutrino-nucleon
scattering (Has 73a),
\begin{eqnarray}
\nu_\mu{\cal N}     & \rightarrow & \nu_\mu X\\
\bar\nu_\mu{\cal N} & \rightarrow & \bar\nu_\mu X 
\end{eqnarray}
provide an excellent method for the measurement of the electroweak
mixing angle. In the approximation in which the (small) antiquark
content of the nucleon is neglected, the ratios of the ${\cal NC}$
over the ${\cal CC}$ neutrino and antineutrino-nucleon cross sections,
$R_\nu$ and $R_{\bar\nu}$, can be expressed (Seh 73) solely by the
electroweak mixing angle $\sin^2\vartheta_W$:
\begin{eqnarray}
R_\nu       & = & \sigma(\nu_\mu \rightarrow \nu_\mu)\big /
\sigma(\nu_\mu \rightarrow \mu^-) = \frac{1}{2} - \sin^2\vartheta_W +
\frac{20}{27}\sin^4\vartheta_W\\ 
R_{\bar\nu} & = &\sigma(\bar\nu_\mu \rightarrow \bar\nu_\mu)\big /
\sigma(\bar\nu_\mu \rightarrow \mu^+) = \frac{1}{2} - \sin^2\vartheta_W
+ \frac{20}{9}\sin^4\vartheta_W
\end{eqnarray}
Including the corrections due to the antiquarks in the nucleon, the
deep-inelastic neutrino-scattering experiments allow for a high
precision determination of $\sin^2\vartheta_W= 0.2253(22)$ (Cas 98).
Higher-order QCD and electroweak corrections are included in the 
experimental analysis.\\

Besides neutral-current neutrino processes, also deep-inelastic electron
scattering on nucleons is affected by $Z$ exchange at large momentum
transfer. The $Z$ exchange interferes with the $\gamma$ exchange in the
elastic scattering of electrons on quarks:
\begin{equation}
eq \stackrel{\gamma ,Z} \longrightarrow eq
\end{equation}
and the additional $Z$ contributions modify the cross sections as
predicted in quantum electrodynamics. Moreover, since the electroweak
theory is parity-violating, the cross sections for the scattering of
electrons with left-handed and right-handed polarization are different.
This is apparent at the level of the subprocesses,
\begin{eqnarray}
\frac{d\sigma}{dy} \left[ e_L^-q \rightarrow e_L^-q \right] &=&
\frac{4\pi\alpha^2}{Q^4} \left[ Q_{LL}^2+Q_{LR}^2(1-y)^2\right]\\
 \frac{d\sigma}{dy} \left[ e_R^-q \rightarrow e_R^-q \right] &=&
\frac{4\pi\alpha^2}{Q^4} \left[ Q_{RL}^2(1-y)^2+Q_{RR}^2\right] 
\end{eqnarray}
in the usual notation.  The generalized charges in these expressions
are defined by the electric and $Z$ charges of electron and quark; they
also include the $Z$ propagator:
\begin{equation}
Q_{ij} = -Q_q+\frac{\sqrt{2}G_F M_Z^2}{\pi\alpha} C_i^e C_j^q
\frac{Q^2}{Q^2+M_Z^2} \hspace{1cm} (i,j=L,R)
\end{equation}
The $Z$ exchange in deep-inelastic electron scattering has been observed
experimentally at SLAC and at HERA. 

\noindent
{\bf (i)} At the SLAC polarization experiment (Pre 79) the
parity-violating asymmetry between the cross sections for left- and
right-handedly polarized electrons
\[
A=\frac{\sigma_R-\sigma_L}{\sigma_R+\sigma_L} 
\] 
had been studied at $Q^2\ll M_Z^2$. The value of the asymmetry is
predicted in the Standard Model to be
\begin{equation}
A=\frac{3G_FQ^2}{5\sqrt 2\pi \alpha}\left[ \left( -\frac{3}{4}+\frac{5}{3} 
\sin^2\vartheta_W\right)+\left(-\frac{3}{4}+3\sin^2
\vartheta_W\right)\frac{1-(1-y)^2}{1+(1-y)^2}\right]  
\end{equation}
The observation of a non-zero asymmetry proved, in a model-independent
way, the parity violation of the electroweak neutral-current and
$Z$-boson interactions. 

\begin{figure}[hbt]
\begin{center}
\hspace*{0.1cm}
\epsfxsize=7.9cm \epsfbox{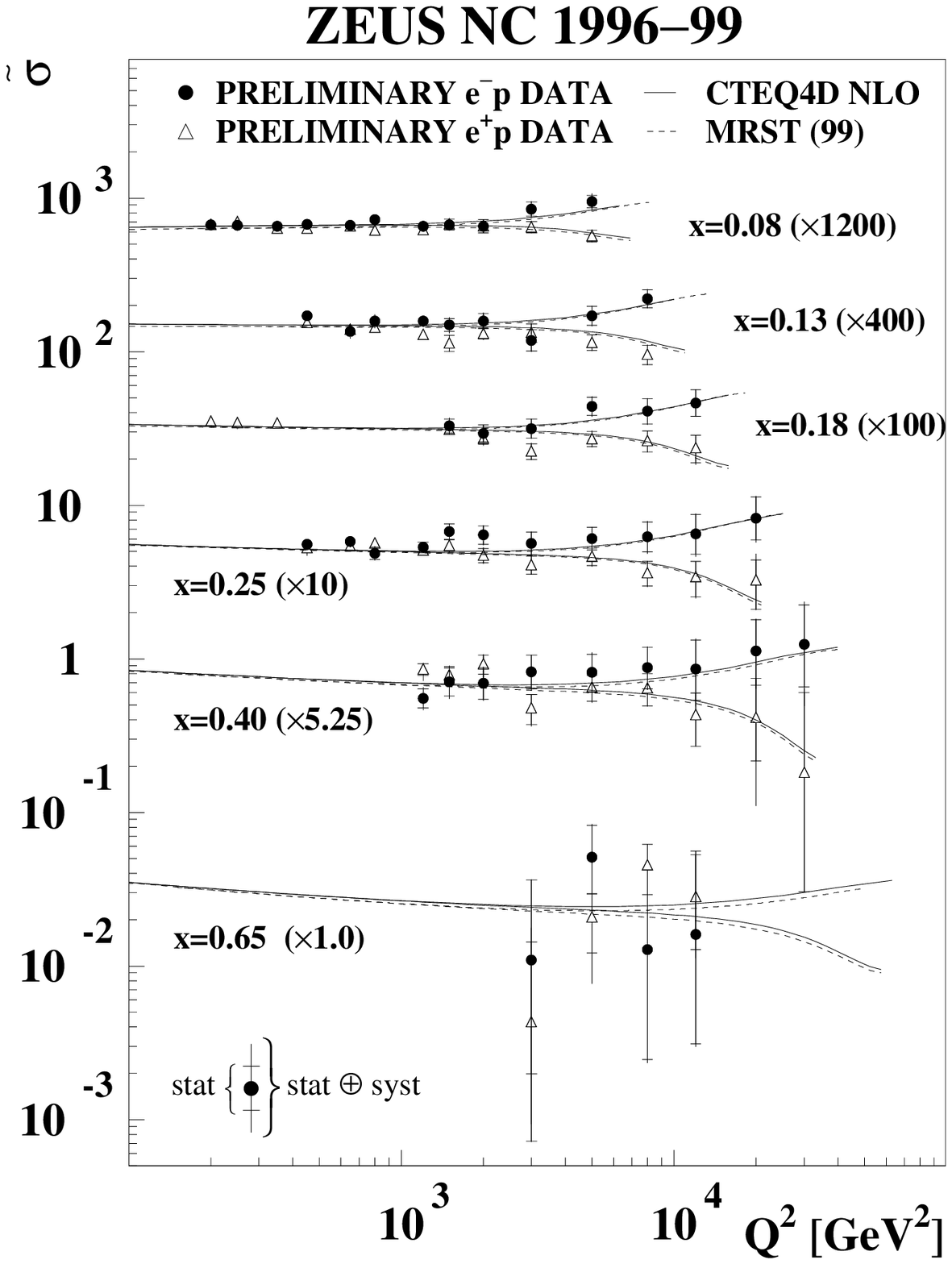}
\vspace*{-1.0cm}
\end{center}
\caption[]{\label{fg:10} \it Effect of $Z$-exchange on deep-inelastic
  ${\cal NC}$ scattering at HERA (Bre 00).}
\end{figure}

\noindent
{\bf (ii)} For momentum transfer $Q^2
{\raisebox{-0.13cm}{~\shortstack{$>$ \\[-0.07cm] $\sim$}}~} M_Z^2$ the
dynamical effect of $Z$-boson exchange in deep-inelastic electron or
positron-proton scattering has been observed at HERA, Fig.\ \ref{fg:10}.
The cross sections deviate in a characteristic way from the prediction
of pure photon exchange.

\subsubsection{\sc Forward-backward asymmetry of leptons in $e^+e^-$
  annihilation}

\begin{figure}[hbt]
\begin{center}
\begin{picture}(100,110)(0,0)
\ArrowLine(0,0)(50,20)
\ArrowLine(50,20)(100,0)
\Photon(50,20)(50,80){3}{6}
\ArrowLine(50,80)(0,100)
\ArrowLine(100,100)(50,80)
\put(-20,0){$e^-$}
\put(-20,90){$\mu^-$}
\put(105,90){$\mu^+$}
\put(105,0){$e^+$}
\put(60,48){$\gamma$}
\end{picture}
\begin{picture}(100,100)(-50,0)
\ArrowLine(0,0)(50,20)
\ArrowLine(50,20)(100,0)
\Photon(50,20)(50,80){3}{6}
\ArrowLine(50,80)(0,100)
\ArrowLine(100,100)(50,80)
\put(-20,0){$e^-$}
\put(-20,90){$\mu^-$}
\put(105,90){$\mu^+$}
\put(105,0){$e^+$}
\put(60,48){$Z$}
\end{picture} \\
\end{center}
\caption[]{ \label{fg:11} \it Muon-pair production in $e^+e^-$ collisions.}
\end{figure}
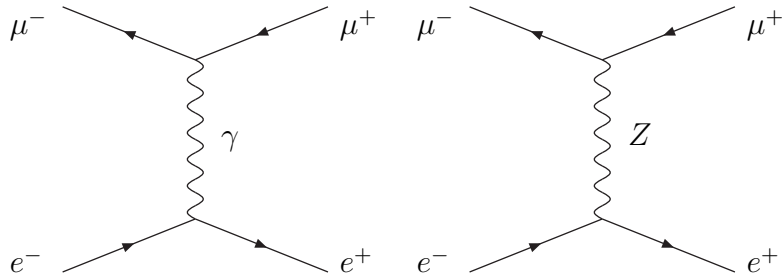

At low energies, the production of charged muon pairs 
\begin{equation}
e^+e^- \rightarrow \mu^+\mu^-
\end{equation}
is mediated to leading order by the exchange of a photon in the
$s$-channel. With rising cms-energy also $Z$ exchange becomes effective,
cf.\ Fig.\ \ref{fg:11}. The value of the annihilation cross section is
therefore modified with respect to the QED prediction. In addition, one
expects a non-zero value for the forward-backward asymmetry of the
observed (negatively) charged leptons with respect to the flight
direction of the electron in the laboratory frame:
\begin{equation}
A_{FB}=\frac{\sigma_F-\sigma_B}{\sigma_F+\sigma_B}
\end{equation}

\noindent
Even though a non-zero value of the forward-backward asymmetry does not
probe parity violation in a model-independent way, the observable is
nevertheless of great interest. $A_{FB}$ vanishes at low energies where
the $Z$ exchange is suppressed and only the photon is exchanged between
initial and final state particles.  However $A_{FB}$ is non-zero for
$Z$-exchange contributions, reflecting, indirectly though, the parity
violating coupling of the $Z$ boson to a lepton pair. In leading order
at energies squared $s\ll M_Z^2$, $A_{FB}$ may be written for muon-pair
production as
\[
A_{FB}\left[ e^+e^- \rightarrow \mu^+\mu^-\right] = \frac{G_Fs}{2
  \pi\alpha}a_e a_\mu 
\] 
A non-zero value of this asymmetry had been measured at the early
low-energy $e^+e^-$ colliders PETRA, PEP and TRISTAN. \\


\subsubsection{\sc The production of $W^\pm$ and $Z$ bosons in hadron
  collisions}

While the observation of ${\cal NC}$ reactions $\nu_\mu e \rightarrow
\nu_\mu e$ and $\nu_\mu{\cal N} \rightarrow \nu_\mu X$ had been a
tremendously important element in the understanding of the structure
of the fundamental forces, the weak interactions in particular, these
phenomena could still be interpreted as effective low-energy phenomena
without the detailed knowledge of the microscopic dynamics.\\

The first crucial step in establishing gauge theories as the basic
theories of the electroweak forces, has been the direct observation of
the heavy gauge particles $W^\pm$ and $Z$.\\

These experiments were performed in colliding proton/antiproton beams at
the CERN Sp$\bar{\rm p}$S:
\begin{eqnarray}
p\bar p& \rightarrow & W^\pm X\\
p\bar p& \rightarrow & Z X  
\end{eqnarray}
Protons and antiprotons simply act in these processes as sources of
quarks and antiquarks and single $W^\pm$ and $Z$ bosons are generated in
Drell-Yan type subprocesses, cf.\ Fig.\ \ref{fg:13},
\begin{eqnarray}
u+\bar d\: ,\: \bar u +d & \rightarrow & W^{\pm} \\
u+\bar u\: ,\: d+\bar d  & \rightarrow & Z 
\end{eqnarray}
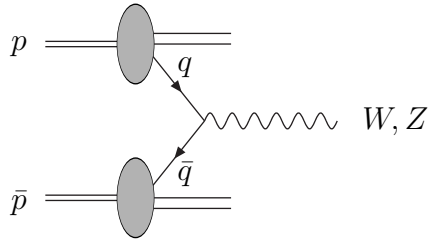
\begin{figure}[hbt]
\begin{center}
\begin{picture}(100,90)(0,10)
\ArrowLine(30,75)(50,50)
\ArrowLine(50,50)(30,25)
\Photon(50,50)(100,50){3}{6}
\Line(30,79)(60,79)
\Line(30,83)(60,83)
\Line(-10,78)(30,78)
\Line(-10,80)(30,80)
\GOval(24,79)(15,7)(0){0.7}
\Line(30,21)(60,21)
\Line(30,17)(60,17)
\Line(-10,22)(30,22)
\Line(-10,20)(30,20)
\GOval(24,21)(15,7)(0){0.7}
\put(40,69){$q$}
\put(40,28){$\bar q$}
\put(110,47){$W,Z$}
\put(-23,77){$p$}
\put(-23,17){$\bar{p}$}
\end{picture}
\end{center}
\caption[]{ \label{fg:13} \it The Drell-Yan subprocess $q + \bar q
  \rightarrow W,Z$.}
\end{figure}

\noindent
The cross sections for $p\bar p$ collisions are given by the
Breit-Wigner cross sections of the subprocesses\footnote{$BW$ denotes
  the normalized Breit-Wigner function $BW(\hat{s} - M^2) = M\Gamma /
  \pi[(\hat{s}-M^2)^2 + M^2\Gamma^2]$.}
\begin{eqnarray}
\hat\sigma [q\bar q^\prime \rightarrow W^\pm] & = & \frac{\sqrt{2}G_F
  \pi \hat s}{3} \times BW(\hat s - M_W^2) \\
\hat\sigma [q\bar q \rightarrow Z] & = & \frac{\sqrt{2}G_F \pi \hat
  s}{12} (v_q^2 + a_q^2) \times BW(\hat s - M_Z^2) 
\end{eqnarray}
convoluted with the number of quark-antiquark pairings generating the
invariant energy $\sqrt{\hat s}=M_{W,Z}$:
\begin{equation}
\frac{d{\cal L}}{d\tau}=\int_\tau^1\frac{dx}{x}q(x,M^2_{W,Z})\,
\bar{q}(\tau/x,M^2_{W,Z})
\end{equation}
with the scaling variable defined as $\tau =M^2_{W,Z}/s$ in the
narrow-width approximation. QCD corrections modify these predictions
slightly. Virtual gluon exchange between quark and antiquark in the
initial state affect the $q\bar q W^\pm$ and $q\bar q Z$ vertices;
moreover, gluons may be radiated off the quarks and antiquarks, the
leading part of which can be taken into account by scale dependent quark
densities; electroweak bosons can also be generated in the inelastic
Compton-like processes $gq \rightarrow q^\prime W^\pm$ and $gq
\rightarrow qZ$. These QCD corrections can be summarized globally in a
$K$ factor which turns out to be $K\approx 1.4$ for Sp$\bar{\rm p}$S
energies of 630 GeV. The final form of the cross sections may therefore
be written as
\begin{eqnarray}
\sigma(p\bar p \rightarrow W^\pm) & = & \frac{\sqrt{2}G_F\pi}{3} 
  K \tau \frac{d{\cal L}^{ud}}{d\tau} \\ 
\sigma(p\bar{p} \rightarrow Z) & = & \frac{\sqrt{2}G_F\pi}{12}
  \sum_{q = u,d} (v_q^2 + a_q^2) K \tau \frac{d{\cal L}^{qq}}{d\tau} 
\end{eqnarray}
The numerical values of the cross sections can easily be determined 
from these expressions. \\

\begin{figure}[hbt]
\begin{center}
\hspace*{0.0cm}
\epsfxsize=12cm \epsfbox{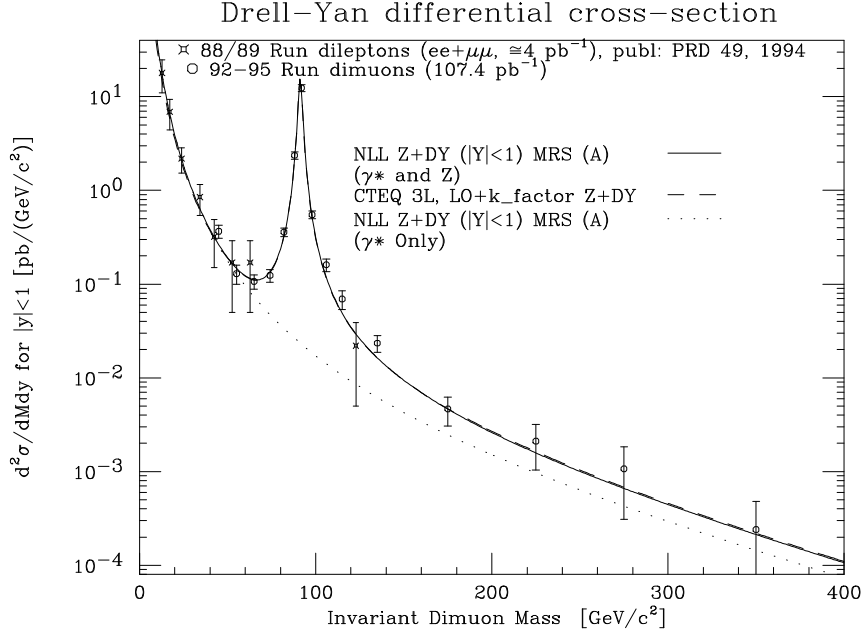}
\vspace*{-1.0cm}
\end{center}
\caption[]{\label{fg:14} \it Drell-Yan cross section for $\mu$ pairs
  (Abe 99).}
\end{figure}

In the experiments (Arn 83, Ban 83, Arn 83a, Bag 83), the $W^\pm$ and
$Z$ bosons have to be identified by their decay products. The leptonic
decays,
\begin{eqnarray}
W^\pm& \rightarrow &\ell\bar\nu_\ell\quad {\rm and} \quad \bar
\ell\nu_\ell\\ 
Z{\phantom{\pm}} & \rightarrow &\ell^+ \ell^- \quad{\rm for}\quad
\ell=e,\nu   
\end{eqnarray}
provided a small but very clean sample of events which have been used
to study the properties of the electroweak gauge bosons: 

\noindent
{\bf (i)} $Z$ decays generate a Breit-Wigner distribution in the
invariant mass $M_{\ell\ell}$ of the $\ell^+ \ell^-$ pairs,
\begin{equation}
Z \rightarrow \ell^+\ell^-:\qquad \frac{d\rho}{dM^2_{\ell\ell}} =
\frac{1}{\pi} 
\frac{M_Z\Gamma_Z} {[M_{\ell\ell}^2-M_Z^2]^2+M_Z^2\Gamma_Z^2}
\end{equation}
giving rise to a narrow peak in the $M_{\ell\ell}$ distribution near
the mass of the $Z$ boson for a small total width $\Gamma_Z\approx
2.49$ GeV, cf.\ Fig.\ \ref{fg:14}. 

\noindent
{\bf (ii)} Due to the escaping neutrino the charged $W^\pm$ bosons
cannot be observed as a leptonic Breit-Wigner peak. However, kinematics
and geometry conspire in such a way that a Jacobian peak is generated in
the transverse momentum of the observed charged lepton.  When the
protons and antiprotons split into quarks and antiquarks, these partons
move parallel to the hadrons with negligible transverse momenta. As a
result, also the $W^\pm$ bosons move parallel to the $p\bar p$ beam
axis. Since, on the other hand, the change of area is singular when the
poles are approached on a sphere, the transverse momentum distribution
of the charged lepton, Lorentz-invariant for boosts along the $p \bar p$
axis, must also be singular near its maximum.

\begin{figure}[hbt]
\begin{center}
\hspace*{-1.0cm}
\epsfxsize=9.5cm \epsfbox{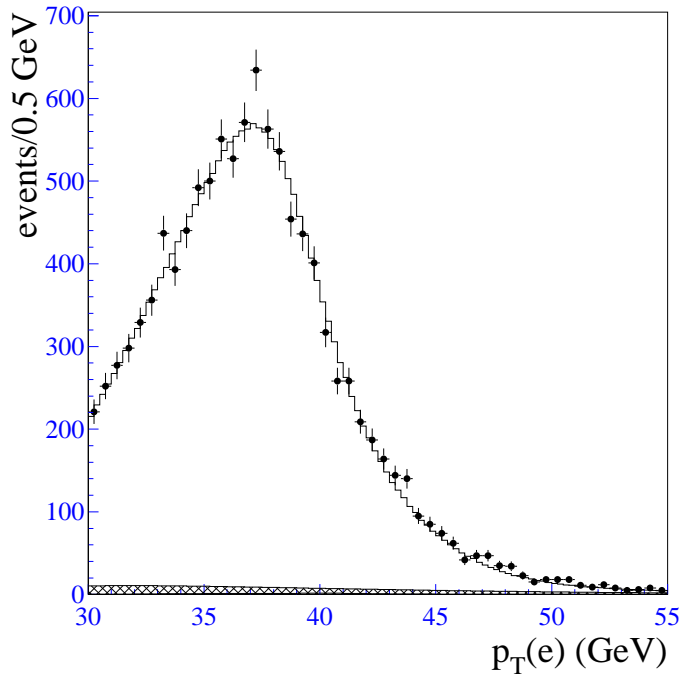}
\vspace*{-1.0cm}
\end{center}
\caption[]{\label{fig:wpte} \it Transverse momentum distribution of
  electrons originating from the decay of a $W$ boson in hadron
  collisions as observed by the D0 detector at the Fermilab (Abb 00).}
\end{figure}

Joining the two kinematical and geometrical arguments, the distribution
of the transverse momentum of the charged leptons in the $W^\pm$ decays
along the $p \bar p$ axis can be derived as
\begin{equation}
W \rightarrow \ell {\nu_\ell}:\qquad\frac{d\rho}{dp_\perp}=
\frac{p_\perp}{\sqrt{p_\perp^2-(M_W/2)^2}}
\end{equation}
{\it In praxi} this distribution is slightly smeared out due to the
radiation of gluons off the initial-state quarks and antiquarks which
kicks the partons out of the $p\bar p$ axis. However, this effect does
not spoil the basic characteristics, and it can be predicted
quantitatively in perturbative QCD calculations. From Fig.\ 
\ref{fig:wpte} it could therefore be concluded that the observed
isolated charged electrons and muons signal the original production of
the $W^\pm$ bosons.

The measurements of the $Z$ mass at the maximum of the Breit-Wigner
distribution and of the $W^\pm$ mass in the Jacobian peak of the lepton
transverse-momentum distribution led to values in the range expected
theoretically. The $W$ mass could be predicted from the Fermi coupling
and the measured electroweak mixing angle $\sin^2\vartheta_W$, based on
the low-energy relation (\ref{eq:mwgf}); the $Z$ mass is directly
related to the $W$ mass. Without taking into account radiative
corrections, their values can be estimated as:
\begin{eqnarray}
M_W & \simeq & \left[ \pi\alpha\Big/ \sqrt 2\sin^2\vartheta_W G_F
\right]^{1/2} 
\approx 79~{\rm GeV}\\ 
M_Z & \simeq & M_W\big/ \cos\vartheta_W\approx 90~{\rm GeV}
\end{eqnarray}
Radiative corrections add shifts of about 1.5 and 1 GeV to $M_W$
and $M_Z$.  The observation of the $W^\pm$ and $Z$ bosons in the UA1
and UA2 experiments has been an essential first step in establishing
the electroweak theory of forces as a gauge theory, strongly supported
moreover by the correct prediction of the $W^\pm$ and $Z$ masses in
this framework.\\


\subsection{\sc High-precision electroweak scattering}

In the preceding sections the Standard Model has been introduced by
means of intuitive arguments.  However, it can be shown that this
non-Abelian gauge theory is mathematically well-defined and that
observables can be calculated to an arbitrarily high precision in a
systematic expansion after a few basic parameters are fixed
experimentally. A series of fundamental papers (tHo 71, tHo 72) in which
the Standard Model has been proven to be a renormalizable theory, played
a key role in establishing the Standard Model as the basic theory of the
electroweak interactions.\\ 

The high precision of the predictions on the theoretical side is matched
by an equivalently high precision on the experimental side. Besides
refinements of the basic scattering experiments which at early times
supported the foundation of the theory, the precision achieved in
$e^+e^-$ experiments at high energies, in particular at LEP and SLC, has
allowed us to perform tests of the theory at the level of quantum
corrections.  Accuracies in general at the per-cent level, in some cases
down to the per-mille level, have been achieved. The most exciting
consequences of this development have been the prediction of the top
quark mass which has nicely been confirmed by the direct observation at
the Tevatron, and the prediction of the Higgs mass --- yet to be
confirmed at the time of writing.\\

The great potential of theoretical and experimental high-precision
analyses is the sensitivity to energy scales beyond those which can be
reached directly.  This method may be even more important in the future
when extrapolations to scales are necessary that may never be accessed
by experiments directly. \\[2ex]


\subsubsection{\sc The renormalizability of the Standard Model}

\begin{figure}[hbt]
\begin{center}
\begin{picture}(100,110)(0,0)
\Line(0,50)(100,50)
\CArc(50,50)(25,0,180)
\put(45,35){$\delta m$}
\put(-20,95){a)}
\end{picture}
\begin{picture}(100,100)(-50,0)
\Line(0,100)(50,50)
\Line(0,0)(50,50)
\Line(25,25)(25,75)
\Line(50,50)(100,50)
\put(47,35){$\delta g$}
\put(-20,95){b)}
\end{picture} \\
\end{center}
\caption[]{\label{fg:loops} \it Quantum corrections modifying mass and
  coupling parameters.}
\end{figure}
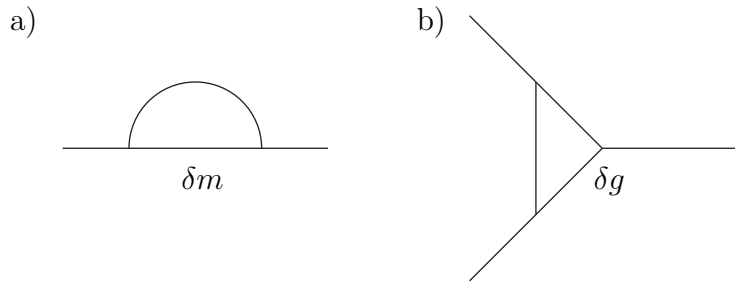

In interacting field theories the emission and re-absorption of quanta
after a short time of splitting, compatible with the time--energy
uncertainty, alters the masses of particles and their couplings, i.e.\ 
the interactions renormalize the fundamental parameters. These effects
can be described by Feynman diagrams including loops. Two characteristic
examples are shown in Fig.\ \ref{fg:loops}.

The self-energy corrections (cf.\ Fig.\ \ref{fg:loops}a) and the vertex
corrections (cf.\ Fig.\ \ref{fg:loops}b) are divergent for pointlike
couplings, leading to integrals of the type $\int d^4k/k^4\sim
\log\Lambda_{cut}^2$ where $\Lambda_{cut}^{-1}$ denotes the small scale
up to which the interaction appears pointlike.  These contributions add
to the unobservable bare mass $m_0$ and to the bare coupling $g_0$ to
generate the observable physical mass $m$ and the physical coupling $g$,
\begin{eqnarray*}
  m_0+\delta m&=&m \\
  g_0+\delta g&=&g
\end{eqnarray*}
If this renormalization prescription is sufficient to absorb all
divergences and to render all other observables finite in the limit
$\Lambda_{cut}^{-1} \rightarrow 0$, the theory is renormalizable and
well-defined. After the masses and couplings are fixed experimentally,
all other observables can be calculated, in principle to arbitrarily
high precision. \\

Non-Abelian gauge theories, like the Standard Model, have been proven
renormalizable. By fixing the electric charge $e$ and strong coupling
$g_s$, the gauge boson masses, the fermion masses, and the Higgs mass,
\[
\Re:\, e,\, g_s,\, M_W,\, M_Z,\, m_f,\, M_H 
\]
the values of all other observables can be predicted theoretically. {\it
  In praxi} the set $\Re$ may be replaced by the set $\Re_{exp}$,
\[
\Re_{exp}:\, \alpha,\, \alpha_s,\, G_F,\, M_Z,\, m_f,\, M_H  
\]
where $\alpha =e^2/4\pi$ and $\alpha_s = g_s^2/4\pi$, to take maximal
advantage of the parameters determined with the highest experimental
accuracy\footnote{ In the early analyses the electroweak mixing
  parameter $\sin^2\vartheta_W$ had been introduced instead of $M_Z$,
  measured in low-energy $\nu$ experiments. The heavy $W$ and $Z$ masses
  could successfully be predicted in this way.}. \\

The couplings $G_F,\, \alpha$ and $\alpha_s$ have been determined very
accurately in a long series of experiments (Cas 98). \\

\noindent
{\bf a)} \underline{The Fermi coupling $G_F$:} The Fermi coupling is
defined by the lifetime of the muon:
\begin{equation}
\tau^{-1}_\mu = \frac{G_F^2m_\mu^5}{192\pi^3} 
f\left( \frac{m_e^2}{m_\mu^2}\right) 
\left( 1+\frac35\frac{m_{\mu}^2}{M_W^2} \right) \left[
  1+\frac{\alpha}{2\pi} \left(\frac{25}4-\pi^2\right)+\dots \right]
\label{eq:muqed}
\end{equation}
where $f(x) = 1 - 8x + 8x^3 - x^4 - 12x^2 \log x$. By convention, QED
corrections to the effective Fermi theory (the term in square brackets)
are factored out explicitly. In the above formula we have displayed only
the one-loop corrections. Other electroweak radiative corrections to the
muon decay are absorbed in $G_F$.  Including the two-loop QED
corrections, one finds in the experimental analysis of the muon decay
the value
\begin{equation}
G_F=1.16637(1)\times 10^{-5}\; {\rm GeV}^{-2} 
\end{equation}
with a relative accuracy of $10^{-5}$. \\

\noindent
{\bf b)} \underline{The Sommerfeld fine structure constant $\alpha$:}
The fine structure constant is generally introduced as $\alpha = e^2 /
4\pi$, defined for on-shell electrons and photons in the $ee\gamma$
vertex.  Proper methods to determine this fundamental parameter are the
measurements of the anomalous magnetic moment of the electron, leading
to
\[
\alpha^{-1}=137.03599976(50)
\]
or the quantum Hall effect, with a significantly larger error though.\\ 

This definition however is not well suited for high-energy analyses.
Since vacuum polarization effects screen the electric charge, the
coupling increases when evaluated at a high scale of the $\gamma$
momentum transfer, $\mu_R=M_Z$ for instance: $\alpha (M_Z^2)=\alpha
/(1-\Delta\alpha)$.\\

The shift $\Delta\alpha$, induced by screening effects due to lepton and
hadron loops, can be determined analytically for leptons and by a
dispersion integral over the $e^+e^-$ annihilation cross section for
hadrons:
\begin{center}
\begin{picture}(100,40)(0,30)
\Photon(0,50)(100,50){3}{12}
\GOval(50,50)(10,20)(0){0.7}
\Line(-5,55)(5,45)
\Line(-5,45)(5,55)
\Line(95,55)(105,45)
\Line(95,45)(105,55)
\put(40,25){$\ell,q,\ldots$}
\end{picture} \\
\end{center}
\begin{eqnarray}
\Delta\alpha_{lept} & = & \sum_{\ell = e,\mu,\tau}
\frac\alpha{3\pi}\left(
  \log\frac{M_Z^2}{m_\ell^2} -\frac{5}{3}\right) +\dots \nonumber \\  
\Delta\alpha_{had}& = & - \frac\alpha{3\pi} 
\int\limits^\infty_{4m^2_\pi}\frac
{M_Z^2ds^\prime}{s^\prime[s^\prime-M_Z^2]}
\frac{\sigma(e^+e^- \rightarrow \gamma^* 
      \rightarrow {\rm hadrons};s^\prime)} 
{\sigma(e^+e^- \rightarrow \gamma^* \rightarrow \mu^+\mu^-;s^\prime)}
\label{eq:dalpha} 
\end{eqnarray}
Evaluating the dispersion integral by making use of the measured
annihilation cross section, the value of the coupling shifts to
\[
\alpha^{-1}(M_Z)=128.934(27)
\]
This shift affects the high-precision electroweak analyses in a drastic
way.\\

\noindent
{\bf c)} \underline{The strong coupling $\alpha_s$:} Since quantum
chromodynamics is an asymptotically free theory, the renormalized QCD
coupling is small at high energies. Perturbative expansions can
therefore be used to perform high-precision tests in processes which
involve quarks and gluons.  In general, the reference value of the
coupling $\alpha_s(\mu_R)$ is defined at the renormalization scale
$\mu_R=M_Z$ in the $\overline{MS}$ scheme in which the coupling is
renormalized by subtracting just the singularity in $D-4$ dimensions and
a few finite constants. At the scale $M_Z$ five quark flavors are active
in the polarization of the vacuum; they reduce the gluon-induced
anti-screening of the color charge.

A large variety of experimental methods can be used at high energies to
extract the QCD coupling: the cross section for $e^+e^-$ annihilation
into hadrons; the hadronic decay of $\tau$ leptons; the number of jets
in $Z$ decays; the hadronic event shapes; scaling violations of the
structure functions; and decays of heavy quarkonia. The coupling is
generally measured at scales different from $M_Z$. However, as long as
the scales are high enough, the coupling can be evolved perturbatively
to the common scale $\mu_R=M_Z$. The coupling has been determined in an
overall fit as 
\[
\alpha_s(M_Z^2)=0.1181\pm 0.002
\]
to next-to-next-to-leading order accuracy.


\subsubsection{\sc $e^+e^-$ annihilation near the $Z$ pole}

After $W^\pm$ and $Z$ bosons had been discovered at the Sp${\rm \bar
  p}$S, the next major step in the understanding of the $Z$ boson and
the electroweak interactions have been the experiments carried out at
the $e^+e^-$ storage ring LEP at CERN and the first $e^+e^-$ linear
collider SLC in Stanford. At LEP about 16 million $Z$ bosons have been
produced, allowing for high-statistics analyses of the electroweak
interactions. SLC, on the other side, could be operated with
longitudinally polarized electron and positron beams which increased
the sensitivity of the electroweak measurements significantly.\\

The precision achieved in the LEP and SLC experiments has allowed tests
of the electroweak theory at the quantum level.  The experimental
results have put the Glashow-Salam-Weinberg theory on very solid ground.
The high-precision quantum analyses led to a tremendously successful
prediction of the top-quark mass confirmed later in the Tevatron
experiments, and to the prediction of a light Higgs boson, the discovery
of which is eagerly awaited in the years to come.  The sensitivity to
quantum fluctuations in the physical observables demands the rigorous
treatment of the electroweak and QCD corrections which will be described
below in several consecutive steps for the basic process of fermion-pair
production near the $Z$ resonance in $e^+e^-$ annihilation. \\

\begin{figure}[hbt]
\begin{center}
\begin{picture}(100,100)(0,10)
\ArrowLine(0,0)(50,50)
\ArrowLine(50,50)(0,100)
\ArrowLine(100,50)(150,100)
\ArrowLine(150,0)(100,50)
\Photon(50,50)(100,50){3}{6}
\put(-15,98){$e^+$}
\put(-15,-2){$e^-$}
\put(65,35){$\gamma,Z$}
\put(155,98){$f$}
\put(155,-2){$\bar f$}
\end{picture}
\end{center}
\caption[]{ \label{fg:zres} \it The annihilation process $e^+e^- \to
  f\bar f$ at leading order.}
\end{figure}
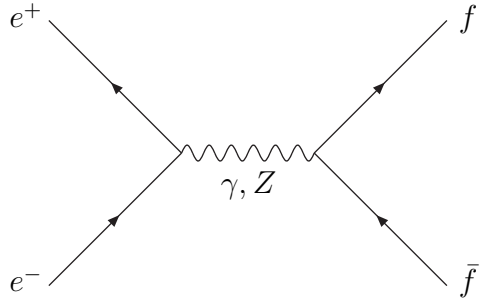

The fundamental process of fermion-pair production in $e^+e^-$
collisions,
\begin{equation}
e^+e^- \rightarrow f\bar{f}
\end{equation}
$f$ denoting leptons and quarks, is mediated by $Z$-boson and $\gamma$
exchange in the $s$-channel, ${\cal A} = {\cal A}_Z + {\cal
  A}_{\gamma}$, cf.\ Fig.\ \ref{fg:zres} (for $f = e$, there are also
$t$-channel contributions). The quantum corrections at next-to-leading
order include two different components: the pure QED corrections, i.e.\ 
virtual photon corrections and real photon radiation, and the genuine
electroweak corrections in loops. \\ 


\noindent
{\bf a)} \underline{The improved Born approximation.}  The basic
amplitudes in lowest order, ${\cal A}^0_{\gamma}$ and ${\cal A}^0_Z$,
are current $\times$ current amplitudes in which the electromagnetic and
electroweak currents are connected by the exchange of a photon and a $Z$
boson:
\begin{equation}
\begin{array}{l} 
\displaystyle
{\cal A}^0_{\gamma} = \frac{4\pi\alpha(s) Q_e Q_f}{s} 
  j_{\mu}^{\rm em}(f) j_{\mu}^{\rm em}(e) \\[3ex]
\displaystyle
{\cal A}^0_{Z} = \frac{\sqrt{2} G_F M_Z^2}{s - M_Z^2 + i M_Z
  \Gamma_Z(s)} j_{\mu}^{Z}(f) j_{\mu}^{Z}(e) 
\end{array}
\end{equation}
The coefficients $Q_e$, $Q_f$ denote the electric charges of the
electron and of the fermion $f$; the electroweak currents
$j_{\mu}^Z(e)$ and $j_{\mu}^Z(f)$ are coherent superpositions of a
vector part proportional to $v_e$ and $v_f$, and an axial-vector part
proportional to $a_e$ and $a_f$, cf.\ Eq.\ (\ref{eq:vfaf}).  By
evaluating the electromagnetic coupling $\alpha(s)$ at the proper
scale $s$ which characterizes the process, large logarithms from
anticipated radiative corrections are incorporated in the {\it
  improved} Born approximation in a natural way. The Fermi coupling
$G_F$ is related in Eq.\ (\ref{eq:mwgf}) to the $SU(2)$ coupling $g_W$
at the scale $M_Z^2$ which is logarithmically equivalent to the proper
scale $s$.

The vector and axial-vector $Z$ charges may be replaced by the partial
widths $\Gamma(Z \rightarrow f \bar{f})$. The total $Z$-boson width in
the Breit-Wigner denominator, interpreted as the imaginary part of the
inverse $Z$ propagator at scale $s$, may be defined with an
energy-dependent coefficient, $\Gamma_Z(s) = (s / M_Z^2) \Gamma_Z$,
while the proper $Z$ width at the pole is denoted by $\Gamma_Z$.

Thus, the leading logarithmic radiative corrections can easily be
incorporated in the cross sections within the improved Born
approximation, resulting in the cross section (Hol 00)
\begin{equation}
\sigma(s) = \frac{12 \pi \Gamma_e \Gamma_f}{(s-M_Z^2)^2 + (s^2/M_Z^2) 
                                             \Gamma_Z^2} 
\left[ 1 + \Delta_Z \right]
+ \frac{4\pi\alpha^2(s)}{3s} Q_f^2 N_c
\label{eq:sigmaz}
\end{equation}
The first part is the Breit-Wigner form of the $Z$ contributions
corrected by the real and imaginary parts
\begin{equation}
\Delta_Z = (1 + R_f) \frac{s-M_Z^2}{M_Z^2} 
         + I_f \frac{\Gamma_Z}{M_Z} 
\end{equation}
of the $\gamma-Z$ interference contribution $R_f$ and $I_f$; the
second part describes the photon-exchange contribution.\\

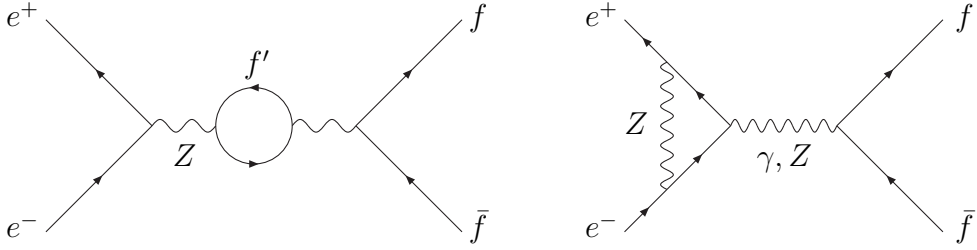
\begin{figure}[tb]
\begin{center}
\SetScale{0.8}
\begin{picture}(145,100)(50,10)
\ArrowLine(0,0)(50,50)
\ArrowLine(50,50)(0,100)
\ArrowLine(146,50)(196,100)
\ArrowLine(196,0)(146,50)
\Photon(50,50)(80,50){3}{2}
\Photon(116,50)(146,50){3}{2}
\ArrowArc(98,50)(18,0,180)
\ArrowArc(98,50)(18,180,360)
\put(-15,78){$e^+$}
\put(-15,-2){$e^-$}
\put(48,25){$Z$}
\put(75,60){$f'$}
\put(160,78){$f$}
\put(160,-2){$\bar f$}
\end{picture}
\begin{picture}(100,100)(-20,10)
\ArrowLine(0,0)(20,20)
\ArrowLine(20,20)(50,50)
\ArrowLine(50,50)(20,80)
\ArrowLine(20,80)(0,100)
\ArrowLine(100,50)(150,100)
\ArrowLine(150,0)(100,50)
\Photon(50,50)(100,50){3}{6}
\Photon(20,20)(20,80){3}{6.5}
\put(-15,78){$e^+$}
\put(-15,-2){$e^-$}
\put(0,37){$Z$}
\put(50,25){$\gamma,Z$}
\put(125,78){$f$}
\put(125,-2){$\bar f$}
\end{picture}
\SetScale{1}
\end{center}
\caption[]{ \label{fg:zres1l} \it Typical diagrams contributing to
  the genuine electroweak corrections to $e^+e^- \to f\bar f$.}
\end{figure}

\noindent
{\bf b)} \underline{Electroweak corrections.} The most important
electroweak corrections for energies near the $Z$ resonance are
self-energy corrections in the $\gamma$ and $Z$ propagators, as well as
vertex corrections due to the exchange of electroweak bosons. Typical
diagrams are depicted in Fig.\ \ref{fg:zres1l}. Additional box diagrams
give relative corrections of less than $10^{-4}$ near the $Z$ resonance
and can safely be neglected.

The $\rho$ parameter (Vel 77)
\begin{equation} 
\rho_{e,f} = 1 + \Delta \rho + \Delta \rho_{\rm non-univ}
\end{equation}
modifies the neutral-current coupling by universal corrections of the
gauge boson propagators and by flavor-specific vertex corrections.

In the same way the electroweak mixing angle in the currents must be
specified properly.  Defining
\begin{equation}
\sin^2 \vartheta_W = 1 - M_W^2/M_Z^2
\end{equation}
the weak mixing angle entering in the vector $Z$ couplings of Eq.\ 
(\ref{eq:vfaf}) are modified and replaced by effective mixing angles
which are related to the basic definition as
\begin{equation}
\sin^2\vartheta^{e,f}_{\rm eff} = \kappa_{e,f} \sin^2 \vartheta_W
\end{equation}
Again, $\kappa_{e,f}$ can be separated into a universal and a
flavor-specific part
\begin{equation}
\kappa_{e,f} = 1+ \Delta\kappa + \kappa_{\rm non-univ}
\end{equation}
with
\begin{equation}
\Delta \kappa = \cot^2 \vartheta_W \Delta \rho 
\end{equation}
The non-universal contribution is particularly large for the $Zbb$
coupling.  After the replacement ${\cal A}_Z^0 \rightarrow {\cal A}_Z =
\sqrt{\rule{0mm}{2.7mm}\rho_e \rho_f} {\cal A}_Z^0$ and $\sin^2\vartheta_W
\rightarrow \sin^2\vartheta_{\rm eff}^{e,f}$ , the corrections
$\rho_{e,f}$ and $\kappa_{e,f}$ enter the cross sections $\sigma(e^+e^-
\rightarrow f\bar{f})$ and the partial widths $\Gamma(Z \rightarrow
f\bar{f})$ in the same form. The expression Eq.\ (\ref{eq:sigmaz}) for
the cross section therefore can be kept unmodified when the dominant
electroweak corrections are included.

Apart from the explicit form of $\Delta \rho$ which we will give at
the end of this section, we will not discuss the various corrections
in detail here but instead refer the reader to the literature (Bar 99).\\


\noindent
{\bf c)} \underline{QCD corrections.} These corrections affect the
production cross sections for quark pairs and the partial $Z$ decay
widths in the same way, when the quark masses are neglected, by the
additional coefficient (Sch 73)
\begin{equation}
\Delta_{QCD} = 1 + \frac{\alpha_s}{\pi} + \cdots
\end{equation}
Higher-order terms up to order $(\alpha_s / \pi)^3$ are known as well.\\


\noindent
{\bf d)} \underline{QED corrections.} Numerically the QED corrections
(Ren 81) are the most important radiative corrections near the $Z$
resonance.  Since the formation of the $Z$ resonance leads to a sharp
Breit-Wigner peak, the radiation of photons from the initial electrons
and positrons shifts the energy $\sqrt{s} \rightarrow \sqrt{\hat{s}}$
away from the peak, resulting in a large modification of the production
cross section.  The cross section including the QED vertex corrections
and the photon radiation in the initial state can be expressed as a
convolution of the electroweak cross section $\sigma$, as calculated
above, with the radiator function $H$:
\begin{equation}
\sigma_{QED}(s) = \int_0^{x_{max}} dx H(x) \sigma [ (1-x) s ]
\end{equation}
The upper integration limit $x_{max}$ describes the maximal fraction
$x$ of the photon energy not resolved in the detector. The result
therefore depends strongly on possible cuts on the energies of
observed photons.  Without any cuts, the kinematical limit $x_{max} =
2 E^{\gamma}_{max} / \sqrt{s} \leq 1 - 4 m_f^2/s$ must be inserted.
Including the resummation of soft photons, the radiator function can
be written in leading logarithmic order as
\begin{equation}
\begin{array}{c}
H(x) = \beta x^{\beta - 1} \quad {\rm with} \quad
\beta = \frac{2 \alpha}{\pi} \left( \log \frac{s}{m_e^2} - 1 \right)
\end{array}
\end{equation}

The complete QED corrections reduce the peak value of the resonance
cross section by about $30\,\%$ and shift the position of the peak
upward by about $+100$ MeV.  Given the high precision of the LEP and SLC
experiments, both these effects are crucial for the correct
interpretation of these measurements. \\ 

\begin{figure}[hbt]
\begin{center}
\hspace*{-0cm}
\epsfxsize=10cm 
\epsfbox{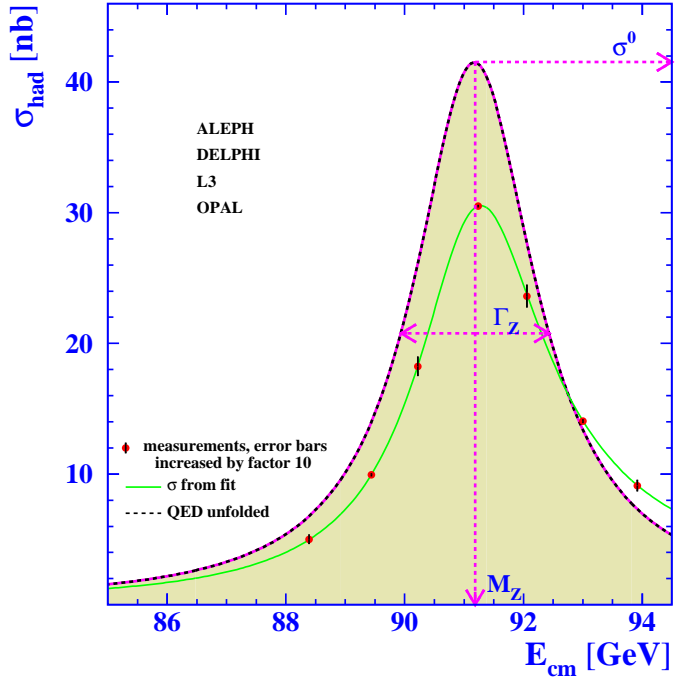}
\vspace*{-1.0cm}
\end{center}
\caption[]{\label{fig:Zresonance} \it $Z$-peak cross section observed by 
  LEP in $e^+e^- \rightarrow hadrons$ and compared with the
  complete Standard Model prediction (LEP 00).}
\end{figure}

The final result for the \underline{\it total cross section} $\sigma [
e^+e^- \rightarrow hadrons]$ compared with experimental
data is shown in Fig.\ \ref{fig:Zresonance} for $e^+e^-$ energies near
the $Z$ resonance. After adjusting the free parameters of the
electroweak theory, e.g.\ the $Z$-boson mass $M_Z$ and the electroweak
mixing angle $\sin^2 \vartheta_W$, the theoretical prediction of the
cross section is in wonderful agreement with the data. The experimental
analysis supports the validity of the Glashow-Salam-Weinberg model as
the theory of the
electroweak interactions at a very high level of accuracy. \\

Two other important observables are the \underline{\it
  forward-backward asymmetry} $A_{FB}$, which describes the difference
of the production cross sections for leptons and quarks in the forward
and backward hemispheres with respect to the direction of the incoming
electron, and the \underline{\it left-right asymmetry} for
longitudinally polarized electron and positron beams. These
asymmetries can be expressed in terms of the electroweak parameters as
\begin{equation}
\begin{array}{ll}
\displaystyle
A_{FB}  = \frac{3}{4}\frac{2 v_f a_f}{v_f^2 + a_f^2} 
\frac{2 v_e a_e}{v_e^2 + a_e^2} 
& \displaystyle \quad \Rightarrow \quad
\frac{3}{4}\left[ \frac{1 - 4\sin^2\vartheta_{\rm eff}^l}
                {1 + (1 - 4\sin^2\vartheta_{\rm eff}^l)^2} \right]^2
\\ \displaystyle
A_{LR}  = \frac{2 v_e a_e}{v_e^2 + a_e^2} 
& \displaystyle \quad \Rightarrow \quad
\frac{1 - 4\sin^2\vartheta_{\rm eff}^l}%
                     {1 + (1 - 4\sin^2\vartheta_{\rm eff}^l)^2}
\end{array}
\end{equation}
The explicit form of $A_{FB}$ as function of $\sin^2 \vartheta_{\rm
  eff}^l$ is valid for lepton asymmetries $f = l = e$, $\mu$, $\tau$.
Since for leptons $\sin^2 \vartheta_{\rm eff}^l$ is near 1/4, it is
apparent that the sensitivity to the value of $\sin^2 \vartheta_{\rm
  eff}^l$ is maximal for the left-right asymmetry. \\

\begin{table}[tb]
\hspace*{5.0cm}
\epsfysize=13cm \epsfbox{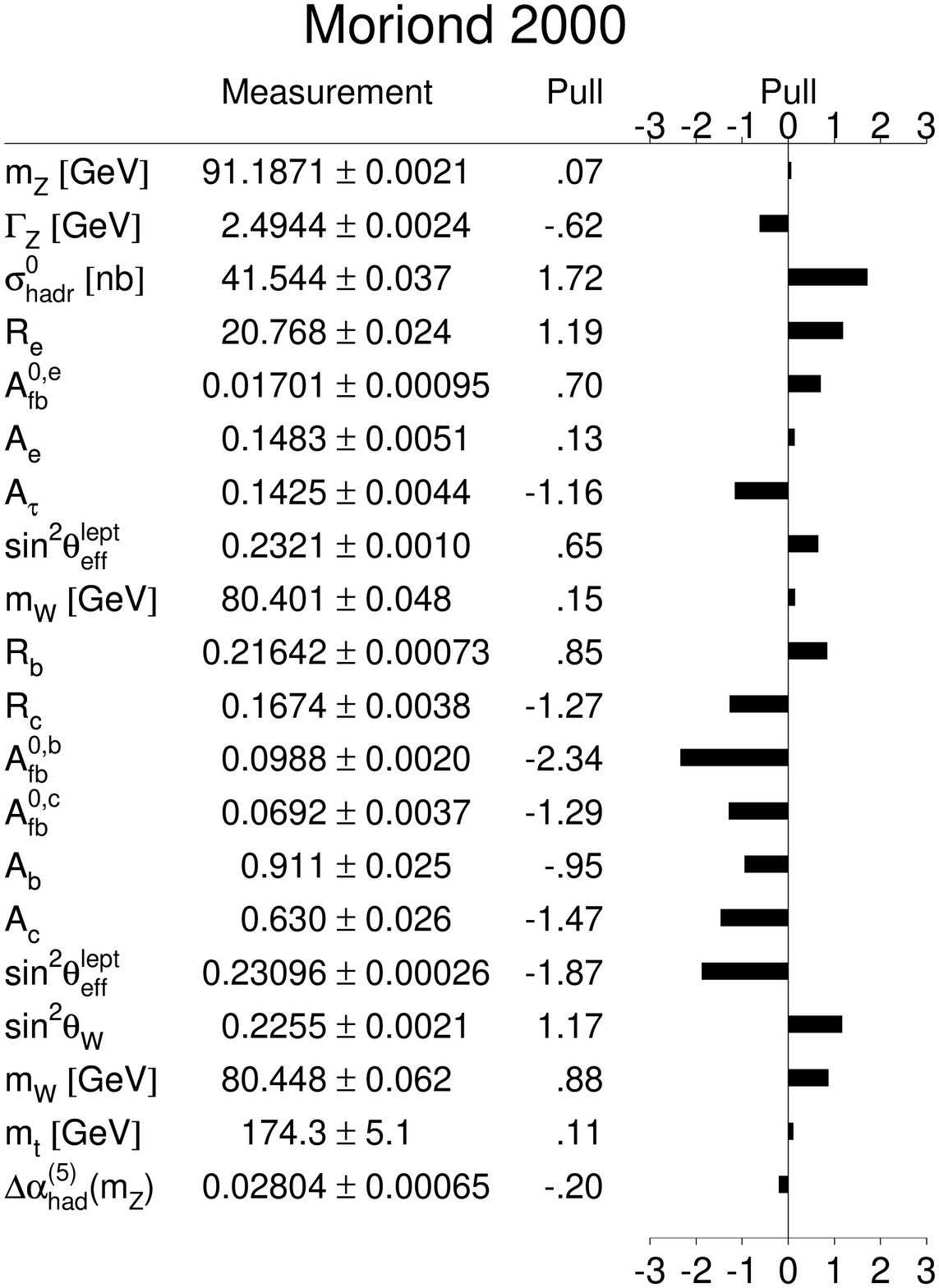}
\vspace*{-1.8cm}
\caption[]{\label{tb:elwdata} \it Experimental results for several
  precision observables at LEP1. The pull is defined as the deviation
  from the theoretical prediction in units of the corresponding
  one-standard deviation experimental uncertainty.}
\end{table}

Many observables have been evaluated to scrutinize the validity of the
electroweak Standard Model. The measurements of a large set of
observables is compared with their best values within the Standard
Model in Table \ref{tb:elwdata}, most noticeable: 
\begin{equation}
M_Z = 91.1882 \pm 0.0022 \, {\rm GeV}
\end{equation}
\begin{equation}
\Gamma_Z = 2.4952 \pm 0.0026 \, {\rm GeV}
\end{equation}
\begin{equation}
\sin^2 \vartheta_{\rm eff}^{l} = 0.23096 \pm 0.00026
\end{equation}
In several cases the agreement of the data with the predictions is at
the per-mille level --- a triumph of field theory as the proper
formulation of electroweak interactions.


\subsubsection{\sc $W^+W^-$ gauge-boson pair production in $e^+e^-$
  annihilation}  

The second process which could be exploited to perform precision tests
of the Standard Model, is the production of $W^{\pm}$ pairs in $e^+e^-$
annihilation:
\begin{equation}
e^+ e^- \rightarrow W^+ W^-
\label{eq:eeWW}
\end{equation}
This is a much cleaner production channel than for proton colliders
since no additional spectator hadrons are generated in the final state.
Compared with the other processes for gauge-boson pair production,
$e^+e^- \rightarrow \gamma \gamma$, $\gamma Z$, $ZZ$, the process
(\ref{eq:eeWW}) constitutes the most important reaction expected to
provide a detailed knowledge of the mass and the couplings of the $W$
boson.

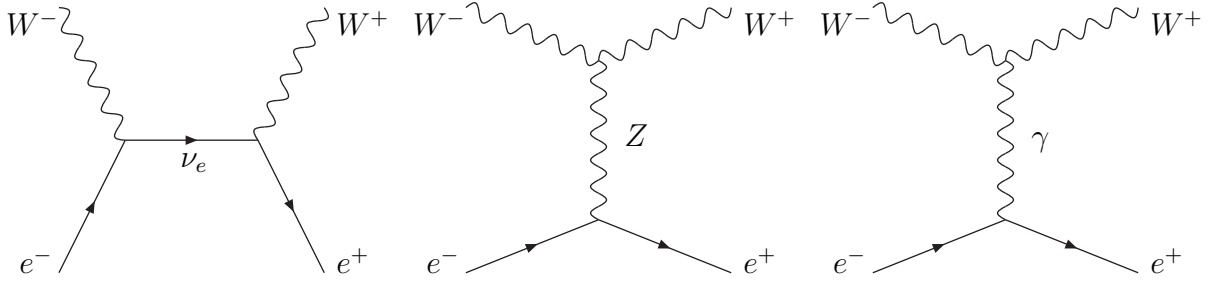
\begin{figure}[hbt]
\begin{center}
\begin{picture}(100,100)(50,0)
\ArrowLine(0,0)(25,50)
\ArrowLine(25,50)(75,50)
\ArrowLine(75,50)(100,0)
\Photon(25,50)(0,100){3}{5}
\Photon(100,100)(75,50){3}{5}
\put(-15,0){$e^-$}
\put(-20,90){$W^-$}
\put(105,90){$W^+$}
\put(105,0){$e^+$}
\put(46,40){$\nu_e$}
\end{picture} 
\begin{picture}(100,100)(0,0)
\ArrowLine(0,0)(50,20)
\ArrowLine(50,20)(100,0)
\Photon(50,20)(50,80){3}{6}
\Photon(50,80)(0,100){3}{5}
\Photon(100,100)(50,80){3}{5}
\put(-15,0){$e^-$}
\put(-20,90){$W^-$}
\put(105,90){$W^+$}
\put(105,0){$e^+$}
\put(60,48){$Z$}
\end{picture} 
\begin{picture}(100,110)(-50,0)
\ArrowLine(0,0)(50,20)
\ArrowLine(50,20)(100,0)
\Photon(50,20)(50,80){3}{6}
\Photon(50,80)(0,100){3}{5}
\Photon(100,100)(50,80){3}{5}
\put(-15,0){$e^-$}
\put(-20,90){$W^-$}
\put(105,90){$W^+$}
\put(105,0){$e^+$}
\put(60,48){$\gamma$}
\end{picture}
\end{center}
\caption[]{ \label{fig:eeWW} \it $W^+W^-$ pair production in $e^+e^-$
  annihilation.} 
\end{figure}

\begin{figure}[thb]
\begin{center}
\hspace*{0.0cm}
\epsfxsize=10cm \epsfbox{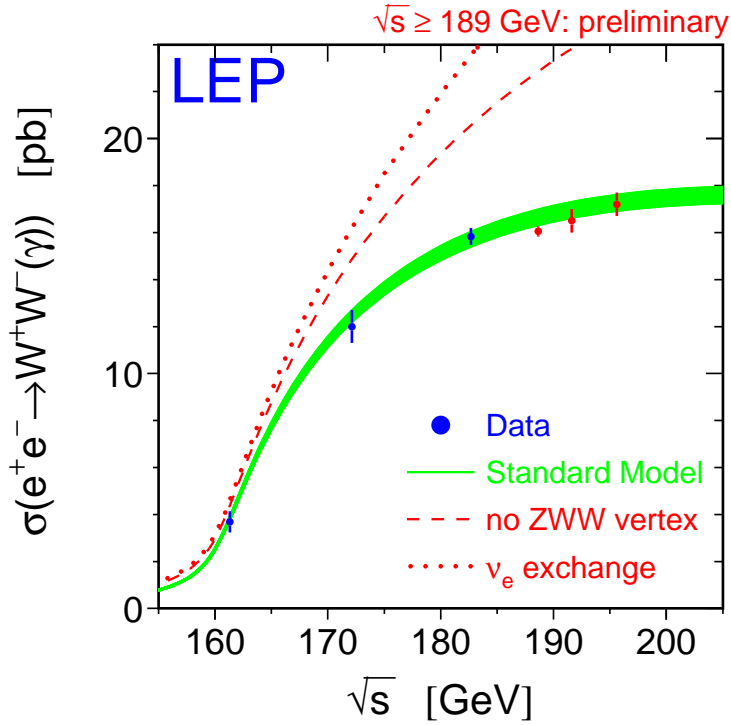}
\vspace*{-1.5cm}
\end{center}
\caption[]{\label{fig:eeWWxsec} \it Total cross section for $W^+W^-$
  production in $e^+e^-$ annihilation as predicted from the Standard
  Model (grey band), without the $s$-channel $Z$-exchange diagram
  (dashed line), without the $s$-channel $\gamma$, $Z$-exchange diagrams
  (dotted line), and compared with data from LEP2 (LEP 00a).}
\end{figure}

The $W^+W^-$ pair production (All 77) is described by the Feynman
diagrams shown in Fig.\ \ref{fig:eeWW}.  Both $s$- and $t$-channel
exchanges are needed in order to generate a high-energy behavior
compatible with unitarity. The separate $s$- and $t$-channel
contributions grow strongly with $\sqrt{s} \rightarrow \infty$ as seen
in Fig.\ \ref{fig:eeWWxsec}. The sum of all contributions which
interfere destructively, leads to a reduced increase above threshold but
decreases proportional to $\log(s)/s$ at large $s$. The result (All 77)
valid for all $s$ reads
\begin{equation}
\begin{array}{l}
\displaystyle
\sigma[e^+e^- \rightarrow W^+ W^-] 
\\[2ex] \quad \displaystyle
= \frac{\pi \alpha^2}{2s_W^4} \frac{\beta}{s} 
\Biggl\{ \left[1 + \frac{2M_W^2}{s} + \frac{2M_W^4}{s^2}\right]
        \frac{1}{\beta} \log\frac{1+\beta}{1-\beta} - \frac{5}{4} 
 \\[3ex]
\quad \quad \displaystyle
+ \frac{M_Z^2(1 - 2s_W^2)}{s - M_Z^2} 
  \left[ 2 \left(\frac{M_W^4}{s^2} + \frac{2M_W^2}{s} \right)
           \frac{1}{\beta} \log \frac{1+\beta}{1-\beta} 
        - \frac{s}{12M_W^2} - \frac{5}{3} - \frac{M_W^2}{s} \right]
\\[3ex]
\quad \quad \displaystyle
+ \frac{M_Z^4 \left(8s_W^4 - 4s_W^2 + 1\right) \beta^2}{48(s-M_Z^2)^2}
\left[ \frac{s^2}{M_W^4} + \frac{20}{M_W^2} + 12\right]
\Biggr\}
\end{array}
\end{equation}
with $s_W = \sin \vartheta_W$, $c_W = \cos \vartheta_W$ and $\beta =
\sqrt{1 - 4 M_W^2/s}$.  The cancellation amounts to a suppression by one
order of magnitude at $\sqrt{s} = 400$ GeV and two orders of magnitude
at $\sqrt{s} = 1$ TeV, induced by the interplay of the $\gamma WW$, $ZWW$
and $e\nu W$ couplings which are related to each other by the gauge
symmetry of the Standard Model.

The measurements of the $W$-pair production cross section at LEP2
therefore provide the first determination of the 3-gauge-boson couplings
$\gamma WW$ and $ZWW$. These couplings will be tested with higher
accuracy at higher energies since even small anomalous couplings will
upset the gauge cancellations, quickly leading to sizable deviations
from the Standard Model predictions. \\

The detailed study of the excitation curve $\sigma_{\rm tot}(s)$ close
to the threshold at $\sqrt{s} = 2M_W$ provides a precise
model-independent measurement of the $W$ mass. In terms of the velocity
$\beta$, the cross section close to threshold, $\beta \ll 1$,
differential in the scattering angle $\theta$, reads
\begin{equation}
\frac{d\sigma}{d\cos\theta} = \frac{\pi \alpha^2}{s}
\frac{1}{2\sin^2\vartheta_W} \beta \left(1 + 4 \beta\cos\theta 
\frac{3\cos^2\vartheta_W-1}{4\cos^2\vartheta_W-1} + O(\beta^2) \right)
\end{equation}
The first, $\theta$-independent term is due to the $\nu$-exchange
diagram; the second term due to the two $s$-channel diagrams vanishes at
threshold. As a result, the total cross section
\begin{equation}
\sigma_{\rm tot} = \frac{\pi\alpha^2}{s} \frac{1}{\sin^2\vartheta_W}
\beta + O(\beta^3)
\end{equation}
rises linearly with $\beta$, and it is determined by the kinematics
and the well-established $\nu e W$ coupling alone.

The dominance of the neutrino $t$-channel exchange contribution is a
consequence of angular momentum and $CP$ conservation: the $s$-channel
exchange of spin-1 bosons restricts the total angular momentum of the
final state to $J \leq 1$. On the other hand, since at threshold $J$
is equal to the total spin of the two $W$ bosons, one can have only $J
= 0,$ 1, or 2. The first of these values is forbidden because of
fermion-helicity conservation in the initial state, the second by
$CP$ conservation. The $s$-channel contribution, as well as its
interference with the $t$-channel diagram thus has to vanish at
threshold.

Finite-width effects and higher-order corrections modify the above
formula, but the general behavior is not altered. 

Since the $W$ boson is unstable and decays into either a charged lepton
and the corresponding antineutrino or into two hadron jets, the analysis
of $W$-pair production requires the proper understanding of all Standard
Model processes leading to four fermions in the final state $e^+ e^-
\rightarrow f_1 \bar{f}_2 f_3 \bar{f}_4$. The three Feynman diagrams
shown in Fig.\ \ref{fig:eeWW} for $e^+e^- \rightarrow W^+W^-$ completed
by subsequent decays of the $W$ bosons constitute only a small subset of
possible Feynman diagrams for this more general process. They dominate
however, if two fermions with appropriate quantum numbers have an
invariant mass close to the $W$-boson mass.  Investigations of the
invariant mass spectrum of pairs of final state fermions therefore
provide another means to measure $M_W$. \\

From the decay products of the $W^{\pm}$ bosons in the final state,
$W^{\pm} \rightarrow \ell \nu, ~~ q\bar{q}^{\prime}$, the vector bosons
can be reconstructed directly. This is a particularly useful method at
energies well above the threshold in the continuum.  Mixed lepton-jet
pairs $\ell \nu j j$ where jets $j$ emerge as hadronization products
from the original quarks, provide a very clean event sample.  Four-jet
final states can also be used in the analysis.

Combining all methods, from LEP as well as from the reconstruction of
$W$ bosons in proton collisions, leads to the final value (LEP 00a)
\begin{equation}
M_W = 80.382 \pm 0.026~ {\rm GeV}
\end{equation}
for the mass of the charged $W^{\pm}$ boson.


\subsubsection{\sc Physical interpretation of the measurements} 

{\bf 1.} A most important conclusion can be drawn from the
high-precision measurements of the $Z$-boson width. By observing
$\Gamma_Z$ in the resonance excitation curve, the decay width into
invisible final states of neutrino pairs can be derived by subtracting
the visible charged-lepton and hadron channels with
\begin{eqnarray}
\Gamma_{\rm invis} & = & N_{\nu} \Gamma(Z \rightarrow \nu_{\ell}
                         \bar{\nu}_{\ell})
\nonumber \\
                   & = & N_{\nu} \frac{G_F M_Z^3}{12\sqrt{2}\pi} 
\end{eqnarray}
the number of families in the Standard Model can be counted by measuring
$N_{\nu}$, the number of light neutrinos:
\begin{equation}
N_{\nu} = 2.994 \pm 0.012
\end{equation}
The measurement confirms the existence of three Standard Model families,
the minimum number necessary for incorporating $CP$ violation in the
theory. 

\noindent
{\bf 2.} In the canonical form of the Standard Model, the precision
observables measured at the $Z$ peak are affected by quantum
fluctuations; they give access to two high mass scales in the model: the
top-quark mass $m_t$, and the Higgs-boson mass $M_H$. These particles
enter as virtual states in the loop corrections to various relations
between the electroweak observables. For instance, the radiative
corrections to the relation between the $W$ and the $Z$ mass, and
between the $Z$ mass and $\sin^2 \vartheta_{\rm eff}^l$, have a strong
quadratic dependence on $m_t$ and a logarithmic dependence on $M_H$.

More generally, quantum fluctuations from scales of physics beyond the
Standard Model, e.g.\ supersymmetric or technicolor extensions, may also
affect the electroweak observables. The modifications can either be
exploited to scrutinize hypothetical extensions or, at least, to
constrain the new scales characterizing the extended theories.

A second area of new physics phenomena are possible deviations of the
interactions between the electroweak gauge bosons from the predictions
of the Standard Model. The form and the strength of the trilinear
couplings are predicted by the non-Abelian gauge symmetry.  They can be
measured in the production of $W^+W^-$ pairs in $e^+e^-$ annihilation.
These experiments therefore serve to verify one of the
most important symmetry concepts in Nature. \\

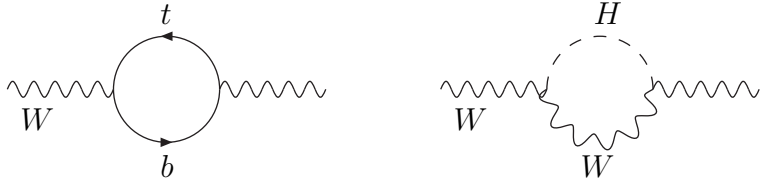
\begin{figure}[tb]
\begin{center}
\begin{picture}(100,100)(30,10)
\Photon(0,50)(40,50){3}{5}
\Photon(80,50)(120,50){3}{5}
\ArrowArc(60,50)(20,0,180)
\ArrowArc(60,50)(20,180,360)
\put(5,35){$W$}
\put(58,75){$t$}
\put(58,17){$b$}
\end{picture}
\begin{picture}(100,100)(-30,10)
\Photon(0,50)(40,50){3}{5}
\Photon(80,50)(120,50){3}{5}
\DashCArc(60,50)(20,0,180){5}
\PhotonArc(60,50)(20,180,360){3}{6}
\put(5,35){$W$}
\put(58,75){$H$}
\put(53,17){$W$}
\end{picture}
\end{center}
\caption[]{ \label{fg:tbloop} \it Top-bottom and Higgs-boson loops
  contributing to the $W$ boson self-energy.}
\end{figure}

\noindent
{\bf a)} \underline{The masses of the top quark and the Higgs boson.}
Eliminating either $M_W$ or $\sin^2 \vartheta_W$ from the basic
low-energy connection between the Fermi constant, the $W$ mass and the
electroweak mixing angle, the connection may be written in two forms. In
the first case,
\begin{equation}
\sin^2 \vartheta_{\rm eff}^{\ell} \left( 1 - \sin^2 \vartheta_{\rm
    eff}^{\ell} \right) = 
\frac{\pi \alpha}{\sqrt{2} G_F M_Z^2 \left(1 - \Delta r_Z \right)}
\end{equation}
The correction $\Delta r_Z$ includes the shift $\alpha \rightarrow
\alpha(M_Z^2)$ and the contribution\footnote{Here and in the following,
  the ellipses denote higher-order corrections not shown explicitly.}
from top-bottom and Higgs fluctuations in the propagators of the
electroweak vector bosons, cf.\ Fig.\ \ref{fg:tbloop},
\begin{equation}
\Delta r_Z = \Delta \alpha - \Delta \rho^t + \Delta r_Z^H + \cdots
\end{equation}
The shift $\Delta \alpha$ of the electromagnetic coupling has been
discussed earlier. The leading top contribution to the $\rho$ parameter
(Vel 77) is quadratic in $m_t$:
\begin{equation}
\Delta \rho^t = \frac{3G_F m_t^2}{8\pi^2\sqrt{2}} + \cdots
\end{equation}
The Higgs contribution is screened, depending only logarithmically on
the Higgs-boson mass for large $M_H$:
\begin{equation}
\Delta r_Z^H = \frac{G_F M_W^2}{8\pi^2\sqrt{2}} 
               \frac{1 + 9 \sin^2 \vartheta_W}{3\cos^2\vartheta_W}
               \log\frac{M_H^2}{M_W^2} + \cdots
\end{equation}

Alternatively, the relation can be written in terms of $M_W$:
\begin{equation}
\frac{M_W^2}{M_Z^2} \left(1 - \frac{M_W^2}{M_Z^2} \right) = 
\frac{\pi\alpha}{\sqrt{2}G_F M_Z^2 \left(1 - \Delta r_W \right)}
\end{equation}
where the quantum correction $\Delta r_W$ is composed of 
\begin{equation}
\Delta r_W = \Delta \alpha - \cot^2 \vartheta_W \Delta \rho^t + \Delta
r_W^H + \cdots
\end{equation}
with
\begin{equation}
\Delta r_W^H = \frac{G_F M_W^2}{8\pi^2\sqrt{2}} 
               \frac{11}{3}
               \log\frac{M_H^2}{M_W^2} + \cdots
\end{equation}
and $\Delta \alpha$ and $\Delta \rho^t$ as introduced above. 

\begin{figure}[thb]
\begin{center}
\vspace*{-1.0cm}
\hspace*{0.0cm}
\epsfxsize=10cm \epsfbox{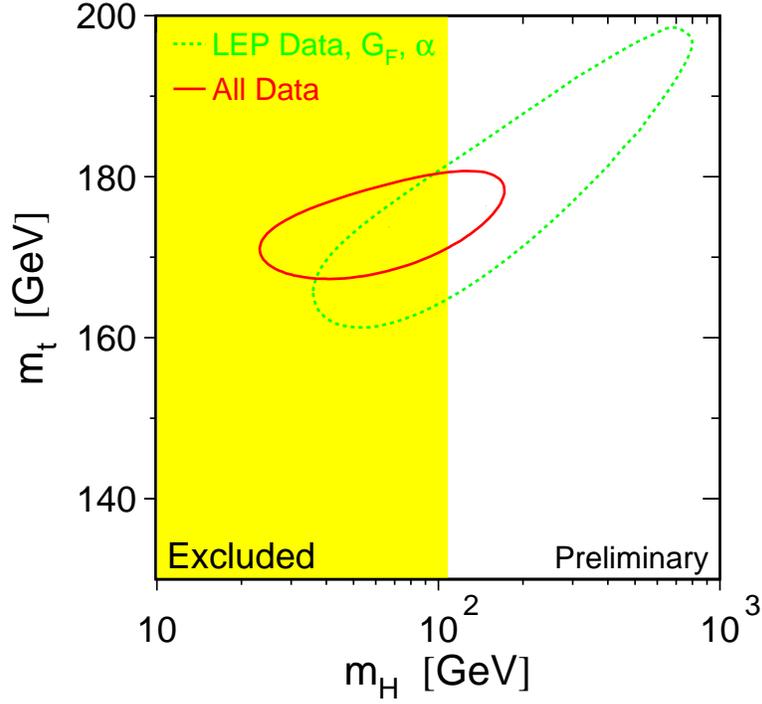}
\vspace*{-1.0cm}
\end{center}
\caption[]{\label{fig:mtmhlep} \it Allowed region in the $m_t-M_H$ plane 
  derived from a fit of LEP results to the Standard Model predictions
  (LEP 00a).} 
\end{figure}

Inserting the precision data for $\alpha$, $G_F$, $M_Z$ and $\sin^2
\vartheta_{\rm eff}^l$ in the first case or $M_W$ in the second case,
these relations can be solved for $m_t$ and $M_H$.  The final results
for all the precision observables have been used to determine the top
and the Higgs-boson masses as demonstrated in Fig.\ \ref{fig:mtmhlep}.
The value of the top mass obtained from the global fit (LEP 00a)
\begin{equation}
{\it indirect~ precision~ measurements:~~ } m_t =
169.7^{+9.8}_{-7.0}~{\rm GeV} 
\end{equation}
is in excellent agreement with the value measured directly in top-quark
production at the Tevatron (Abb 98, Yao 98):
\begin{equation}
{\it direct~ experimental~ measurements:~~ } m_t = 174.3 \pm 5.1~{\rm
  GeV} 
\end{equation}
In fact, the \underline{prediction} of $m_t$ from the electroweak
precision analyses was in agreement with the directly measured value
within 15 GeV.

The correctly predicted value of the top-quark mass raises hope that
the second prediction may also be realized in Nature. Fixing the top
mass to the experimental value, the evaluation of the precision data
leads to the following values for the Higgs-boson mass (LEP 00a):
\begin{equation}
\begin{array}{l}
M_H = 77 ^{+69}_{-39} ~ {\rm GeV} \\
M_H < 215 ~ {\rm GeV ~~ at} ~~ 95\,\% {\rm CL}
\end{array}
\end{equation}
Due to yet unknown two-loop corrections, the accuracy of the analysis is
expected in the range of 100 GeV. This prediction is exciting for
several reasons: (i) The numbers are compatible with a light Higgs-boson
mass in the characteristic range of the scales of the electroweak
symmetry breaking between $M_W \simeq 81$ GeV and $v / \sqrt{2} \simeq
174$ GeV; (ii) The small Higgs-boson mass is a prerequisite for
extending the Standard Model to grand unification scales; in this way
the value of the electroweak mixing angle is predicted correctly in the
range observed experimentally. Moreover, the restricted mass range of
the Higgs boson is nicely compatible with the low mass value expected in
supersymmetric theories which are attractive extensions of the Standard
Model. \\


\noindent
{\bf b)} \underline{$S$, $T$, $U$ parameters.} Generalizing the previous
analysis, a special class of ``new physics'' contributions can be
described by the $S$, $T$, $U$ parameters (Pes 90). They account for the
quantum fluctuations of the electroweak bosons into new heavy states
which have negligible couplings to the light leptons and quarks.
Expanding the self energies of the electroweak bosons generated by the
new physics contributions, about $Q^2 = 0$,
\begin{equation}
\begin{array}{ll}
\displaystyle
\Pi_{\gamma \gamma}(Q^2) = Q^2 \Pi^{\prime}_{\gamma \gamma}(0) + \cdots
\, , \quad
& 
\displaystyle
\Pi_{ZZ}(Q^2) = \Pi_{ZZ}(0) + Q^2 \Pi^{\prime}_{ZZ}(0) + \cdots
\\ 
\displaystyle
\Pi_{\gamma Z}(Q^2) = Q^2 \Pi^{\prime}_{\gamma Z}(0) + \cdots
\, , \quad
&
\displaystyle
\Pi_{WW}(Q^2) = \Pi_{WW}(0) + Q^2 \Pi^{\prime}_{WW}(0) + \cdots
\end{array}
\end{equation}
three of the six parameters are absorbed in the renormalized input
parameters $\alpha$, $G_F$ and $M_Z$, whereas the other three are new
observables:
\begin{equation}
\begin{array}{l}
\displaystyle
S = \frac{\sin^2 2\vartheta_W}{\alpha} \left( \Pi^{\prime}_{ZZ}(0) - 2
  \cot 2 \vartheta_W 
  \Pi^{\prime}_{Z \gamma}(0) - \Pi^{\prime}_{\gamma \gamma}(0) \right) 
\\[3ex] \displaystyle
T = \frac{1}{\alpha} \left( \frac{\Pi_{WW}(0)}{M_W^2} -
  \frac{\Pi_{ZZ}(0)}{M_Z^2} \right) 
\\[3ex] \displaystyle
U = \frac{4\sin^2\vartheta_W}{\alpha} \left( \Pi^{\prime}_{WW}(0) -
  \cos^2 \vartheta_W 
  \Pi^{\prime}_{ZZ}(0) - \sin 2 \vartheta_W \Pi^{\prime}_{Z \gamma}(0) -
  \sin^2 \vartheta_W \Pi^{\prime}_{\gamma \gamma}(0) \right) 
\end{array}
\end{equation}
The parameters have been determined in a global fit to the electroweak
precision data. After subtracting the contributions from the Standard
Model at $M_H = M_Z$, the result (Cas 98) is
\begin{equation}
\begin{array}{l}
S = - 0.16 \pm 0.14 \\
T = -0.21 \pm 0.16 \\
U = + 0.25 \pm 0.24
\end{array}
\end{equation}

The parameter $T$ corresponds to $\Delta \rho$ and measures the weak
isospin breaking, and so does $U$.  In fact, the contributions of any
new isodoublet with masses $m_U$ and $m_D$ to the three parameters are
given by
\begin{equation}
\begin{array}{l}
\displaystyle
S = \frac{1}{6\pi} \left(1 - Y \log \frac{m_U^2}{m_D^2} \right) 
\\[3ex] 
\displaystyle
T = \frac{1}{4\pi\sin^2 2\vartheta_W M_Z^2} \left(m_U^2 + m_D^2 -
  \frac{2m_U^2 m_D^2}{m_U^2 - m_D^2} \log \frac{m_U^2}{m_D^2}
  \right) \\[3ex]
\displaystyle
U = \frac{1}{6\pi} \left(- \frac{5m_U^4 - 22m_U^2 m_D^2 + 5m_D^4}%
 {3(m_U^2 - m_D^2)^2} + \frac{m_U^6 - 3m_U^4 m_D^2 - 3m_U^2 m_D^4 +
   m_D^6}{(m_U^2 - m_D^2)^3} \log \frac{m_U^2}{m_D^2}
  \right) \\ 
\end{array}
\end{equation}
where $Y$ represents the hypercharge of the doublet. Since $T$ is
measured to be very small, the new up and down masses must be nearly
degenerate. In this limit, the expressions simplify:
\begin{equation}
\begin{array}{l}
\displaystyle
S = \frac{1}{6\pi} \simeq 0.05 \\[2ex]
\displaystyle
T = \frac{1}{3\pi\sin^2 2\vartheta_W} \frac{(m_U - m_D)^2}{M_Z^2}\\[2ex]
\displaystyle
U = \frac{2}{15\pi} \frac{(m_U - m_D)^2}{m_U^2}
\end{array}
\end{equation}
While the contributions to $T$ and $U$ are suppressed for nearly
degenerate isodoublet masses, this is different for $S$; for a multiplet
of heavy degenerate chiral fermions:
\begin{equation}
S = \frac{N_c}{3\pi} \sum_i \left( I_{3L}^i - I_{3R}^i \right)^2
\end{equation}

A very important application of these measurements scrutinizes
technicolor theories of electroweak symmetry breaking. In these
approaches the doublets add up to a value as large as $S \simeq 1.6$
if they are formulated as QCD-type theories, in disagreement with the
experimentally observed value for $S$. This discrepancy can be avoided
only if the non-perturbative dynamics is extended to high scales as
formulated in walking-technicolor scenarios.

Supersymmetric theories however pass this control gate in a natural way:
Since the new degrees of freedom decouple for asymptotic values of the
masses, the contribution to the $S$, $T$, $U$ parameters can be reduced
to negligible size.\\


\noindent
{\bf c)} \underline{$W$ interactions and non-Abelian gauge symmetry.}
The non-Abelian gauge symmetry $SU(2) \times U(1)$ of the electroweak
interactions predicts the form and the strength of the trilinear $\gamma
W^+W^-$ and $ZW^+W^-$ vertices. In the most general scenarios these
couplings are described each by seven parameters (Gae 79).  Assuming
$C$, $P$ and $T$ invariance in the pure electroweak boson sector, the
number of static parameters is reduced to three,
\begin{equation}
{\cal L}_k/ig_k = g^1_k W^{\ast}_{\mu\nu} W_{\mu} A_{\nu}^k + 
\kappa_k W^{\ast}_{\mu} W_{\nu} F_{\mu\nu}^k + 
\frac{\lambda_k}{M_W^2} W^{\ast}_{\rho\mu} W_{\mu\nu} F_{\nu\rho}^k
\end{equation}
for $k = \gamma$, $Z$ with $g_{\gamma} = e$ and $g_Z = e \cot
\vartheta_W$. The parameters $g_{\gamma}^1 = 1$, $g_Z^1 = 1 + \Delta
g_Z^1$, $\kappa_k = 1 + \Delta \kappa_k$ and $\lambda_k$ can be
identified with the $\gamma$ and $Z$ charges of the $W$ bosons and the
related magnetic dipole moments $\mu_k$ and electric quadrupole
moments $Q_k$:
\begin{equation}
\begin{array}{rl}
\displaystyle
\mu_{\gamma} = \frac{e}{2M_W} \left[ 2 + \Delta \kappa_{\gamma} +
  \lambda_{\gamma} \right] 
\quad & \quad \displaystyle
\mu_{Z} = \frac{e \cot \vartheta_W}{2M_W} \left[ 2 + \Delta g_Z^1 + 
  \Delta \kappa_{Z} + \lambda_{Z} \right] 
 \\[2ex]
\displaystyle
Q_{\gamma} = -\frac{e}{M_W^2} \left[ 1 + \Delta \kappa_{\gamma} -
  \lambda_{\gamma} \right] 
\quad & \quad \displaystyle
Q_{Z} = -\frac{e}{M_W^2} \left[ 1 + \Delta \kappa_Z - \lambda_Z \right] 
\end{array}
\end{equation}
The gauge symmetries of the Standard Model demand $g_k^1 = 1$,
$\kappa_k = 1$ and $\lambda_k = 0$. The magnetic dipole and the
electric quadrupole moments can be measured {\it directly} in the
production of $W\gamma$ and $WZ$ pairs at $p\bar{p}$ and $pp$
colliders and $WW$ pairs at $e^+e^-$ and $\gamma \gamma$ colliders.

Detailed experimental analyses have been carried out for the reaction
$e^+e^- \rightarrow W^+W^- \rightarrow (l\nu_l)(q\bar{q}')$. Presently
the bounds from LEP2 are of the order of a few percent, those from the
Tevatron slightly larger. Bounds on $\Delta \kappa$ and $\lambda$ which
are expected at $e^+e^-$ colliders (Zer 99) at $\sqrt{s} = 500$ GeV, 
\begin{equation}
\begin{array}{rl}
\quad  & \quad \Delta g_Z^1 = 2.5 \times 10^{-3} \\
\Delta \kappa_{\gamma} = 4.8 \times 10^{-4} 
\quad  & \quad \Delta \kappa_{Z} = 7.9 \times 10^{-4} \\ 
\lambda_{\gamma} = 7.2 \times 10^{-4} 
\quad  & \quad \lambda_{Z} = 6.5 \times 10^{-4} 
\end{array}
\end{equation}
are significantly better than the bounds expected from the LHC.
High-precision tests can therefore be performed in the near future on
one of the most basic symmetries of the electroweak interactions.


\subsection{\sc Dynamics of the Higgs sector}

The Higgs sector (Hig 64, Eng 64, Gur 64) is the cornerstone for the
consistent field-theoretic description of the electroweak interactions
which include non-zero masses for the gauge bosons. The Higgs
mechanism manifests itself in the existence of a spin-zero particle,
the Higgs-boson, the properties of which have been discussed earlier.
The discovery of the Higgs-particle is the {\it experimentum crucis}
of the Standard Model.  The most important scattering processes in
which this particle can be searched for will therefore be summarized
briefly.

A surplus of events has recently been observed in LEP experiments (Acc
00, Bar 00) in the 4-jet channel and in the channel 2-jet $+$ missing
energy when the machine was operated at the highest possible energy
above 206 GeV. Since clearly identified $b$ quarks have been isolated in
the jets, these events are compatible with the expected Higgs signal for
$M_H = 115.0 {}^{+1.3}_{-0.9}$ GeV. However, with $2.9\,\sigma$, the
rejection of the surplus as a possible statistical fluctuation or the
increase of the significance are reserved for future experiments.


\subsubsection{\sc Higgs production channels at $e^+e^-$ colliders}

The main production mechanisms (Spi 97) for Higgs bosons in $e^+e^-$
collisions are
\begin{eqnarray}
\mbox{\it Higgs-strahlung} &  & e^+e^- \to Z^* \to ZH \\ 
\mbox{\it $WW$ fusion}     &  & e^+e^- \to \bar \nu_e \nu_e (WW) 
\to \bar \nu_e \nu_e H
\label{eq:wwfusion}
\end{eqnarray}
In the Higgs-strahlung process the Higgs boson is emitted from the
$Z$-boson line, while $WW$ fusion is a formation process of Higgs bosons
in the collision of two quasi-real $W$ bosons radiated off the electron
and positron beams.

In the LEP2 experiments, the SM Higgs has been searched for,
unsuccessfully, in the mass range up to about 112 GeV. The high-energy
$e^+e^-$ linear colliders can cover the entire Higgs mass range in the
second phase of the machines in which they will reach a total energy of
about 2 TeV.\\ 


\noindent
{\bf a)} \underline{Higgs-strahlung.} The cross section for
Higgs-strahlung can be written in a compact form as
\begin{equation}
\sigma (e^+e^- \to ZH) = \frac{G_F^2 M_Z^4}{96\pi s} \left[ v_e^2 + a_e^2
\right] \lambda^{1/2} 
\frac{\lambda + 12 M_Z^2/s}{\left[ 1- M_Z^2/s \right]^2}
\end{equation}
where $v_e$ and $a_e$ were defined in Eq.\ (\ref{eq:vfaf}) and
$\lambda = [1-(M_H+M_Z)^2/s] [1-(M_H-M_Z)^2/s]$ is the usual
two-particle phase-space function. The cross section is of the size
$\sigma \sim g_W^4/s$, i.e.\ of second order in the weak coupling, and
it scales in the squared energy.

\begin{figure}[hbt]
\vspace*{-4.0cm}
\hspace*{-2.0cm}
\epsfxsize=20cm \epsfbox{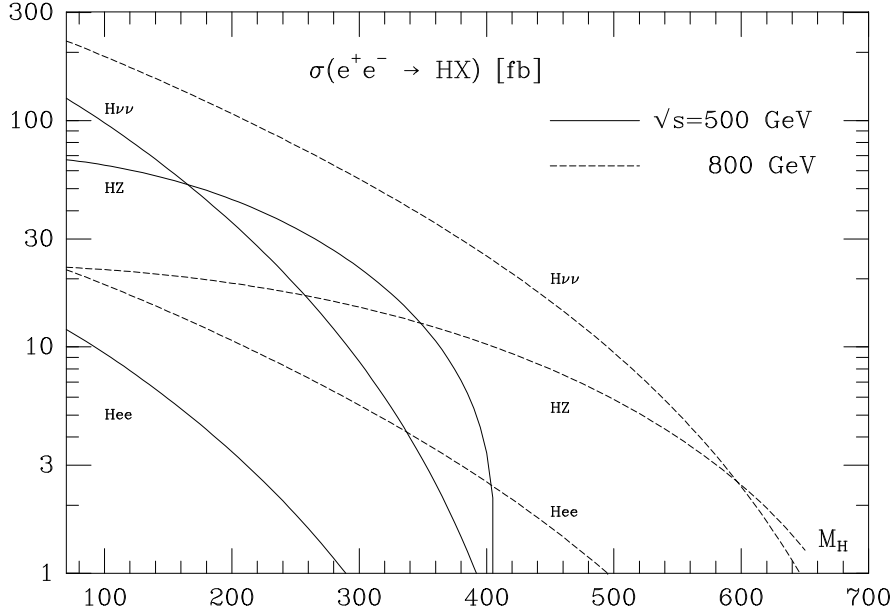}
\caption[]{\label{fg:eehx} \it The cross section for the production of
  SM Higgs bosons in Higgs-strahlung $e^+e^-\to ZH$ and $WW/ZZ$ fusion
  $e^+e^- \to \bar \nu_e \nu_e/e^+e^- H$; solid curves: $\sqrt{s}=500$
  GeV, dashed curves: $\sqrt{s}=800$ GeV; (Spi 97).}
\end{figure}

Since the cross section vanishes for asymptotic energies, the
Higgs-strahlung process is most useful for the search for Higgs bosons
in the range where the collider energy is of the same order as the
Higgs mass, $\sqrt{s} {\raisebox{-0.13cm}{~\shortstack{$>$ \\[-0.07cm]
      $\sim$}}~} {\cal O} (M_H)$.  The size of the cross section is
illustrated in Fig.\ \ref{fg:eehx} for an $e^+e^-$ linear collider
with the energy $\sqrt{s}=500$ GeV as a function of the Higgs mass.
Since the recoiling $Z$ in the two-body reaction $e^+e^- \to ZH$ is
mono-energetic, the mass of the Higgs boson can be reconstructed from
the energy of the $Z$ boson, $M_H^2 = s -2\sqrt{s}E_Z + M_Z^2$,
without any need to analyze the decay products of the Higgs boson. For
leptonic
$Z$ decays, missing-mass techniques provide a very clear signal. \\


\noindent
{\bf b)} \underline{$WW$ fusion.}  Also the cross section for the fusion
process (\ref{eq:wwfusion}) can be cast implicitly into a compact form:
\begin{eqnarray}
\sigma (e^+e^-\to\bar \nu_e \nu_e H) & = & \frac{G_F^3
  M_W^4}{4\sqrt{2}\pi^3} 
\int_{\kappa_H}^1 dx \int_x^1 dy \frac{1}{[1+(y-x)/\kappa_W ]^2}f(x,y) 
\\[3ex]
f(x,y) & = & \left( \frac{2x}{y^3} - \frac{1+3x}{y^2} + \frac{2+x}{y} -1
\right) 
\left[ \frac{z}{1+z} - \log (1+z) \right] + \frac{x}{y^3}
\frac{z^2(1-y)}{1+z}  
\nonumber
\end{eqnarray}
with $\kappa_{H,W} = M_{H,W}^2/s$ and $z = y(x - \kappa_H) /
(\kappa_Wx)$.

Since the fusion process is a $t$-channel exchange process, the size
is set by the Compton wavelength of the $W$, suppressed however with
respect to Higgs-strahlung by the third order of the electroweak
coupling, $\sigma \sim g_W^6/M_W^2$.  As a result, $WW$ fusion becomes
the leading production process for Higgs particles at high energies.
At asymptotic energies the cross section simplifies to
\begin{equation}
\sigma (e^+e^- \to \bar \nu_e \nu_e H) \to \frac{G_F^3
  M_W^4}{4\sqrt{2}\pi^3} \left[ \log\frac{s}{M_H^2} - 2 \right] 
\end{equation}
In this limit, $WW$ fusion to Higgs bosons can be interpreted as a
two-step process: the $W$ bosons are radiated as quasi-real particles
from electrons and positrons, $e \to \nu W$, with the Higgs bosons
formed subsequently in the colliding $W$ beams.

The size of the fusion cross section is compared with Higgs-strahlung in
Fig.\ \ref{fg:eehx}. At $\sqrt{s}=500$ GeV the two cross sections are of
the same order, yet the fusion process becomes increasingly important
with rising energy.\\


\noindent
{\bf c)} \underline{$\gamma \gamma$ fusion.}  The production of Higgs
bosons in $\gamma\gamma$ collisions can be exploited to determine
important properties of these particles, in particular the two-photon
decay width. The $H\gamma\gamma$ coupling is mediated by loops of
charged particles.  If the mass of the loop particle is generated
through the Higgs mechanism, the decoupling of the heavy particles is
lifted and the $\gamma\gamma$ width reflects the spectrum of these
states with masses possibly far above the Higgs mass.

The two-photon width is related to the production cross section for
polarized $\gamma$ beams by
\begin{equation}
\sigma (\gamma\gamma \to H) = \frac{16 \pi^2 \Gamma(H\to \gamma\gamma)}
{M_H} \times BW(s_{\gamma \gamma} - M_H^2)
\end{equation}
where $BW$ denotes the Breit-Wigner resonance factor as a function of
the $\gamma \gamma$ energy. For narrow Higgs bosons the observed cross
section is found by folding the $\gamma \gamma$ cross section with the
invariant $\gamma\gamma$ energy flux $\tau d{\cal L}^{\gamma\gamma} /
d\tau$ at $\tau=M_H^2/s_{ee}$.

The event rate for the production of Higgs bosons in $\gamma\gamma$
collisions of Weizs\"acker-Williams photons is too small to play a
role in practice. However, the rate is sufficiently large if the
photon spectra are generated by Compton back-scattering of laser
light. The $\gamma\gamma$ invariant energy in such a Compton collider
is nearly of the same size as the parent $e^+e^-$ energy and the
luminosity is expected to be only slightly smaller than the luminosity
in $e^+e^-$ collisions. In the Higgs mass range between 100 and 150
GeV, the final state consists primarily of $b\bar b$ pairs.  The large
$\gamma\gamma$ continuum background is suppressed in the
$J_z^{\gamma\gamma}=0$ polarization state. For Higgs masses above 150
GeV, $WW$ final states become dominant, supplemented by $ZZ$ final
states above the $ZZ$ decay threshold in the ratio 1:2.  While the
continuum $WW$ background in $\gamma\gamma$ collisions is very large,
the $ZZ$ background appears controllable for masses up to order 300 
GeV.\\


\subsubsection{\sc Higgs production at hadron colliders}

Several processes can be exploited to produce Higgs particles in hadron
colliders (Spi 97): \\

\begin{tabular}{lll}
\hspace*{21mm} 
& {\it gluon fusion} & $gg\to H$ \\ 
& {\it $WW,ZZ$ fusion}           & $W^+ W^-, ZZ \to H$ \\ 
& {\it Higgs-strahlung off $W,Z$} & $q\bar q \to W,Z \to W,Z + H$ \\ 
& {\it Higgs bremsstrahlung off top} & $q\bar q, gg \to t\bar t + H$
\end{tabular} \\[0.5cm]
While gluon fusion plays a dominant role throughout the entire Higgs
mass range of the Standard Model, the $WW$ and $ZZ$ fusion processes
become increasingly important with rising Higgs mass. The last two
radiation processes are relevant only for light Higgs masses.

The production cross sections at hadron colliders, at the LHC in
particular, are quite sizeable so that a large sample of SM Higgs
particles can be produced in this machine. Experimental difficulties
arise from the huge number of background events that come along with the
Higgs signal events.  This problem will be tackled by either triggering
on leptonic decays of $W,Z$ and $t$ in the radiation processes or by
exploiting the resonance character of the Higgs decays $H\to
\gamma\gamma$ and $H\to ZZ \to 4\ell^\pm$.  In this way, the Tevatron is
expected to search for Higgs particles in the mass range above that of
LEP2, extended to $\sim 180$ GeV only for high-luminosity operation.
The LHC is expected to cover the entire canonical Higgs mass range $M_H
{\raisebox{-0.13cm}{~\shortstack{$<$ \\[-0.07cm] $\sim$}}}~ 700$ GeV of
the Standard Model (ATL 99, CMS 94).\\

\begin{figure}[hbt]
\vspace*{0.5cm}
\begin{center}
\setlength{\unitlength}{1pt}
\begin{picture}(180,90)(0,0)
\Gluon(0,20)(50,20){3}{5}
\Gluon(0,80)(50,80){3}{5}
\ArrowLine(50,20)(50,80)
\ArrowLine(50,80)(100,50)
\ArrowLine(100,50)(50,20)
\DashLine(100,50)(150,50){5}
\put(155,46){$H$}
\put(25,46){$t,b$}
\put(-15,18){$g$}
\put(-15,78){$g$}
\end{picture} 
\vspace*{-0.5cm}
\setlength{\unitlength}{1pt}
\caption[ ]{\label{fg:gghlodia} \it The formation of Higgs bosons in
  gluon-gluon collisions.} 
\end{center}
\end{figure}
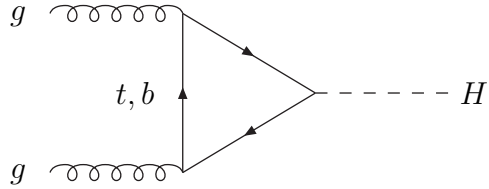

\noindent
{\bf a)} \underline{Gluon fusion.}  The gluon-fusion mechanism
\begin{displaymath}
pp \to gg \to H
\end{displaymath}
provides the dominant production mechanism of Higgs bosons at the LHC in
the entire relevant Higgs mass range up to about 1 TeV. The gluon
coupling to the Higgs boson in the SM is mediated by triangular loops of
top and bottom quarks as shown in Fig.\ \ref{fg:gghlodia}.  Since the
Yukawa coupling of the Higgs particle to heavy quarks grows with the
quark mass, thus balancing the decrease of the amplitude, the form
factor approaches a non-zero value for infinitely heavy quarks in the
loop. If the masses of heavy quarks beyond the third generation were
generated solely by the Higgs mechanism, these particles would add the
same amount to the form factor as the top quark in the asymptotic
heavy-quark limit.

The partonic cross section, Fig.\ \ref{fg:gghlodia}, can be expressed by
the gluonic width of the Higgs boson at lowest order:
\begin{equation}
\hat \sigma_{LO} (gg\to H) = \frac{\pi^2}{8} 
\frac{\Gamma(H \to gg)}{M_H} \times BW(\hat{s} - M_H^2)
\end{equation}
where $\hat s$ denotes the partonic cms-energy squared. In the
narrow-width approximation, the hadronic cross section can be cast
into the form
\begin{equation}
\sigma_{LO} (pp\to H) = \frac{\pi^2}{8s} \frac{\Gamma(H \to gg)}{M_H}
\frac{d{\cal L}^{gg}}{d\tau} 
\end{equation}
with $d{\cal L}^{gg}/d\tau$ denoting the $gg$ luminosity of the $pp$
collider, evaluated for the Drell-Yan variable $\tau = M_H^2/s$, where
$s$ is the total hadronic energy squared.  

The size of the radiative QCD corrections can be parametrized by
defining the $K$ factor as $K = \sigma_{NLO} / \sigma_{LO}$. The
virtual corrections $K_{virt}$ and the real corrections $K_{gg}$ for
the $gg$ collisions are of the same size, and both are large and
positive; the corrections for $q\bar q$ collisions and the $gq$
inelastic Compton contributions are less important.  After including
these higher-order QCD corrections, the theoretical uncertainty in the
prediction of the cross section is significantly reduced from a level
of ${\cal O}(100\%)$ down to a level of about 20\%.

\begin{figure}[th]
\vspace*{0.5cm}
\hspace*{1.0cm}
\begin{turn}{-90}%
\epsfxsize=9cm \epsfbox{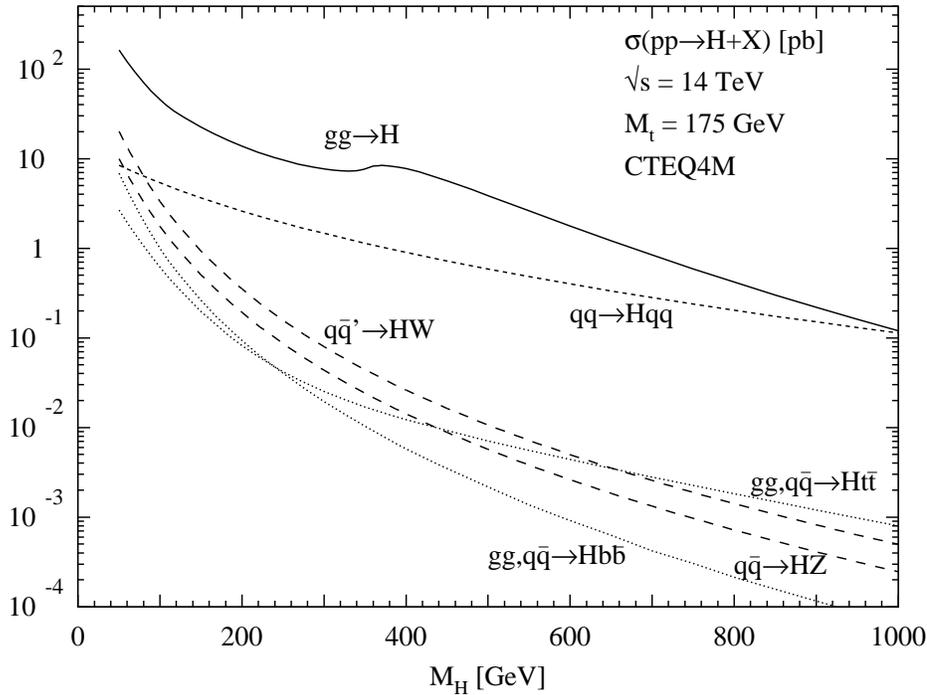}
\end{turn}
\caption[]{\label{fg:lhcpro} \it Higgs-production cross sections at the
  LHC for various production mechanisms as a function of the Higgs mass.
  QCD corrections are included except for Higgs bremsstrahlung; cf.
  (Spi 97).}
\end{figure}

The theoretical prediction for the production cross section of Higgs
particles is presented in Fig.\ \ref{fg:lhcpro} for the LHC as a
function of the Higgs mass.  The cross section decreases with increasing
Higgs mass.  This is, to a large extent, a consequence of the sharply
falling $gg$ luminosity for large invariant masses. The bump in the
cross section is induced by the $t\bar t$ threshold in the top triangle.
The overall theoretical accuracy of this calculation is expected to be
at a level of 20 to 30\%. \\


\noindent
{\bf b)} \underline{Vector-boson fusion.}  The second important channel
for Higgs production at the LHC is vector-boson fusion, $W^+W^- \to H$.
For large Higgs masses this mechanism becomes competitive with gluon
fusion; for intermediate masses the cross section is smaller by about
an order of magnitude.

For large Higgs masses, the two electroweak bosons $W$, $Z$ which form
the Higgs boson, are predominantly polarized longitudinally. At high
energies, the equivalent particle spectra of the longitudinal $W$, $Z$
bosons in quark beams are given by
\begin{eqnarray}
f^W_L (x) & = & \frac{G_F M_W^2}{2\sqrt{2}\pi^2} \frac{1-x}{x} 
 \label{eq:xyz} \\[3ex]
f^Z_L (x) & = & \frac{G_F M_Z^2}{8\sqrt{2}\pi^2}
\left[v_q^2 + a_q^2\right] \frac{1-x}{x}
\nonumber 
\end{eqnarray}
where $x$ is the fraction of energy transferred from the quark to the
$W$, $Z$ boson in the splitting process $q\to q +W/Z$. Denoting
the parton cross section for $WW,ZZ\to H$ by $\hat \sigma_0$ with
\begin{equation}
\hat \sigma_0(VV\to H) = \sqrt{2} \pi G_F \delta\left(1 - 
M_H^2/M_{VV}^2 \right)
\end{equation}
the cross section for Higgs production in quark-quark collisions is
given by
\begin{equation}
\hat \sigma(qq\to qqH) = \sqrt{2} \pi G_F \frac{d{\cal L}^{VV}}{d\tau_{VV}}
\label{eq:vvhpart}
\end{equation}
with $\tau_{VV} = M_{VV}^2 /\hat{s}$.  The hadronic cross section is
finally obtained by summing the parton cross section (\ref{eq:vvhpart})
over the flux of all possible pairs of quark-quark and antiquark
combinations:
\begin{equation}
\sigma (qq'\to VV \to H) = \int_{M_H^2/s}^1 d\tau \sum_{qq'}
\frac{d{\cal L}^{qq'}}{d\tau} \hat \sigma (qq'\to qq'H; \hat s = \tau s) 
\end{equation}

Since to lowest order the proton remnants are color singlets in the
$WW$, $ZZ$ fusion processes, no color will be exchanged, at
next-to-leading order, between the two quark lines from which the two
vector bosons are radiated. As a result, the leading QCD corrections to
these processes are already accounted for by the corrections to the
quark parton densities.

The sum of the $WW$ and $ZZ$ fusion cross sections for Higgs bosons at
the LHC is shown in Fig.\ \ref{fg:lhcpro}. The process is apparently
most important in the upper range of Higgs masses, where the cross
section approaches values
close to gluon fusion. \\


\noindent
{\bf c)} \underline{Higgs-strahlung off vector bosons.}  Higgs-strahlung
$q\bar q \to V^* \to VH~(V=W,Z)$ is a very important mechanism (cf.\ 
Fig.\ \ref{fg:lhcpro}) for the search of light Higgs bosons at the
hadron colliders Tevatron and LHC. Though the cross section is smaller
than for gluon fusion, leptonic decays of the electroweak vector bosons
are extremely useful to filter Higgs signal events out of the huge
background. Since the dynamical mechanism is the same as for $e^+e^-$
colliders, except for the folding with the quark-antiquark densities,
intermediate steps of the calculation need not be noted in detail, and
the final values of the cross sections for the Tevatron and the LHC are
recorded in Fig.\ \ref{fg:lhcpro}. \\ 


\noindent
{\bf d)} \underline{Higgs bremsstrahlung off top quarks.}  Also the
processes $gg,q\bar q \to t\bar t H$ are relevant only for small
Higgs-boson masses (cf.\ Fig.\ \ref{fg:lhcpro}).  The analytical
expression for the parton cross section, even at lowest order, is quite
involved, so that just the final results for the LHC cross section are
shown in Fig.\ \ref{fg:lhcpro}.

Higgs bremsstrahlung off top quarks is an interesting process for
measurements of the $Htt$ Yukawa coupling. The cross section $\sigma
(pp\to t\bar t H)$ is directly proportional to the square of this
fundamental coupling.\\ 

\noindent
\underline{Summary.}  An overview of the production cross sections for
Higgs particles at the LHC is presented in Fig.\ \ref{fg:lhcpro}.
Three classes of channels can be distinguished. The gluon fusion of
Higgs particles is a universal process, dominant over the entire SM
Higgs mass range. Higgs-strahlung off electroweak $W$ and $Z$ bosons
or top quarks is prominent for light Higgs bosons. The $WW$ and $ZZ$
fusion channels, by contrast, become increasingly important for large
values of the SM Higgs boson.

The signatures for the search of Higgs particles are dictated by the
decay branching ratios. In the lower part of the intermediate mass
range between 100 and 180 GeV, resonance reconstruction in
$\gamma\gamma$ final states and $b\bar b$ jets can be exploited. In
the upper part of the intermediate mass range, decays to pairs of
virtual and real $ZZ$ and $WW$ bosons are important, with the two
electroweak bosons decaying leptonically.  In the mass range above the
on-shell $ZZ$ decay threshold, the charged-lepton decays $H\to ZZ \to
4\ell^\pm$ provide gold-plated signatures. At the upper end of the
classical Standard Model Higgs-mass range, decays to neutrinos and
jets, generated in $W$ and $Z$ decays, complete the search techniques.\\


\section{\sc Summary and Perspectives}

Ingenious theoretical and experimental developments have led to a clear
picture of the structure of Nature, formulated at microscopic distances.
The Standard Model of particle physics is based on quantum field theory.
The fundamental constituents of matter are leptons and quarks. Abelian
and non-Abelian gauge symmetries are the platform on which the
electroweak and strong forces are built up. The particle masses are
generated by interactions with a scalar field. \\

A large variety of experiments in hadron-hadron scattering,
lepton-nucleon scattering and electron-positron annihilation have
established essential elements of this picture in the past decades.  The
ensemble of three lepton-quark families has been completed.  The $SU(3)
\times SU(2) \times U(1)$ gauge symmetry has been established in the
interactions of leptons and quarks with the gauge fields, partly with
accuracies at the per-mille level. The same level of experimental
accuracy can be envisaged for the interactions among the gauge bosons
when the future proton-proton collider LHC and prospective
electron-positron linear colliders will operate in the TeV energy range.
The most pressing experimental problem within the Standard Model however
appears to be the Higgs mechanism which has been introduced to generate
the masses of the fundamental particles in a way consistent with the
gauge symmetries. The production processes for the Higgs particle at the
LHC will cover the entire canonical mass range of the Standard Model in
the near future and $e^+e^-$ colliders can scrutinize the dynamics of
the Higgs mechanism. \\

Even though the Standard Model provides a valid frame for physics from
microscopic to macroscopic scales, the model cannot be considered as
{\it ultima ratio} of Nature. (i) The separate electro--weak and strong
gauge symmetries should be unified in a simple group. Attempts to unify
$SU(3) \times SU(2) \times U(1)$ in $SU(5)$ at scales of order $10^{16}$
GeV predict the electroweak mixing angle approximately in the correct
range $\sin^2 \vartheta_W \sim 0.2$. This signals a step in the right
direction even though the picture must be considered oversimplified for
many reasons. Embedding the Standard Model at the TeV scale in a
supersymmetric theory in which fermionic and bosonic degrees of freedom
are unified, offers a solution for many of the conceptual problems. The
coexistence of vastly different scales, the Planck scale at $M_{\rm Pl}
\sim 10^{19}$ GeV and the electroweak scale $v/\sqrt{2} \sim 174$ GeV,
is stabilized by this theory and the proton decay can be suppressed.
(ii) Gravity is attached {\it ad hoc} to the Standard Model and not
incorporated properly in the theory. Again, supersymmetry formulated as
a local symmetry provides a first {\it raison d'\^etre} for the spin-2
gravitational field. (iii) A consistent quantum theory of gravity may
break the frame of local field theory. Theories of extended superstrings
offer the solution of this problem. Moreover, they unify gravity with
particle physics. This approach should provide answers to many problems
of the Standard Model: the symmetry structure; the family structure; the
prediction of the large number of mass parameters and couplings. Such a
theory to which the Standard Model would be an essential element, could
describe Nature between the Planck length of $10^{-33}$ cm and the scale
$10^{+28}$ cm of the Universe. \\

Future scattering experiments at TeV energies will be crucial for
establishing such a comprehensive theory of matter and forces.\\[1.5ex]


\section{\sc References}

\setlength{\parindent}{0mm}

Abachi S et al (1995) (D0 Collaboration) {\it Phys.\ Rev.\ Lett.}
  {\bf 74} 2632

Abbott B et al (1998) (D0 Collaboration) {\it Phys.\ Rev.} {\bf D58}
052001

Abbott B et al (2000) (D0 Collaboration) {\it Phys.\ Rev.} {\bf D62}
092006

Abe F et al (1995) (CDF Collaboration) {\it Phys.\ Rev.\ Lett.} {\bf
    74} 2626

Abe F et al (1999) (CDF Collaboration) {\it Phys.\ Rev.} {\bf D59}
052002

Acciari M et al (2000) (L3 Collaboration) Report CERN-EP/2000-140 
[hep-ex/0011043] 

Accomando E et al (1998) {\it Phys.\ Rep.} {\bf 299} 1

Adloff C et al (2000) (H1 Collaboration) {\it Eur.\ Phys.\ J.} {\bf
  C13} 609
  
Alles W, Boyer C and Buras A (1977) {\it Nucl. Phys.} {\bf B119} 125

Altarelli G and Isidori G (1994) {\it Phys.\ Lett.} {\bf B337} 141 
  
Anderson C D (1937) {\it Phys.\ Rev.} {\bf 51} 884 
  
Ankenbrandt C M et al (1999) {\it Phys.\ Rev.\ S T Accel.\ Beams} {\bf
  2} 081001

Arnison G et al (1983) (UA1 Collaboration) {\it Phys.\ Lett.} {\bf
  B122} 103

Arnison G et al (1983a) (UA1 Collaboration) {\it Phys.\ Lett.} {\bf
  B126} 398

ATLAS Collaboration (1999) {\it Technical Design Report} Report
CERN-LHCC 99-14

Aubert J J et al (1974) {\it Phys.\ Rev.\ Lett.} {\bf 33} 1404

Augustin J E et al (1974) {\it Phys.\ Rev.\ Lett.} {\bf 33} 1406

Autin B, Blondel A and Ellis J (1999) (eds) Report CERN 99-02

Bagnaia P et al (1983) (UA2 Collaboration) {\it Phys.\ Lett.} {\bf
  B129} 130
  
Banner M et al (1983) (UA2 Collaboration) {\it Phys.\ Lett.} {\bf
  B122} 476

Barate R et al (2000) (ALEPH Collaboration) Report CERN-EP/2000-138 
[hep-ex/0011045] 

Barber D P et al (1979) (MARK-J Collaboration) {\it Phys.\ Rev.\ 
  Lett.} {\bf 43} 830

Bardin D and Passarino C (1999) {\it The Standard Model in the
  Making}, International Series of Monographs on Physics 104, Oxford

Berger C et al (1979) (PLUTO Collaboration) {\it Phys.\ Lett.} {\bf
  B86} 418

Bjorken J D and Paschos E A (1970) {\it Phys.\ Rev.} {\bf D1} 3151

Brandelik R et al (1979) (TASSO Collaboration) {\it Phys.\ Lett.}
{\bf B86} 243

Breitweg J et al (2000) (ZEUS Collaboration) {\it Eur.\ Phys.\ J.}
{\bf C12} 411

Cabibbo N (1963) {\it Phys. Rev. Lett.} {\bf 10} 531

Cabibbo N, Maiani L, Parisi G and Petronzio R (1979) {\it Nucl.\ 
  Phys.} {\bf B158} 295

Caso C et al (1998) (Particle Data Group) {\it Eur.\ J.\ Phys.} {\bf
  C3} 1
  
CMS Collaboration (1994) {\it Technical Proposal} Report CERN-LHCC
94-38

Cornwall J M, Levin D M and Tiktopoulos G (1974) {\it Phys.\ Rev.}
{\bf D10} 1145

Danby G, Gaillard J-M, Goulianos K, Lederman L M, Mistry N, Schwartz M
and Steinberger J (1962) {\it Phys.\ Rev.\ Lett.} {\bf 9} 36
  
Dirac P A M (1927) {\it Proc.\ Roy.\ Soc.\ (London)} {\bf A114} 710

Eichten T et al (1973) (Gargamelle Collaboration) 
{\it Phys.\ Lett.} {\bf B46} 274
  
Einstein A (1905) {\it Ann.\ Phys.} {\bf 17} 132

Englert F and Brout R (1964) {\it Phys.\ Rev.\ Lett.} {\bf 13} 321 

Fermi E (1934) {\it Nuovo Cim.} {\bf 11} 1; {\it Z.\ Phys.}  {\bf 88}
161

Feynman R P (1949) {\it Phys.\ Rev.} {\bf 76} 749
  
Feynman R P (1972) {\it Photon-Hadron Interactions}, Frontiers in
Physics, Benjamin (Reading)

Friedman J I and Kendall H W (1972a) {\it Ann.\ Rev.\ Nucl.\ Sci.} {\bf
  22} 203

Fritzsch H and Gell-Mann M (1972) Proc.\ {\it XVI International
  Conference on High Energy Physics} (Fermilab)

Fritzsch H, Gell-Mann M and Leutwyler H (1973) {\it Phys.\ Lett.} {\bf
  B47} 365

Gaemers K J and Gounaris G J (1979) {\it Z.\ Phys.} {\bf C1} 259

Gell-Mann M (1964) {\it Phys.\ Lett.} {\bf 8} 214

Glashow S L (1961) {\it Nucl.\ Phys.} {\bf 20} 579

Gross D J and Wilczek F (1973) {\it Phys.\ Rev.\ Lett.} {\bf 30} 1343

Guralnik G S, Hagen C R and Kibble T W (1964) {\it Phys.\ Rev.\ Lett.}
{\bf 13} 585
  
Hasert F J et al (1973) (Gargamelle Collaboration) {\it Phys. Lett.}
{\bf B46} 121

Hasert F J et al (1973a) (Gargamelle Collaboration) {\it Phys. Lett.}
{\bf B46} 138

Heisenberg W and Pauli W (1929) {\it Z.\ Phys.} {\bf 56} 1; (1930) {\it
  Z.\ Phys.} {\bf 59} 169
  
Herb S W et al (1977) {\it Phys.\ Rev.\ Lett.} {\bf 39} 252

Higgs P W (1964) {\it Phys.\ Lett.} {\bf 12} 132

Hollik W and Duckeck G (2000) {\it Electroweak Precision Tests at LEP},
Springer Tracts in Modern Physics 162 (Springer Verlag, Heidelberg)

Jordan P and Pauli W (1928) {\it Z.\ Phys.} {\bf 47} 151

Kaufmann W (1897) {\it Ann.\ Phys.} {\bf 61} 544

Kaufmann W and Aschkinass E (1897a) Ann.\ Phys.\ {\bf 62} 588

Kobayashi M and Maskawa T (1973) {\it Prog. Theor. Phys.} {\bf 49} 652

Lee B W, Quigg C and Thacker H B (1977) {\it Phys.\ Rev.} {\bf D16}
1519

LEP Collaborations (2000) LEP-EWWG/LS 2000-01
 
LEP Electroweak Working Group (2000a) {\tt
  http://lepewwg.web.cern.ch/LEPEWWG/Welcome.html}
 
Lindner M (1986) {\it Z.\ Phys.} {\bf C31} 295

Llewellyn Smith C (1973) {\it Phys.\ Lett.} {\bf 46B} 233
  
Lundberg B et al (2000) (DONUT Collaboration) \\ 
{\tt http://fn872.fnal.gov/presentation/complete/2000/WC2000.pdf} 
  
Murayama H and Peskin M I (1996) {\it Ann.\ Rev.\ Nucl.\ Part.\ Sci.}
{\bf 46} 533

Perl M et al (1975) {\it Phys.\ Rev.\ Lett.} {\bf 35} 1489
  
Peskin M E and Takeuchi T (1990) {\it Phys.\ Rev.\ Lett.} {\bf 65} 964
  
Politzer H D (1973) {\it Phys.\ Rev.\ Lett.} {\bf 30} 1346

Prescott C Y et al (1979) {\it Phys.\ Lett.} {\bf B84} 524

Reines F and Cowan C L Jr (1953) {\it Phys.\ Rev.} {\bf 92} 830
  
Renard F M (1981) {\it Basics of Electron Positron Collisions} (Editions
Fronti\`{e}res, Gif sur Yvette) 

Riesselmann K (1997) {\it School of Subnuclear Physics (Erice)} DESY
97-222

Salam A (1968) in {\it Elementary Particle Theory} Svartholm N (ed) 
(Almqvist and Wiksells, Stockholm)

Schwinger J (1948) {\it Phys.\ Rev.} {\bf 73} 416
  
Schwinger J (1973) {\it Particles, Sources and Fields} (Addison-Wesley,
Reading) 

Sehgal L M (1973) {\it Nucl.\ Phys.} {\bf B65} 141

Sher M (1989) {\it Phys.\ Rep.} {\bf 179} 273

Spira M and Zerwas P M (1997) {\it Electroweak Symmetry Breaking and
  Higgs Physics} Lecture Notes in Physics 512 (Springer Verlag,
Heidelberg)

Thomson J J (1897), Talk, The Royal Institution of Great Britain,
London; {\it Phil.\ Mag.} {\bf 44} 269 

't Hooft G (1971) {\it Nucl.\ Phys.} {\bf B33} 173; {\it Nucl.\ Phys.}
{\bf B35} 167

't Hooft G and Veltman M (1972) {\it Nucl.\ Phys.} {\bf B44} 189

Tomonaga S (1946) {\it Progr.\ Theor.\ Phys.} {\bf 1} 27

Veltman M (1977) {\it Acta Phys.\ Polon.} {\bf B8} 475

Weinberg S (1967) {\it Phys.\ Rev.\ Lett.} {\bf 19} 1264

Weyl H (1929) {\it Z.\ Phys.} {\bf 56} 330

Wiechert E (1897) Lecture, Universit\"at zu K\"onigsberg, K\"onigsberg;
Abh.\ Phys.-\"Okon.\ Ges.\ K\"onigsberg {\bf 37} 1

Yang C N and Mills R (1954) {\it Phys.\ Rev.} {\bf 96} 191

Yao W (1998) (CDF Collaboration) Proc.\ {\it 29th International
  Conference on High-Energy Physics, Vancouver} Vol {\bf 2} 1093

Yukawa H (1935) {\it Proc.\ Phys.\ Math.\ Soc.\ Japan} {\bf 17} 48

Zerwas P M (1999) {\it Physics with an $e^+e^-$ Linear Collider at High
  Luminosity} in: {\it Particle Physics: Ideas and Recent Developments}
Aubert J J et al (eds) (Kluwer Academic Publishers, Amsterdam)

Zweig G (1964) CERN-Report 8182/TH401, reprinted in Lichtenberg D B and
Rosen S P (eds) { \it Developments in the Quark Theory of Hadrons}, Vol
I, 1964--1978 (Hadronic Press, Nonatum)

\end{document}